\documentclass[useAMS,usenatbib]{mn2e}

\def \HI {H{\sc \,i}}
\def \HII {H{\sc \,ii}}
\def \DI {D{\sc \,i}}
\def \HeI {He{\sc \,i}}
\def \HeII {He{\sc \,ii}}
\def \HeIII {He{\sc \,iii}}
\def \HeIIthree {$^3$He{\sc \,ii}}
\def \CI {C{\sc \,i}}
\def \CII {C{\sc \,ii}}
\def \Cboth {C{\sc \,i}, C{\sc \,ii}}
\def \NII {N{\sc \,ii}}
\def \OI {O{\sc \,i}}

\usepackage[dvips]{graphicx}
\title[Signals of the cosmological reionization in the radio sky through
C and O fine structure lines]{Signals of the cosmological reionization
in the radio sky through C and O fine structure lines}
\author[M. Kusakabe and M. Kawasaki]{M. Kusakabe$^{1}$ and
M. Kawasaki$^{1,2}$\thanks{E-mail: kusakabe@icrr.u-tokyo.ac.jp} \\
$^{1}$Institute for Cosmic Ray Research, University of Tokyo, Kashiwa,
Chiba 277-8582, Japan\\
$^{2}$Institute for the Physics and Mathematics of the Universe,
University of Tokyo, Kashiwa, Chiba 277-8582, Japan}
\begin{document}

\date{Accepted xxx. Received xxx; in original form xxx}

\pagerange{\pageref{firstpage}--\pageref{lastpage}} \pubyear{20XX}

\maketitle

\label{firstpage}

\begin{abstract}
We study the excitation of fine structure levels of \Cboth\ and
 \OI\ by ultraviolet (UV) photons around strong UV sources which
 are also ionizing sources of the cosmological reionization at redshift
 of $\sim 10$.  The evolutions of ionized regions around a point source
 are calculated by solving rate equations for non-equilibrium chemistry.
 Signals of UV photons through the fine structure lines are considered
 to be stronger at locations of more abundant chemical species of \Cboth\
 and \OI.  Such environments would be realized where
 strong fluxes of non-ionizing UV line photons available for the pumping
 up of fine structure levels exist, and simultaneously ionizing UV
 photons are effectively shielded by dense \HI\ regions.  Signals
 from \HI\ regions of moderately large densities induced by
 redshifted UV photons emitted at the point sources are found to be
 dominantly large over those of others.  We discuss the detectability of
 the signals, and show that signals from idealized environments will be
 possibly detected by radio observations with next-generation arrays to come after the Atacama Large Millimeter/submillimeter Array (ALMA).
\end{abstract}

\begin{keywords}
atomic processes -- hydrodynamics -- ISM: \HII\ regions -- cosmology:
 observations -- dark ages, reionization, first stars -- radio lines:
 general.
\end{keywords}

\section{Introduction}\label{sec1}

At the epoch of Big Bang Nucleosynthesis of cosmic temperature
$T \ga 0.1$~MeV, light elements of
$^2$H,$^{3,4}$He and $^{7}$Li are produced in abundances larger than
$10^{-10}$ times that of hydrogen.  As the temperature decreases, light
elements recombine with electrons, and trace amounts of chemical
molecules form during the Dark Ages between the recombination and
the reionization of the universe~\citep{von2009}.  The universe
is then considered to experience the reionization at temperature
of $\sim 30$~K, i.e., cosmic redshift of $z\sim 10$~\citep{lar2010}.
Sources of the reionization are perhaps ultraviolet (UV) photons from quasi-stellar objects (QSOs),
young galaxies and hypothetical low metallicity objects in the early
universe, i.e., Population III stars, or some exotic energy injection
processes triggered, for example, by the particle decay.  If stars
including Population III have contributed to the reionization,
the metal enrichment in the universe would occur simultaneously.  The
reionization history as well as the metal enrichment of the universe is,
however, not yet determined precisely.  Many suggestions have been made for investigations of
the cosmic reionization using atomic physics with future observations in
radio frequency.

The first method is observations of intergalactic medium (IGM) through
the redshifted 21~cm line of neutral hydrogen~\citep*{mad1997}.  The
neutral gas regions could be quickly preheated to temperature larger than
that of cosmic background radiation (CBR) in the presence of a strong
flux of UV photons through the photoionization heating~\citep[e.g.,][]{che2004}.  In such environments, the spin temperature
determined by the population fractions of the upper and lower states of
hyperfine levels is forced to deviate from the CBR temperature by
level-mixing via Ly$\alpha$ photon scattering, i.e., Wouthuysen-Field
effect~\citep{wou1952,fie1958}.  There would be brief chances of
observing them in
absorption, and long chances in emission against the CBR.  Future radio
observations at meter wavelengths of IGM would
produce plenty of information on the epoch, picture and sources of the
reionization.

Although the 21~cm signal of the reionization of the
universe is expected to be relatively strong, the strong contamination
from Galactic and extragalactic foregrounds makes it difficult to observe
the absolute signal over the whole sky except for fluctuations around it.
Measurements of angular fluctuations in the \HI\ signal are
very difficult, while fluctuations in frequency would be easily
separated and be detectable with future large
arrays~\citep{gne2004}.  As for angular fluctuation,
\citet*{dim2004} have shown that angular fluctuations in 21~cm
emission at $\theta \ga 1\arcmin$ can be detected if efficient
subtraction of foreground sources is performed successfully.  The 21~cm
signal not by Wouthuysen-Field effect but by an effect of collision in
earlier universe of higher densities has also been suggested to provide
a valuable informations in the epoch when the spin temperature traced the
gas kinetic temperature~\citep{loe2004}.  

Long-lived metastable 2$s$ state of hydrogen has been studied, and
signals of the reionization through 3~cm fine structure line (2$s_{1/2}
\rightarrow 2p_{3/2}$) have been found to be undetectable with existing
radio telescopes~\citep{dij2008}.  The reason is that a Ly$\beta$
photon which is the pumping source of the ground state can excite only one hydrogen atom into the 2$s$
state, while a Ly$\alpha$ photon can excite many atoms into the excited
spin state.  The signals through hyperfine transitions of \DI,
\HeIIthree\ \citep{deg1985} also exist.  \citet{sig2006} have
suggested a challenging possibility of determination of D/H abundance
ratio by looking at hyperfine structure line at high redshift of both
the H and D.  Signals from IGMs and dense objects at high redshifts through
the $^3$He hyperfine transition has been studied.  Detailed physics and
suggested observational plans were presented~\citep{mcq2009}, and the
anticorrelation of \HI\ 21 cm maps and \HeIIthree\ 3 cm
maps was suggested~\citep{bag2009}.

The infrared fine structure lines of \Cboth\ and \OI~\citep{bah1968} play
an important role in cooling of neutral regions~\citep{hol1999}.
Astronomical
observations of those lines provide rich information about physical
conditions in observed objects~\citep{kau1999}.  The signatures
of existence of heavy chemical species like C and O through resonant scatterings
with CBR imprinted on the CBR spectrum have been
investigated~\citep*{bas2004,her2006}.  With the
Planck~\footnote{http://www.rssd.esa.int/index.php?project=planck.} spacecraft, the
signatures can be searched for in the temperature anisotropy of CBR, and
constraints on abundances of heavy species will be derived from the
observations in reverse.

Idea of UV pumping through fine structure lines of metals during reionization as a way to distort the CMB during a reionization epoch has been suggested~\citep{her2007}.  They focused on the \OI\ fine structure, and provided estimations for the order of magnitude of the distortion for different levels of UV
fluxes.  The angular power spectrum of the distortion introduced by the \OI\ hosting regions at high redshift has subsequently been studied~\citep{her2008}.

Observations of bright objects in radio frequency have been
performed in search of emission lines of fine structures of heavy
chemical species.  The \CII\ 157$\mu$m line emissions of high
redshift QSOs [J1148+5251 at $z=6.42$~\citep{mai2005},
BR~1202-0725 at $z=4.7$~\citep{ion2006}] and a galaxy
[BRI~0952-0115 at $z=4.43$~\citep{mai2009}] have been detected.  \citet{wal2009} observed the high redshift quasars
(J1148+5251 at $z=6.42$) and derived an upper limit on a \NII\
205$\mu$m line luminosity.  The targets of these observations were, in
fact, star-forming interstellar medium which emits through atomic fine
structure lines.  There are dense systems detected which absorb photon spectra
of QSOs at fine structure transitions of \Cboth\ and
their physical conditions have been
constrained~\citep*{sil2002,qua2002,rei2005}.

In this paper, we focus on cosmological signals of only UV
sources without star forming activities through fine structure lines and
their detectability.  The structure of this paper is as follows.  In Section\ \ref{sec2}, the
pumping mechanism of fine structure levels and the dependence
of the pumping efficiency as a redshift are shown.  In Section\
\ref{sec3}, the model for calculation of reionization around a point
source is introduced.  In Section\ \ref{sec4}, the result of the
calculation is presented.  Intensities of signals of UV radiation from
the source through fine structure lines are estimated, and the
dependence of signals on the redshift is shown.  We discuss the
detectability of signals, and claim that signals may be detected by
future radio measurements in $O(100-1000~{\rm GHz})$ range.  In Section\
\ref{sec5}, we summarize our conclusions.

\section{Ultraviolet Photon Pumping to C\,{\sevensize\bf I},
 C\,{\sevensize\bf II} and O\,{\sevensize\bf I} Metastable States}\label{sec2}

\subsection{Populations of Fine Structure Levels}
\subsubsection{Formulation}

For a pair of two states, i.e., a stable and an unstable states, the
spin temperature $T_{\rm S}$ is defined as
\begin{eqnarray}
\frac{n_{\rm 1}}{n_{\rm
 0}}&=&\frac{g_1}{g_0}\exp\left(-\frac{h\nu_{10}}{kT_{\rm S}}\right) \nonumber\\
&=&\frac{g_1}{g_0}\exp\left(-\frac{T_\ast}{T_{\rm S}}\right),
\label{eq1}
\end{eqnarray}
where 
$n_i$ ($i=1$ for the upper state, $i=0$ for the lower) is the number
density of the state $i$,
$g_i=2S_i+1$ is the spin degrees of freedom of state $i$ with spin $S_i$,
$h$ is the Planck's constant,
$\nu_{10}$ is the frequency corresponding to the transition from the
state 1 to 0,
$k$ is the Boltzmann's constant.  At the second equality, the
temperature $T_\ast\equiv h \nu_{10}/k$ was defined.   The spin
temperature describes the ratio of occupation degree of the upper state
to that of the lower.

For a group of three states, i.e., a ground state, a first excited
state unstable to the decay into the ground state, and a second excited
state unstable to the decay only into the first excited state, the spin
temperature $T_{\rm S2}$ is defined as
\begin{eqnarray}
\frac{n_{\rm 2}}{n_{\rm
 1}}&=&\frac{g_2}{g_1}\exp\left(-\frac{T_{\ast2}}{T_{\rm S2}}\right),
\label{eq2}
\end{eqnarray}
where
the subscripts $i=1$ and $2$ are attached to physical values of the
first and second excited states, respectively,
the temperature $T_{\ast2}\equiv h \nu_{21}/k$ is defined with the
frequency for the transition from the state 2 to 1, i.e., $\nu_{21}$.
Note that the spin temperatures for second excited states are defined
using the number density of the $i=1$ and $2$ states, not $i=0$ and $2$
states.

Fine structure levels of \CI, \CII\ and \OI\ can
be excited by continuum radiation at the frequency corresponding to the
energy levels of excited states, UV radiation at the frequency
corresponding to the energy levels of intermediate bound states, and by
collisions.  For two
discrete fine structure energy levels denoted by 0 and 1, the rate
equation for their population is written as
\begin{eqnarray}
-\frac{dn_0}{dt}=\frac{dn_1}{dt}&\hspace{-0.8em}=&\hspace{-0.8em}A_{10}\left[\frac{c^2 J_{\rm B}
					  }{2h\nu^3} \frac {g_1}{g_0}
					  n_0 -\left(1+\frac{c^2 J_{\rm
						B}}{2h\nu^3}\right)n_1\right]
\nonumber \\
&\hspace{-0.8em}&\hspace{-0.8em}+n_0 P_{01}^{\rm UV}-n_1 P_{10}^{\rm UV}+n_0 C_{01}-n_1 C_{10},
\label{eq3}
\end{eqnarray}
where
$A_{10}$ is the Einstein $A$-coefficient of the $1\rightarrow 0$ transition,
$\nu$ is the frequency of the transition and
$C_{10}$ and $C_{01}$ are the collisional rates of the
$1\rightarrow 0$ and $0\rightarrow 1$ transitions, respectively.
$J_{\rm B}=\int \phi(\nu) I_\nu d\nu d\Omega/(4\pi)$ is the mean
intensity of radiation, where $\phi(\nu)$ is the line profile function and
$I_\nu$ is the specific intensity at the transition frequency.  The
rates of excitation and deexcitation by UV photons, i.e., $P_{01}^{\rm UV}$ and $P_{01}^{\rm UV}$ are given by
\begin{equation}
P_{01}^{\rm UV}=\sum_j \left(\frac{c^2}{2  h \nu_{j0}^3}\right)
 \left(\frac{A_{j0} J_{j0} g_j}{g_0}\right) \frac{A_{j1}}{\sum_i A_{ji}},
\end{equation}
\begin{equation}
P_{10}^{\rm UV}=\sum_j \left(\frac{c^2}{2  h \nu_{j1}^3}\right)
 \left(\frac{A_{j1} J_{j1} g_j}{g_1}\right) \frac{A_{j0}}{\sum_i A_{ji}},
\end{equation}
where
$j$ indicates excited levels above the fine structure multiplets and
$i$ is either 0 or 1.  There are several or more excited levels
of \Cboth\ and \OI\ where transitions from the ground
state fine structure levels by electric dipole interaction are allowed.
Although many lines could contribute to populating excited states of
fine structures, we treat only transitions corresponding to the
lowest energies for respective species for the moment.  All of the permitted lines are considered in Section\ \ref{sec4}.2.2.

\begin{figure*}
\includegraphics[width=160mm]{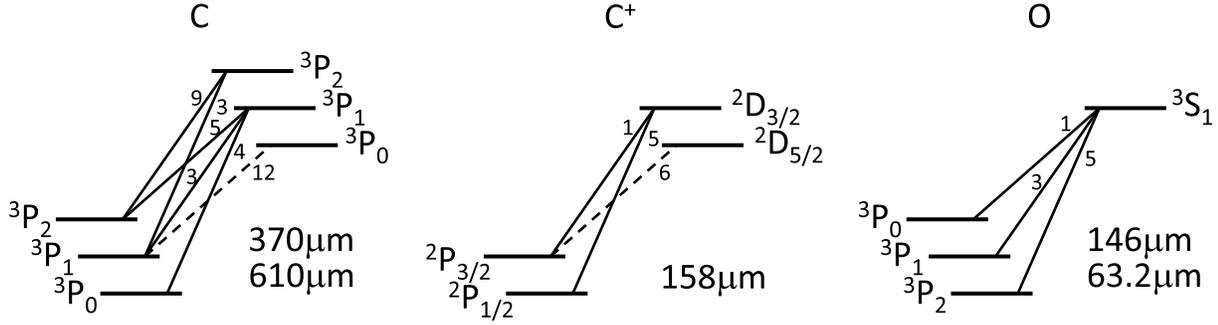}
\caption{Diagrams for ultraviolet permitted transitions which mix the ground
 state fine structures (solid lines) and do not (dashed lines).  The
 attached numbers are relative strengths of spontaneous emission rates $A$.}
\label{fig1}
\end{figure*}

Figure\ \ref{fig1} shows the transitions from the ground state fine
structures to the lowest energy levels which is connected to the fine
structures through strong electric dipole transitions.  Numbers attached
to lines are relative strengths of spontaneous emission rates $A$.

The spin temperature in the steady
state is derived from equation\ (\ref{eq3}).  We assumed the Planck
distribution of the mean intensity $J_{\rm B}$ with the temperature of cosmic
background radiation (CBR), i.e., $T_{\rm
CBR}=2.725(1+z)$~K~\citep{mat1999} with the redshift $z$.  In addition, the thermal energy distribution of gas is assumed.  The collisional transition rates are then given by 
\begin{equation}
C_{ij}=n_{\rm target} \gamma_{ij}(T_{\rm gas}),
\end{equation}
where
$n_{\rm target}$ is the number density of target species and $\gamma_{ij}$ is the collisional (de)excitation coefficient as a function of the gas temperature, i.e., $T_{\rm gas}$.  

For the \CII\ two level states, the spin temperature is
\begin{equation}
T_{\rm S}=-T_\ast\left\{\ln\left[\frac{{\rm e}^{-T_\ast/T_{\rm CBR}}+y_{\rm L}{\rm e}^{-T_\ast/T_{\rm UV}} +y_{\rm C}{\rm e}^{-T_\ast/T_{\rm gas}}}{1+y_{\rm L}+y_{\rm C}}\right]\right\}^{-1},\\
\label{eq19}
\end{equation}
where
$T_{\rm UV}$ is given by the relation
\begin{equation}
\frac{P_{10}^{\rm UV}}{P_{01}^{\rm UV}} = \frac{g_0}{g_1} \exp\left( \frac{T_\ast}{T_{\rm
				UV}}\right),
\end{equation}
~\citep{fie1958,deg1985}, The temperature $T_{\rm gas}$ satisfies
\begin{equation}
\frac{P_{10}^{\rm C}}{P_{01}^{\rm C}} = \frac{g_0}{g_1} \exp\left( \frac{T_\ast}{T_{\rm	gas}}\right),
\end{equation}
and
\begin{equation}
y_{\rm L}= \frac{P_{10}^{\rm UV}}{A_{10}}\left[1-\exp(-T_\ast/T_{\rm CBR})\right],
\end{equation}
\begin{equation}
y_{\rm C}= \frac{P_{10}^{\rm C}}{A_{10}}\left[1-\exp(-T_\ast/T_{\rm CBR})\right]
\end{equation}
were defined.

For three level states, the equations describing the balance between
production and destruction of the first and second excited states are
given by
\begin{equation}
n_0 P_{01}+ n_2 P_{21} = n_1 \left(P_{10}+P_{12} \right),
\end{equation}
\begin{equation}
n_0 P_{02}+ n_1 P_{12} = n_2 \left(P_{20}+P_{21} \right),
\end{equation}
where
$P_{ij}$ includes transitions $i\rightarrow j$ by all of the CBR continuum, UV radiation and particle collision.  The sum of the above two equations is equal to the
balance equation for the ground state.  These equations are solved for
number fractions of the energy states, that is
\begin{equation}
\frac{n_1}{n_0}= 
 \frac{P_{01} P_{20} + P_{21} P_{02} + P_{01} P_{21}}
 {P_{10} P_{20} + P_{12} P_{20} + P_{10} P_{21}},
\end{equation}
\begin{equation}
\frac{n_2}{n_0}= 
 \frac{P_{10} P_{02} + P_{12} P_{02} + P_{01} P_{12}}
 {P_{10} P_{20} + P_{12} P_{20} + P_{10} P_{21}}.
\end{equation}
Spin temperatures, i.e., $T_{\rm S}$ and $T_{\rm S2}$ of the first and
second excited states are calculated using equations (\ref{eq1}) and
(\ref{eq2}).

Table\ \ref{tab1} shows data on the fine structure and UV transitions
taken from~\citet{nis2008} as well as data on the collisional deexcitaion coefficients taken from~\citet{hol1989}.  As for \CI\ the difference between energy levels of $^3P_2$ and $^3P_1$ in the excited states higher than that of the ground state by $\sim 6\times 10^4$~cm$^{-1}$ is $40.51$ cm$^{-1}$~\citep{nis2008}.
\begin{table*}
 \centering
 \begin{minipage}{140mm}
  \caption{Fine structure and UV transitions}\label{tab1}
  \begin{tabular}{@{}cccccccc@{}}
  \hline
            &  \multicolumn{4}{c}{Fine structure transition} &
   \multicolumn{3}{c}{UV transition} \\
    Species & State & $\nu$ (GHz) & $A$ (s$^{-1}$) & $\gamma_{ij}^{\rm H}$ (cm$^3$~s$^{-1}$)\footnote{Deexcitation rate coefficients for collisions with \HI\ taken from~\citet{hol1989}.  $T_2=T_{\rm gas}/(10^2~{\rm K})$.} & State & $\nu$ (cm$^{-1}$) &
   $A$ (s$^{-1}$) \\
 \hline
 \CI\  & $^3$P$_1 - ^3$P$_0$ & 492 & $7.88\times 10^{-8}$ & $1.6\times 10^{-10}T_2^{0.14}$ &
   $^3$P$_1 - ^3$P$_0$ & $6.035\times 10^4$ & $1.13\times 10^8$ \\
              & $^3$P$_2 - ^3$P$_0$ & 1301 & & $9.2\times 10^{-11}T_2^{0.26}$ &
   $^3$P$_1 - ^3$P$_1$ &                    & $8.64\times 10^7$ \\
              & $^3$P$_2 - ^3$P$_1$ & 809 & $2.65\times 10^{-7}$ &  $2.9\times 10^{-10}T_2^{0.26}$ &    $^3$P$_1 - ^3$P$_2$ &                    & $1.44\times 10^8$ \\
              &  &  & & &
   $^3$P$_2 - ^3$P$_1$ &                    & $8.58\times 10^7$ \\
              &  &  & & &
   $^3$P$_2 - ^3$P$_2$ &                    & $2.52\times 10^8$ \\
 \CII\ & $^2$P$_{3/2} - ^2$P$_{1/2}$ & 1901 & $2.30\times
   10^{-6}$ & $8.0\times 10^{-10}T_2^{0.07}$ & $^2$D$_{3/2} - ^2$P$_{1/2}$ & $7.493\times 10^4$ & $2.37\times 10^8$ \\
 & & & & & $^2$D$_{3/2} - ^2$P$_{3/2}$ & & $4.74\times 10^7$ \\
 \OI\  & $^3$P$_1 - ^3$P$_2$ & 4745 & $8.91\times 10^{-5}$ & $9.2\times 10^{-11}T_2^{0.67}$ &
   $^3$S$_1 - ^3$P$_2$ & $7.679\times 10^4$ & $3.41\times 10^8$ \\
              & $^3$P$_0 - ^3$P$_2$ & 6805 & & $4.3\times 10^{-11}T_2^{0.80}$ & $^3$S$_1 - ^3$P$_1$ & & $2.03\times 10^8$ \\
              & $^3$P$_0 - ^3$P$_1$ & 2060 & $1.75\times 10^{-5}$ & $1.1\times 10^{-10}T_2^{0.44}$ & $^3$S$_1 - ^3$P$_0$ & & $6.76\times 10^7$ \\
\hline
\end{tabular}
\end{minipage}
\end{table*}

\subsubsection{Effects of UV pumping and Collisions}\label{sec212}
We compare effects of UV pumping and collisions on populations of fine structure levels.  The effects are not so large as that of CBR if the physical condition which is considered is not dense and not very near to extremely luminous objects.  First, we define $f_{ij}$ as the number ratio
\begin{equation}
f_{ij}(y_{\rm L}, T_{\rm UV}, y_{\rm C}, T_{\rm gas}, T_{\rm CBR})\equiv n_i/n_j.
\end{equation}
Using equation (\ref{eq19}) we can describe the deviation of $f_{10}$ for a two level system under a UV source and/or a collisional environment from that for no such effects by the first order Taylor expansions in $y_{\rm L}$ and $y_{\rm C}$, i.e., 
\begin{eqnarray}
&&\hspace{-1.8em}\frac{\Delta f_{10}(y_{\rm L}, T_{\rm UV}, y_{\rm C}, T_{\rm gas}, T_{\rm CBR})}{f_{10}(0, ^\forall T_{\rm UV}, 0, ^\forall T_{\rm gas}, T_{\rm CBR})}\equiv \frac{f_{10}(y_{\rm L}, T_{\rm UV}, y_{\rm C}, T_{\rm gas}, T_{\rm CBR})}{f_{10}(0, ^\forall T_{\rm UV}, 0, ^\forall T_{\rm gas}, T_{\rm CBR})}-1 \nonumber\\
&&\hspace{-1.8em}=y_{\rm L}\left[\exp\left(-\frac{T_\ast}{T_{\rm UV}}+\frac{T_\ast}{T_{\rm CBR}}\right)-1 \right] \nonumber\\
&&+ y_{\rm C}\left[\exp\left(-\frac{T_\ast}{T_{\rm gas}} +\frac{T_\ast}{T_{\rm CBR}}\right)-1 \right]\nonumber\\
&&\hspace{-1.8em}\equiv (\Delta f_{10}/f_{10})_{\rm L} +(\Delta f_{10}/f_{10})_{\rm C}.
\label{eq20}
\end{eqnarray}
The first and second terms in the left hand side are contributions of UV pumping and collisions, respectively.  The explicit expression for $f_{10}$ value is
\begin{eqnarray}
&&\hspace{-2.em}f_{10}(y_{\rm L}, T_{\rm UV}, y_{\rm C}, T_{\rm gas}, T_{\rm CBR})\nonumber\\
&=&\frac{g_1}{g_0}\left[{\rm e}^{-T_\ast/T_{\rm CBR}}+y_{\rm L}\left({\rm e}^{-T_\ast/T_{\rm UV}}-{\rm e}^{-T_\ast/T_{\rm CBR}}\right) \right.\nonumber\\
&&\hspace{1.5em}\left. +y_{\rm C}\left({\rm e}^{-T_\ast/T_{\rm gas}} -{\rm e}^{-T_\ast/T_{\rm CBR}}\right)\right].
\end{eqnarray}

\begin{figure}
\includegraphics[width=84mm]{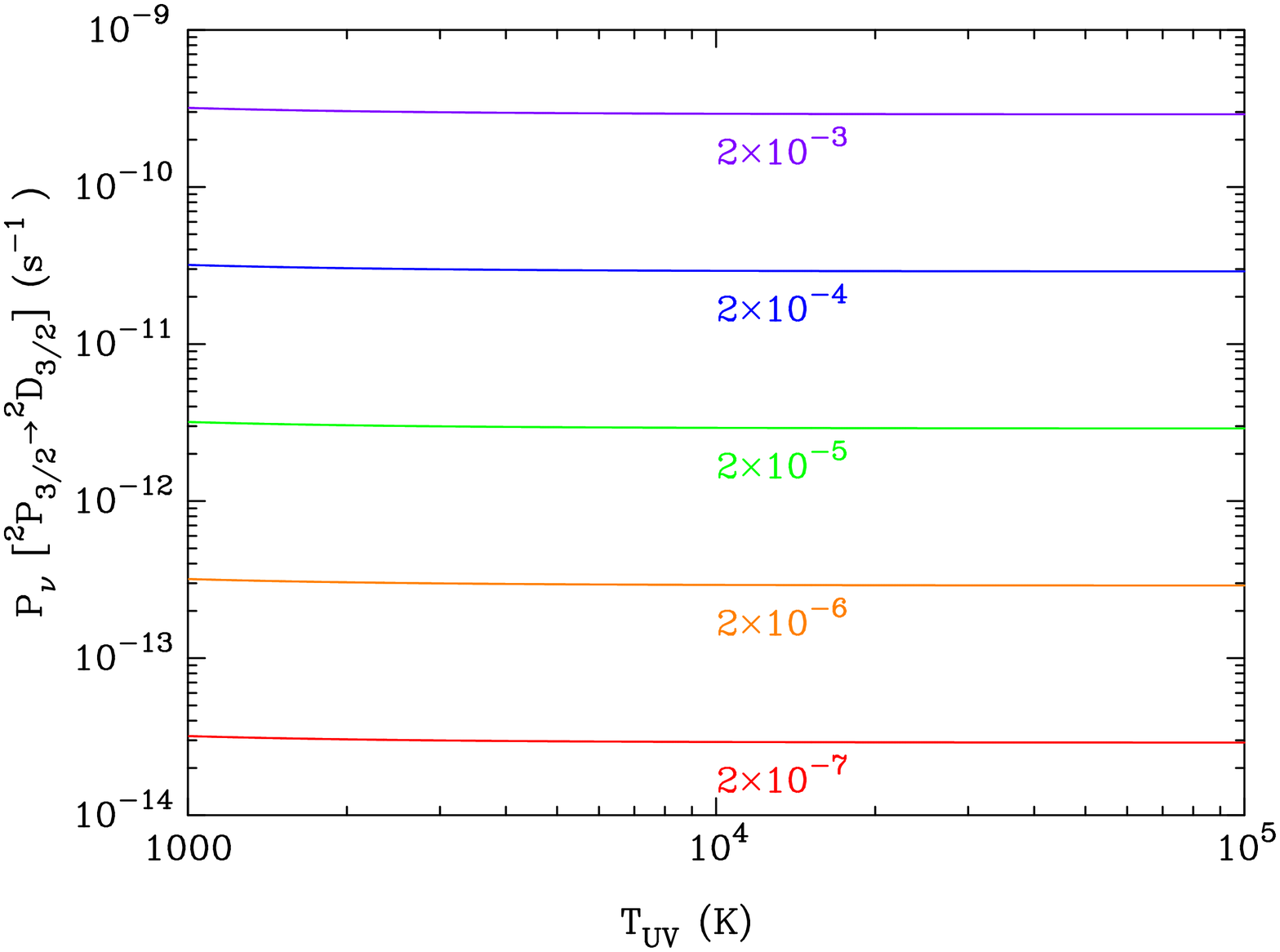}
\includegraphics[width=84mm]{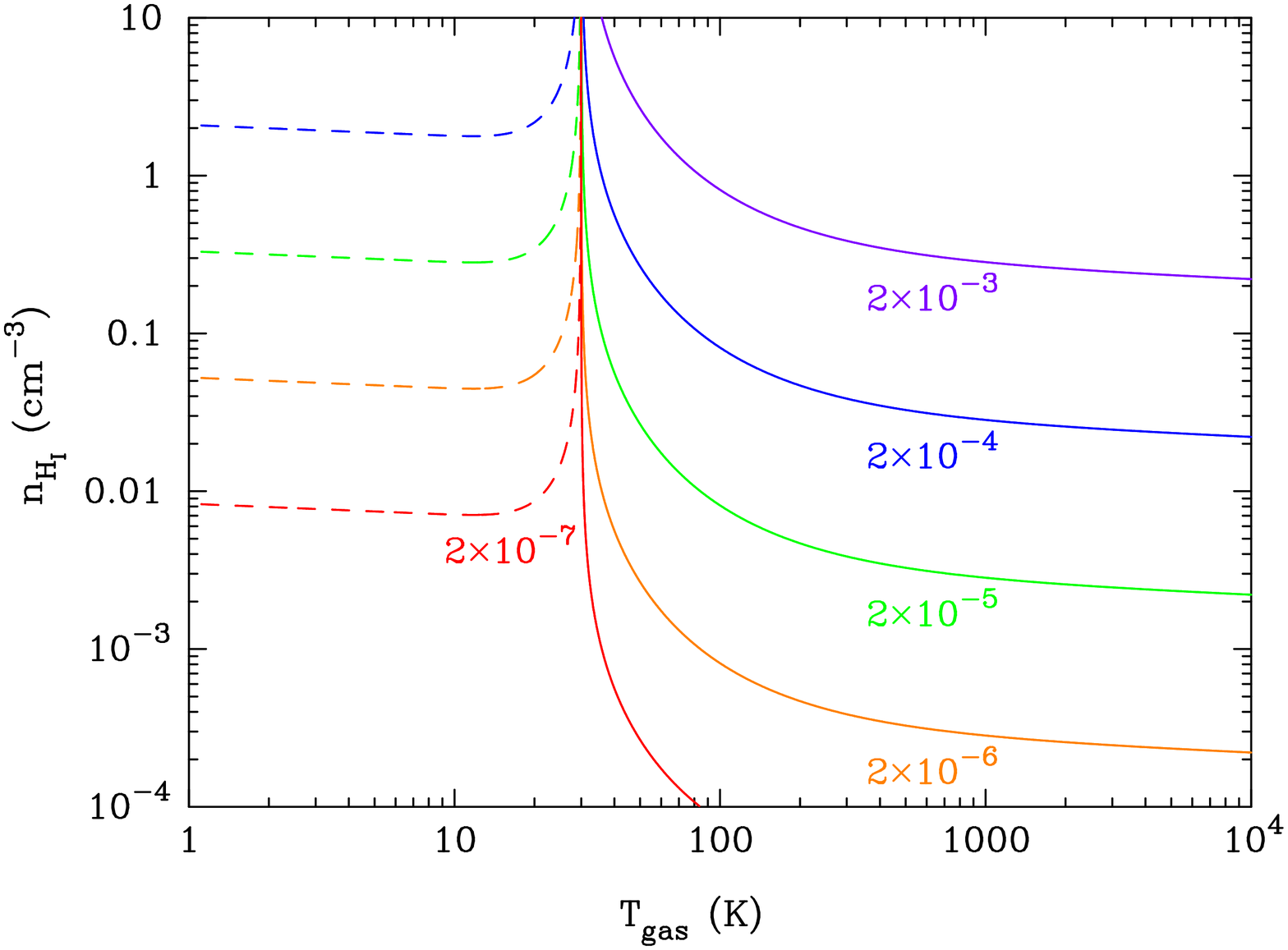}
\caption{Contours of deviations in number ratio $n_1/n_0$ from the values for only CBR contribution.  The upper panel shows the contribution of UV pumping on the parameter plane of UV color temperature $T_{\rm UV}$ and scattering rate $P_\nu$ for the UV transition $^2P_{3/2}\rightarrow ^2D_{3/2}$.  The lower panel shows the contribution of collisions on the parameter plane of the gas temperature $T_{\rm gas}$ and the number density of \HI\ $n_{\rm H_I}$.  Solid and dashed lines correspond to positive and negative values for indicated amplitudes, respectively.}
\label{fige3}
\end{figure}

Figure\ \ref{fige3} shows contours of $(\Delta f_{10}/f_{10})_{\rm L}$ (upper panel) on the parameter plane of UV color temperature $T_{\rm UV}$ and scattering rate $P_\nu$ for the UV transition $^2P_{3/2}\rightarrow ^2D_{3/2}$, and also contours of $(\Delta f_{10}/f_{10})_{\rm C}$ on the parameter plane of the gas temperature $T_{\rm gas}$ and the number density of \HI\ $n_{\rm H_I}$ (lower panel).  For this figure, the redshift was assumed to be $z=10$, and the target in collisional (de)excitation was assumed to be \HI.  Solid and dashed lines correspond to positive and negative values for indicated amplitudes, respectively.  Because of $T_{\rm UV} \gg T_\ast$ the effect of UV pumping is not sensitive to $T_{\rm UV}$ [cf. equation (\ref{eq20})].   $(\Delta f_{10}/f_{10})_{\rm C}$ is positive in the parameter region of $T_{\rm gas} > T_{\rm CBR}$ while it is negative in the parameter region of $T_{\rm gas} < T_{\rm CBR}$ [cf. equation (\ref{eq20})].

\begin{figure}
\includegraphics[width=84mm]{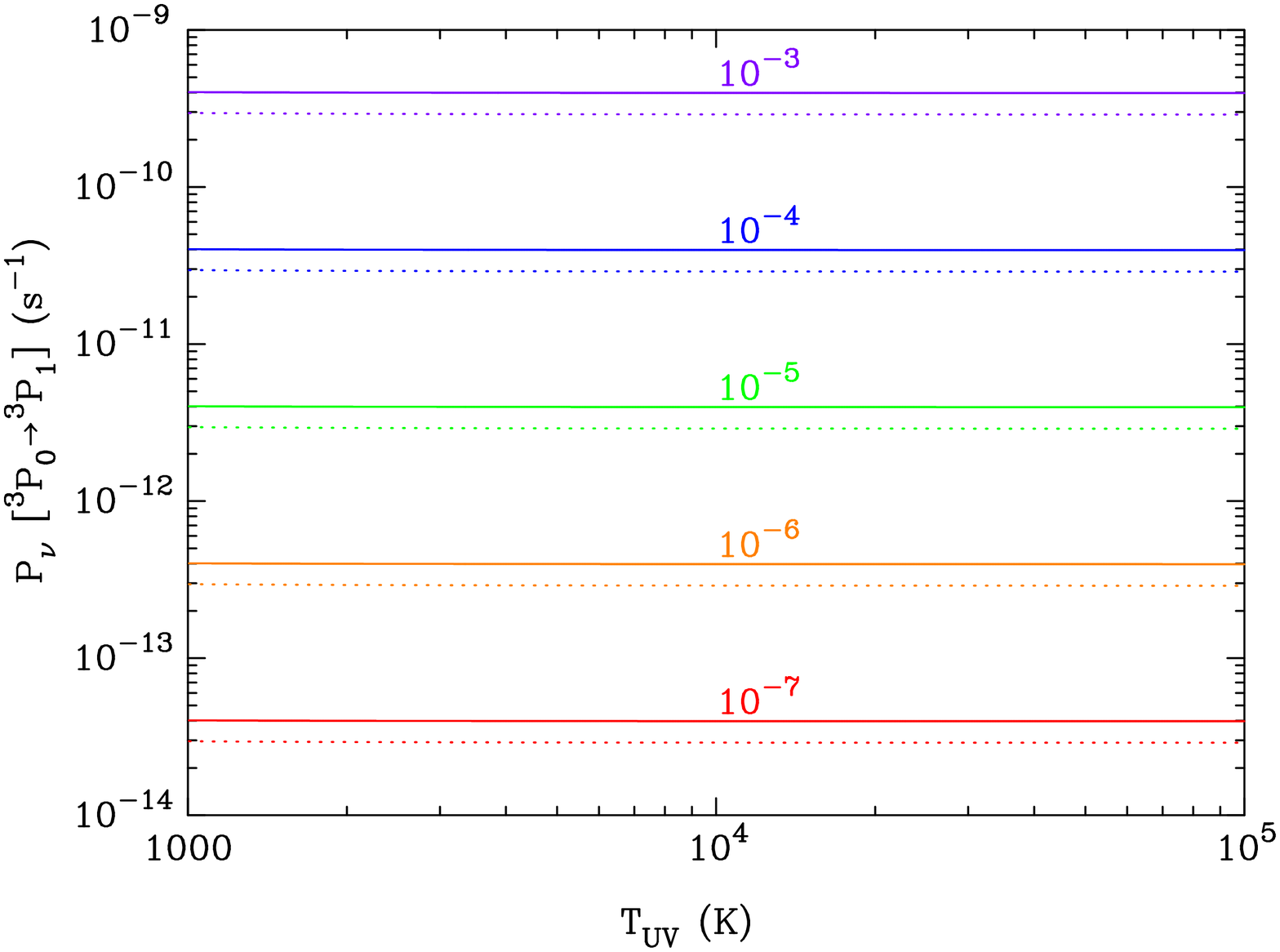}
\includegraphics[width=84mm]{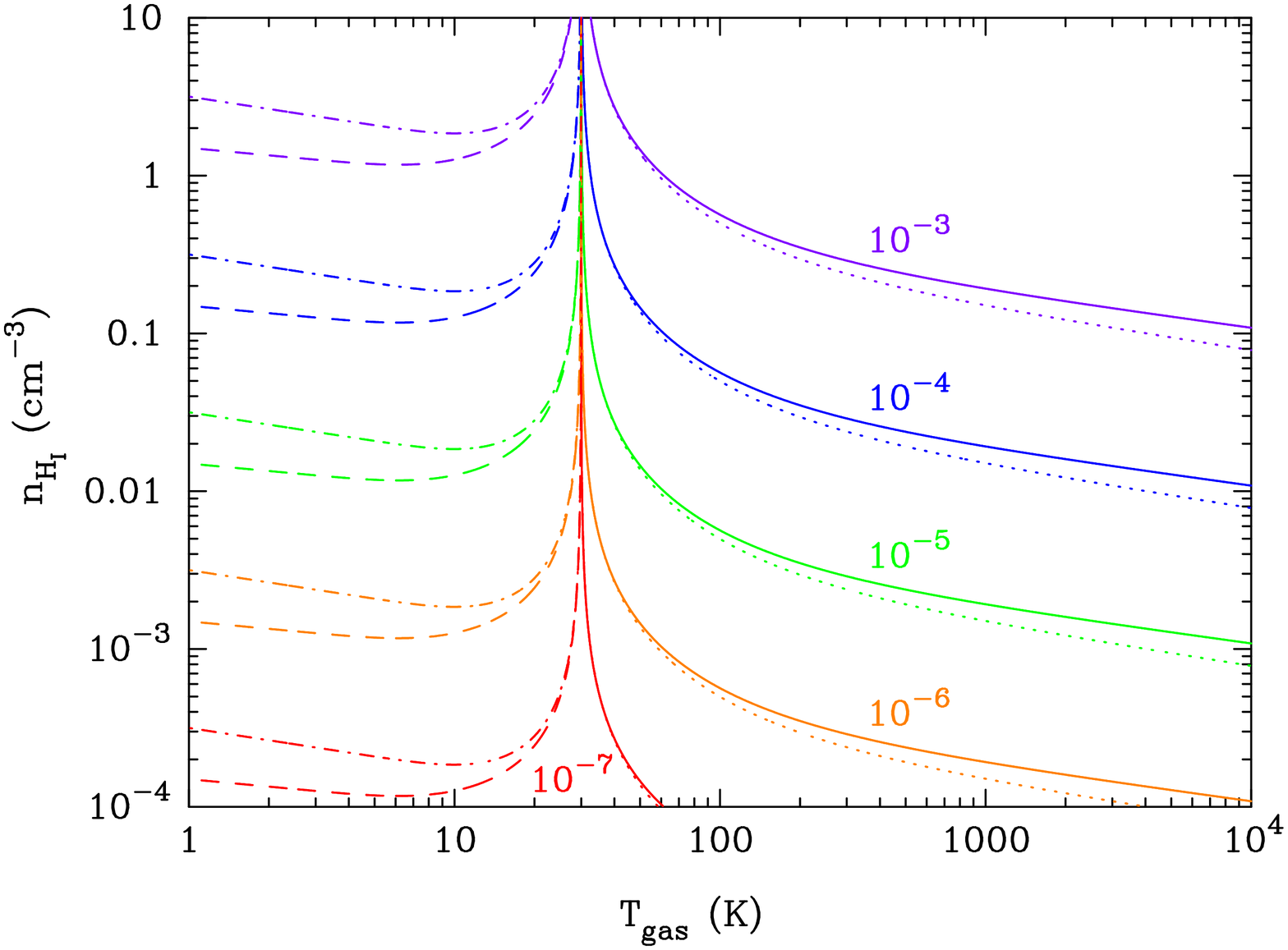}
\caption{Same as in Fig\ \ref{fige3} but for $n_1/n_0$ [solid (for positive values) and dashed lines (for negative values)] and $n_2/n_1$ [dotted (positive) and dot-dashed lines (negative)] of \CI.  The upper panel shows the contribution of UV pumping on the parameter plane of $T_{\rm UV}$ and $P_\nu$ for the UV transition $^3P_{0}$ (ground state) $\rightarrow ^3P_{1}$.  The lower panel shows the contribution of collisions on the plane of $T_{\rm gas}$ and $n_{\rm H_{I}}$.}
\label{fige4}
\end{figure}

Figure\ \ref{fige4} shows contours for \CI\ fine structure lines of $(\Delta f_{10}/f_{10})_{\rm L}$ (solid lines) and $(\Delta f_{21}/f_{21})_{\rm L}$ (dotted) (upper panel) on the parameter plane of $T_{\rm UV}$ and $P_\nu$ for the UV transition $^3P_{0}$ (ground state) $\rightarrow ^3P_{1}$, and also contours of $(\Delta f_{10}/f_{10})_{\rm C}$ [solid (for positive values) and dashed lines (for negative values)] and $(\Delta f_{21}/f_{21})_{\rm C}$ [dotted (positive) and dot-dashed lines (negative)] on the parameter plane of $T_{\rm gas}$ and $n_{\rm H_{I}}$ (lower panel).  The redshift was assumed to be $z=10$, and the target in collisional (de)excitation was assumed to be \HI.

\begin{figure}
\includegraphics[width=84mm]{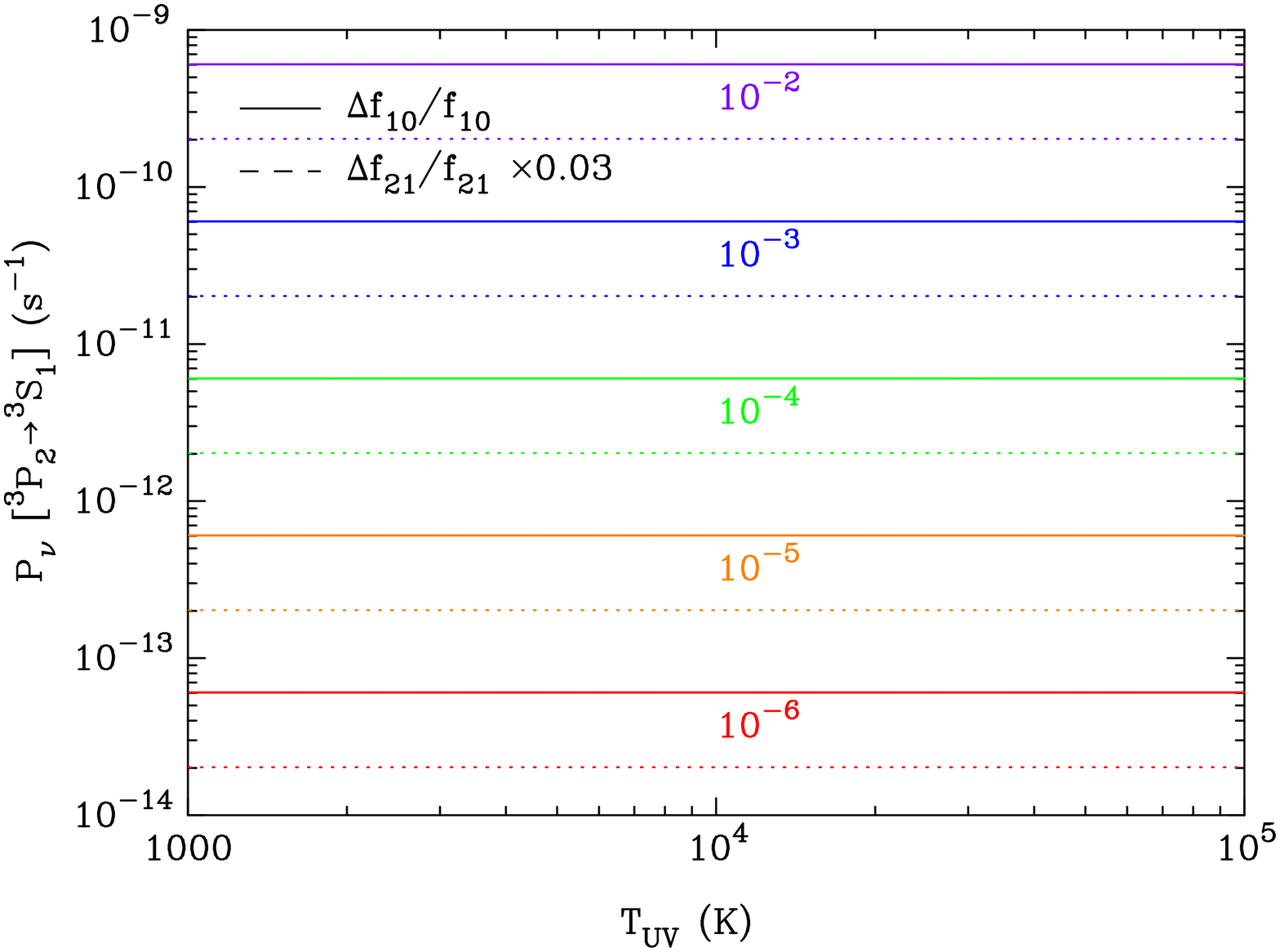}
\includegraphics[width=84mm]{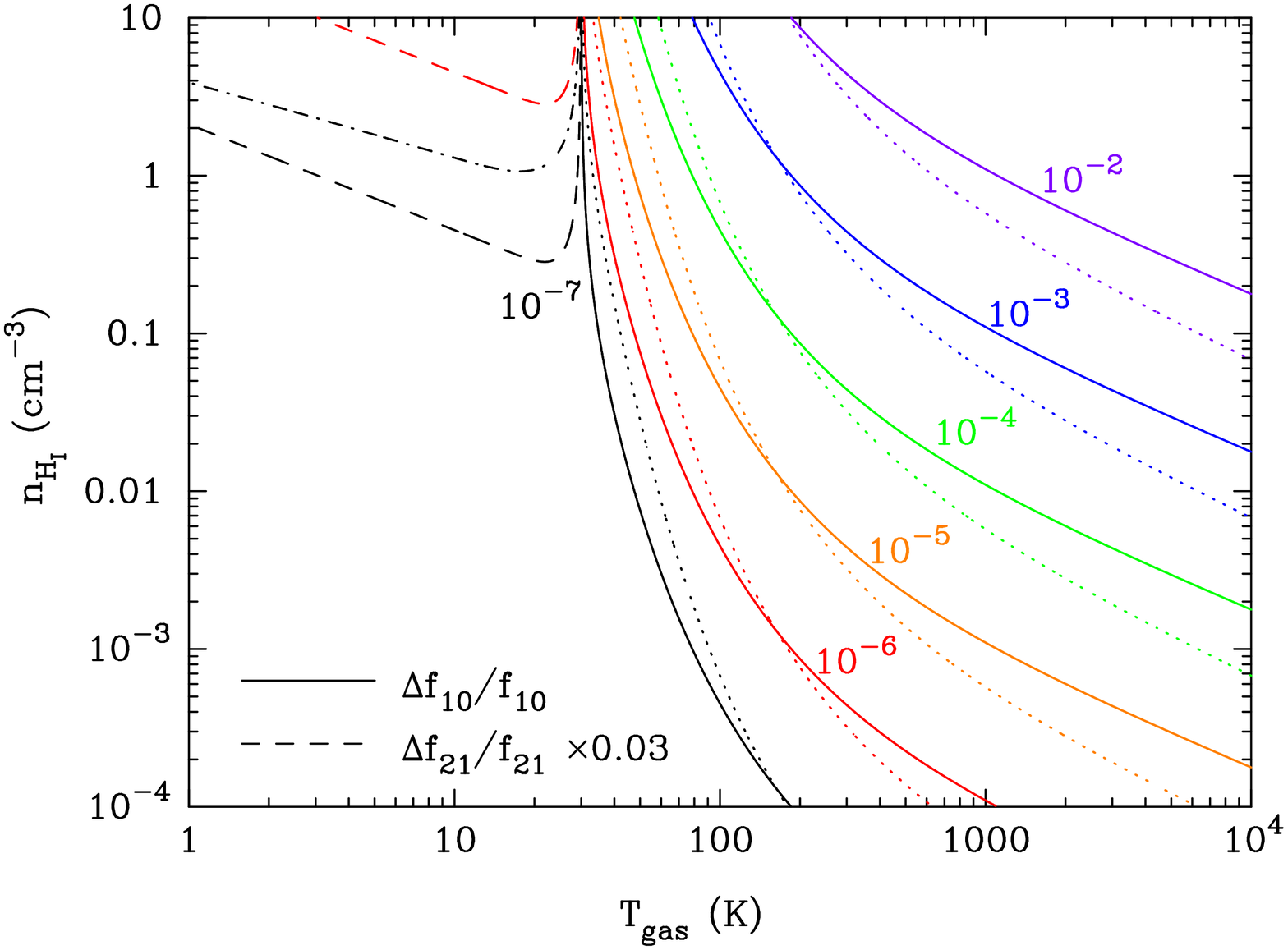}
\caption{Same as in Fig\ \ref{fige4} but for \OI.  The upper panel shows the contribution of UV pumping on the parameter plane of $T_{\rm UV}$ and $P_\nu$ for the UV transition $^3P_{2}\rightarrow ^3S_{1}$.  The lower panel shows the contribution of collisions on the plane of $T_{\rm gas}$ and $n_{\rm H_{I}}$.}
\label{fige5}
\end{figure}

Figure\ \ref{fige5} shows contours for \OI\ fine structure lines of $(\Delta f_{10}/f_{10})_{\rm L}$ (solid lines) and $(\Delta f_{21}/f_{21})_{\rm L}$ (dotted) (upper panel) on the parameter plane of $T_{\rm UV}$ and $P_\nu$ for the UV transition $^3P_{2}\rightarrow ^3S_{1}$, and also contours of $(\Delta f_{10}/f_{10})_{\rm C}$ [solid (for positive values) and dashed lines (for negative values)] and $(\Delta f_{21}/f_{21})_{\rm C}$ [dotted (positive) and dot-dashed lines (negative)] on the parameter plane of $T_{\rm gas}$ and $n_{\rm H_{I}}$ (lower panel).  The redshift was assumed to be $z=10$, and the target in collisional (de)excitation was assumed to be \HI.

From Figs. \ref{fige3}, \ref{fige4} and \ref{fige5}, it is found that under environments of intense UV fluxes and relatively low densities and low gas temperatures considered in this study, the effect of UV pumping is typically larger than that of collisions.  We, therefore, neglect the effect of collisions on the population of fine structures in what follows.

\subsection{Intensities of Redshifted Fine Structure Radiation}

We estimate a signal from a region including abundant \CI\ or \CII, or
\OI\ of angular size on the sky that is larger than a beam width,
and of radial velocity range that is larger than the bandwidth.  For a
fine structure line of frequency $\nu_0$, a
intergalactic optical depth at frequency $\nu_0/(1+z)$ along the line
of sight is~\citep{her2007}
\begin{equation}
 \tau=\frac{A_{ij}\lambda_{ij}^3 n_{A_N}(z)
  \left(g_i f_j/g_j -f_i\right)}{8\pi H_0 \sqrt[]{\mathstrut
  \Omega_{\rm m}(1+z)^3 + 1-\Omega_{\rm m}}},
\label{eq10}
\end{equation}
where
$(i,j)=(1,0)$ or $(2,1)$,
$\lambda_{ij}$ is the wavelength for transition $i \rightarrow j$,
$n_{A_N}(z)$ is the number density of chemical species $A_N$ (\CI, \CII\ or \OI) at redshift $z$,
$f_i=n_i/\left(\sum_i n_i \right)$ is the population fraction of energy state $i$,
$H_0=71.9^{+2.6}_{-2.7}$~km~s$^{-1}$ Mpc$^{-1}$~\citep{dun2009} is the Hubble constant,
$\Omega_{\rm m}$ is the energy density parameters of matter~\citep[e.g.,][]{mad1997,pad1993}.  The optical depth in the homogeneous universe is typically much less
than unity.  In the optically thin case, the intensity of the signal
$\Delta I_\nu$ with respect to that of the CBR radiation $B_\nu(T_{\rm
CBR})$ is
\begin{equation}
 \frac{\Delta I_\nu}{B_\nu (T_{\rm CBR})} =
  \left[\frac{\exp(T_\ast/T_{\rm CBR})-1}{\exp(T_\ast/T_S)-1}-1
  \right] \tau,
\label{eq9}
\end{equation}
which is identical to equation (9) in~\citet{her2007} under the assumption that the background radiation field is that of thermal CBR.
In this study, we investigate signals of the UV continuum
radiation left on the radio background via excitations of fine
structure lines.  The UV source is assumed to have a power law
luminosity spectrum, i.e., $L_\nu \propto \nu^{-\alpha_S}$.  When the UV
photon scattering occurs close to the
emission redshift, the total scattering rate $P_\nu$ for a point
source is given by
\begin{eqnarray}
 P_\nu&\hspace{-0.8em}=&\hspace{-0.8em}\frac {\lambda_{lu}^2}{8 \pi} \frac{g_u}{g_l} A_{ul}
  \frac{L_\nu/(h\nu)}{4\pi r^2} \nonumber\\
&\hspace{-0.8em}=&\hspace{-0.8em} 7.6\times 10^{-9}~{\rm s}^{-1}~\frac{g_u}{g_l}
 \left(\frac{\lambda_{lu}}{\lambda_\alpha}\right)^{\alpha_S+3} \left(
  \frac{A_{ul}}{A_\alpha}\right) \frac{\left(\nu L_\nu\right)_{\alpha,
  47}}{(r_{\rm Mpc}/0.1)^2},
\label{eq13}
\end{eqnarray}
\citep[cf.][]{mad1997} 
where
the subscripts $l$ and $u$ mean the low energy state of fine structure
and the upper energy state to which the transition from the state $l$
is allowed, respectively,
$r$ is the proper distance between the emission and scattering.  In the
second equality,
$\lambda_\alpha=1.215\times 10^{-5}$~cm is the wavelength for Ly$\alpha$
transition of \HI,
$A_\alpha=6.265\times 10^8$~s$^{-1}$ is the spontaneous emission rate
for Ly$\alpha$,
$\left(\nu L_\nu \right)_{\alpha, 47}$ is the product of the luminosity of the point
source and the frequency of photon at Ly$\alpha$ frequency in units of
$10^{47}$~ergs~s$^{-1}$, and
$r_{\rm Mpc}$ is the distance $r$ in Mpc.

\begin{figure}
\includegraphics[width=84mm]{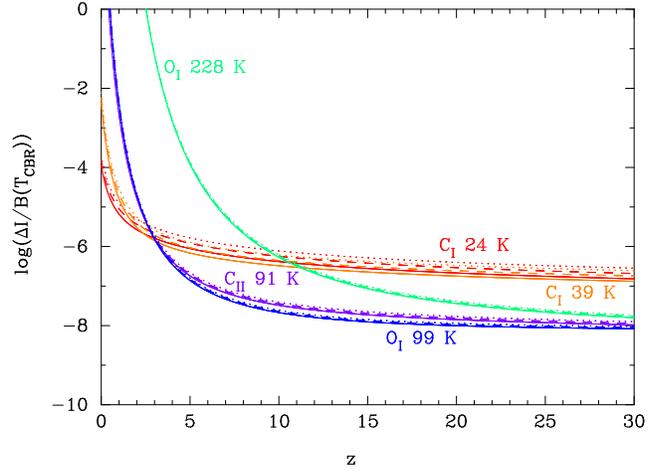}
\caption{Signals through fine structure lines in the homogeneous universe
 relative to corresponding cosmic background radiation as a function of
 the redshift.  A QSO of luminosity of
 $L_\nu=4.1\times 10^{31}(\nu/\nu_\alpha)^{-\alpha_S}$~ergs~s$^{-1}$~Hz$^{-1}$ with
 $\alpha_S=1/2$ (solid lines), $\alpha_S=3/2$ (dashed) and
 $\alpha_S=5/2$ (dotted) is assumed
 to shine at a distance of 1 Mpc.  The respective lines are derived under
 the assumption that the elements (C and O) are completely in the
 chemical species (\CI, \CII\ and \OI), and therefore
 maximum values.}
\label{fig2}
\end{figure}

Figure\ \ref{fig2} shows signals through fine structure lines relative
 to corresponding cosmic background radiation as a function of the
 redshift under the following assumptions:  The universe is homogeneous,
 and the number density of hydrogen is $n_{\rm H}=1.9\times
 10^{-7}$~cm$^{-3}(1+z)^3$.  The abundances of C and O in the universe at
 redshift $z$ are $y_{\rm C}=1.4\times 10^{-4}$ and $y_{\rm O}=3.2\times
 10^{-4}$, respectively, which correspond to local ISM
 values~\citep{mai2005}\footnote{Although the abundances of C and O
 are thought to be smaller in the earlier epoch of the universe, their
 chemical evolution as a function of the redshift is rather uncertain.  We,
 then, adopt here the present values for the abundances which would
 be upper limits on the old day abundances.}.  The elements (C and O) are
 assumed to be completely in the chemical species (\CI, \CII\ and \OI),
 i.e., $n_{\rm C_I}=n_{\rm H} y_{\rm C}$, $n_{\rm C_{II}}=n_{\rm H} y_{\rm C}$ and
 $n_{\rm O_I}=n_{\rm H} y_{\rm O}$,
 respectively.  A QSO of luminosity of
 $L_\nu=4.1\times 10^{31}(\nu/\nu_\alpha)^{-\alpha_S}$~ergs s$^{-1}$ Hz$^{-1}$ 
 [corresponding to $\left(\nu L_\nu \right)_{\alpha, 47}=1$] with
 $\alpha_S=1/2$ (solid lines), $\alpha_S=3/2$ (dashed) and
 $\alpha_S=5/2$ (dotted) is assumed
 to shine at a distance of 1 Mpc ($r_{\rm Mpc}=1$).  A decrease in flux
 of UV photons associated with continuous scattering with \Cboth\ and \OI\ during their propagation is not considered.  This effect is found
 very important in Section\ \ref{sec4} below.  In drawing this figure we confirmed that the spin temperature of \OI\ 228~K line calculated with equations given in Sec\ \ref{sec2}.1 is exactly equal to that described by an approximate equation (7) of~\citet{her2007}.

At high redshift of $z\ga 10$, all signals are small at the distance of
$r_{\rm Mpc}=1$.  As the redshift decreases, spin temperatures become smaller with the CBR temperature decreasing by redshift.
Optical depths decrease since the number density of particles decrease
with the cosmic expansion [see equation\ (\ref{eq10})].  Since the
number density of the CBR is smaller and its energy is lower at low redshift, the UV pumping effect gets relatively
dominant over the CBR effect below some critical redshifts for
respective fine structure lines.  Spin temperatures then decouple from the
CBR temperature at the critical redshifts in the present setting of
fixed radius.  Since the CBR temperature at low redshift is less than
the line temperatures, i.e., ($T_\ast > T_{\rm CBR}$), the
decoupling of the spin temperature ($T_{\rm S} > T_{\rm CBR}$) leads to
the large difference of line intensity from that of CBR.  The factor in
the angle bracket in the right hand side of equation (\ref{eq9}) is,
therefore, large.  The signal to CBR ratios are enhanced at low
redshift for these reasons.

\subsection{UV Color Temperature}

We mention an important point on the UV color temperature, for example,
of the ground state of \CI.

In a time scale for UV photons to scatter \CI\ particles by UV pumping,
i.e., the mean free time, the color temperature of the UV photon around
the line center relaxes to the kinetic temperature of the gas (i.e.,
\CI\ here)~\citep{fie1959,ryb1994}.  The mean free time of a line
center photon is $\tau^{\rm line}=1/(n_{\rm C_I} \sigma_\nu^{\rm C})$ with
$\sigma_\nu^{\rm C}$ the UV line absorption cross section at the line
center.  The line absorption cross section is
\begin{equation}
 \sigma_\nu = \frac{\lambda_{ul}^2}{8 \pi} \frac{g_u}{g_l} A_{ul}
  \phi(\nu).
\label{eq6}
\end{equation}
Neglecting the microturbulence and the very small effect of natural
broadening, the line center cross section is
\begin{equation}
 \sigma_\nu^{\rm C} = \frac{\lambda_{ul}^2}{8 \pi} \frac{g_u}{g_l} A_{ul}\frac{1}{\sqrt[]{\mathstrut \pi}\Delta \nu_{\rm D}},
\end{equation}
where
\begin{equation}
\Delta \nu_{\rm D}=\frac{\nu_0}{c} \sqrt[]{\mathstrut \frac{2kT_{\rm gas}}{m}}
\end{equation}
is the Doppler width with $m$ the mass of an atom~\citep{ryb1979}.

Using this
equation and $n_{\rm C{\sc \,I}}=n_{\rm H}\delta_{\rm b}y_{\rm C}$, the mean free time is found
to be
\begin{equation}
\tau^{\rm line} = 4.2~{\rm yr}~T_{\rm gas, K}^{1/2} \left( \frac{1+z}{10} \right)^{-3}
 \delta_{\rm b}^{-1} \left( \frac{y_{\rm C}}{1.4\times 10^{-4}}\right)^{-1}.
\end{equation}
where
$T_{\rm gas, K}$ is the gas temperature in units of K,
$\delta_{\rm b}$ is the local density excess in units of the universal average
value of baryonic matter.  The color temperature around the line
frequency, therefore, could approach to the gas temperature depending
on the temperature of the gas, redshift, density excess, and the abundance
of the scattered species.  Since the Doppler width
$\Delta \nu_{\rm D}=2.2\times 10^8 {\rm s}^{-1} T_{\rm gas, K}^{1/2}$ is much smaller than
the frequencies corresponding to the hyperfine transitions of
$\nu=O(10^{11}-10^{12} {\rm s}^{-1})$, line photons of
different frequencies are completely distinguished at scattering
processes with C and O.  In other words, a photon with frequency for
one line transition can not induce scatterings corresponding to other transitions.
Furthermore, strong lines
which enable changes in photon energy do not exist around the
frequencies of the UV pumping lines.   The color
temperature is, therefore, given by an original color temperature of the
source $T_{\rm UV}=h\nu_{\rm line}/[k(3+\alpha_S)]$ before relaxing to
the spin temperature by scattering with C or O.  This situation is
completely different from the case of the Ly$\alpha$ pumping in which Ly$\alpha$ photon can mix the
hyperfine levels of the \HI\ ground state many times~\citep{mad1997}.

\section{Numerical Calculation of 1D Hydrodynamics}\label{sec3}

We calculated a hydrodynamics coupled to a chemical reaction network
assuming the spherical symmetry of space in order to show an example of
geometric structure of ionized region.

\subsection{Hydrodynamics}

\subsubsection{Fluid Equations}

The hydrodynamic equations to be solved in this study are as follows:

\begin{equation}
 \frac{\partial \rho }{\partial t}+{\bf \nabla}\cdot \left(\rho {\bf
						      v}\right)=0,
\end{equation}
\begin{equation}
 \frac{\partial {\bf v}}{\partial t}+ \left({\bf v} \cdot {\bf
				       \nabla}\right) {\bf v} =
 -\frac{{\bf \nabla}p}{\rho}-{\bf \nabla}\Phi,
\end{equation}
\begin{equation}
 \frac{\partial \left( \rho E\right)}{\partial t}+{\bf \nabla}\cdot
  \left[\left(\rho E+p\right){\bf v}\right]= S,
\end{equation}
where 
$t$ and ${\bf r}$ are the time and the position, respectively,
$\rho$, ${\bf v}$, $p$ and $\Phi$ are the density, velocity, pressure
and gravitational potential as a function of $t$ and ${\bf r}$,
respectively,
$E\equiv v^2/2+\epsilon$ is the fluid energy per unit mass with
$\epsilon=p/\left[(\gamma-1)\rho\right]$ the internal energy and
$\gamma$ the ratio of specific heat,
$S=\Gamma-\Lambda$ is the energy gain with $\Gamma$ and $\Lambda$ the
heating and cooling terms, respectively.
The ratio of specific heat is fixed to be $\gamma=5/3$ in this study.

We here introduce the comoving coordinates~\citep{sha1980,
sha1985} using comoving variables to describe deviations of the
fluid quantities from those of the expanding homogeneous universe.  New
variable denoted by primes are
\begin{equation}
 dt^\prime=a^{-2} dt ,
\end{equation}
\begin{equation}
 {\bf r^\prime}=a^{-1}{\bf r} ,
\end{equation}
\begin{equation}
 \rho^\prime=a^3 \rho ,
\end{equation}
\begin{equation}
 {\bf v^\prime}=a\left[{\bf v}-H(t){\bf r}\right] ,
\end{equation}
\begin{equation}
 p^\prime=a^5 p ,
\end{equation}
\begin{equation}
 \epsilon^\prime=a^2 \epsilon =p^\prime/\left[\left(\gamma-1\right)\rho^\prime \right] ,
\end{equation}
\begin{equation}
 S^\prime=a^7 S ,
\end{equation}
where
$a(t)=1/(1+z)$ is the scale factor of the universe with $z$ the redshift, and
$H(t)=\dot{a}/a$ is the Hubble expansion rate.  The peculiar
gravitational potential is assumed to be negligible, and only the
potential in a homogeneous isotropically expanding universe is
considered here, i.e.,
\begin{equation}
 \Phi(t,r)=\frac{2\pi G \bar{\rho}(t) r^2}{3} ,
\end{equation}
where
$G$ is the gravitational constant,
and $\bar{\rho}(t)$ is the mean density at time $t$.

We assume that the cosmic expansion is described by $\Lambda$CDM model.
The cosmological parameters are taken from a fit to the Wilkinson
Microwave Anisotropy Probe (WMAP) 5 year data
(Model
$\Lambda$CDM+SZ+lens).~\citep{dun2009}\footnote{http://lambda.gsfc.nasa.gov.}\footnote{Adopted parameters are almost the same as those of WMAP 7 year data~\citep{lar2010}.}
The energy density parameters of baryon, matter and dark energy are
$\Omega_{\rm b}=0.0441\pm 0.0030$, $\Omega_{\rm m}=0.258\pm 0.030$ and
$\Omega_\Lambda=0.742\pm 0.0030$ respectively.  In this study we
concentrate on the early epoch of the universe, i.e., $z>1$.  The model,
therefore, is effectively the CDM dominated universe.

Under the assumption of the CDM dominated universe and of the ratio of
specific heat $\gamma=5/3$,  the fluid equations are transformed into
the following ones:

\begin{equation}
 \frac{\partial \rho^\prime }{\partial t^\prime}+{\bf
  \nabla^\prime}\cdot \left(\rho^\prime {\bf v^\prime}\right)=0 ,
\end{equation}
\begin{equation}
 \frac{\partial \left( \rho^\prime {\bf v^\prime }\right)}{\partial
  t^\prime}+ {\bf \nabla^\prime}\left(\rho^\prime {v^\prime}^2 +
				 p^\prime \right)  = 0 ,
\end{equation}
\begin{equation}
 \frac{\partial}{\partial t^\prime}\left(\rho^\prime E^\prime \right)+{\bf
 \nabla^\prime}\cdot \left[\rho^\prime
		      \left(E^\prime+ \frac{p^\prime}{\rho^\prime}\right){\bf
		      v^\prime}\right]= S^\prime.\nonumber \\
\end{equation}
where $E^\prime \equiv {v^\prime}^2/2+\epsilon^\prime$ was defined.
These equations do not include the cosmic expansion
explicitly
.  They are solved simultaneously with the time evolution of the scale
factor $a(t)$.

\subsubsection{Heating and Cooling Rates}

As a heating term, the photoionization is included.  The heating rate from
kinetic energy of photoelectrons and protons produced by photoionization
is given by
\begin{eqnarray}
 \Gamma&=&\sum_{i={\rm H_I}, {\rm He_I}, {\rm He_{II}}} n_i \times \nonumber \\
&& \int_{E_{{\rm th}, i}}^\infty dE_\gamma~\left(E_\gamma-E_{{\rm th},
					    i}\right) n_\gamma
(E_\gamma) \sigma^{\rm ion}_i (E_\gamma) c ,
\label{heating}
\label{eq16}
\end{eqnarray}
where
$E_\gamma$ is the energy of nonthermal photon,
$E_{{\rm th}, i}$ is the threshold energy for photoionization of species $i$,
$n_\gamma (E_\gamma)$ is the energy spectrum of the photon number
density as a function of time $t$ and position $r$,
$\sigma^{\rm ion}_i (E_\gamma)$ is the photoionization cross section, and
$c$ is the light speed. 
The production of secondary electrons at the photoionization of H and He
is followed, and additional energy inputs by these electrons are taken
into account in the estimation of the heating rate.  For the photoionization
cross sections we used the FORTRAN subroutine phfit2 by
D.~A.~Verner\footnote{Subroutines by D.~A.~Verner are available at
http://www.pa.uky.edu/\~{}verner/fortran.html.}.  We adopted the fitted
functions from~\citet{shu1985} for heat generation fractions
associated with secondary electrons produced by the \HI\ and
\HeI\ photoionization.

As cooling terms, radiative coolings due to collisional
recombination, excitation, ionization and free-free transitions as well
as the Compton cooling off the CBR are
included.  We adopt the cooling rates for Case B (optically thick
case) recombinations of \HII\ and \HeIII\
by~\citet{hum1994} and that of \HeII\ by~\citet{hum1998}.
The cooling from the dielectronic recombination of \HeII\ is also
included with its rate taken from~\citet{sha1987}.  The cooling
rates for the collisional line excitation are calculated using the
excitation rates given in~\citet*{sob1981} for excitations of
\HI, \HeI\ and \HeII\ from the ground states of
main quantum number $n=1$ to excited states of $n=2, 3$ and 4.

Below $10^4$~K, metastable transitions or fine structure transitions of dominant metal species contribute to the cooling.  We, therefore, include important line transitions of \CI, \CII\ and \OI~\citep{hol1989}.  The cooling rate for fine structure transitions of species $i$ is given by
\begin{eqnarray}
 \Lambda_{\rm fs}&=&n_i \sum_{l<u} f_l \frac{g_u}{g_l}\left[\exp(-E_{ul}/T_{\rm gas})-\exp(-E_{ul}/T_{\rm CBR})\right] \nonumber \\
&&\times \sum_{j={\rm H_I}, e} n_j \gamma_{ul}^j E_{ul},
\end{eqnarray}
where
$\gamma_{ul}^j$ is the deexcitation rate of transition $u\rightarrow l$ at collisions with species $j$ whose value was taken from \citet{hol1989}.  
For this equation, the detailed balance between forward and backward reactions is assumed, and the spin states of species are assumed to be described by the CBR temperature.  (Note that spontaneous emission rates, i.e., $A_{ul}$ are typically larger than collisional rates, i.e., $\gamma_{ul}$ in low density environments considered in this paper.)  The cooling rate for (semi)forbidden transitions of species $i$ with transition temperatures of $O(10^4~{\rm K})$ is given by
\begin{eqnarray}
 \Lambda_{\rm for}&=&n_i \sum_{u} \frac{g_u}{g_0}\left[\exp(-E_{u0}/T_{\rm gas})-\exp(-E_{u0}/T_{\rm CBR})\right] \nonumber \\
&&\times \sum_{j={\rm H_I}, e} n_j \gamma_{u0}^j E_{u0},
\end{eqnarray}
where
it is assumed that almost all particles of species $i$ are in its ground state, i.e., $f_0=1$, and
the term $\exp(-E_{u0}/T_{\rm CBR})$ is negligible.  These metal transitions are, however, found not to affect this calculation significantly.

The cooling rate for the
collisional ionization is given by
\begin{equation}
 \Lambda_{\rm ion}= n_e \sum_{i={\rm H_I}, {\rm He_I}, {\rm He_{II}}} n_i
  I_i \langle \sigma_i v \rangle^{\rm col} ,
\end{equation}
where
$I_i$ is the ionization energy of species $i$, and
$\langle \sigma_i v\rangle^{\rm col}$ is the thermal average value of
the cross section times velocity for collisional ionization.  The $\langle
\sigma_i v \rangle^{\rm col}$ values are calculated with a Verner's
subroutine, i.e., cfit.  The cooling rates for the free-free transitions and
the Compton scattering of CBR by free electrons are taken
from~\citet{sha1987}.

\subsubsection{Numerical Method}

We use the Monotone Upstream-centered Schemes for Conservation Laws
(MUSCL) with the Roe's method and the second order Runge-Kutta method in
time integration.  The computation is then second order accurate in
space and time.  The number of grid point is 400 and the spacing is
$\Delta r=0.5$~kpc at the redshift of $z=8.7$. The
computational region is thus $0\lid r \lid 0.2$~Mpc.

The calculation domain is assumed to be uniform in density initially.  A
point source is put on the origin and starts lighting at $t=0$, whose
luminosity $L_\nu$ is given by
\begin{equation}
L_\nu=4.1\times 10^{31}~{\rm ergs}~{\rm s}^{-1}~{\rm Hz}^{-1} \left(\frac{h \nu}{10.2~{\rm
					eV}}\right)^{-\alpha_S} ,
\label{eq5}
\end{equation}
where
$\alpha_S$ is the spectral index fixed to be 1.5, and
10.2~eV is the energy corresponding to the  Ly$\alpha$ transition of
\HI\ atom.  
The amplitude and spectral index of the point source are the same as those
in \citet{mad1997}.  This point source emits total ionizing photon for
hydrogen of $S_\gamma=\int_{13.6~{\rm eV}/h}^\infty L_\nu/(h\nu)~d\nu=1.6\times
10^{58}$~s$^{-1}$.

\subsection{Chemical Reaction Network}

We constructed a chemical reaction network code including 22 chemical species
and 55 chemical reactions.  The species are atoms and all ions of H, He,
C and O as well as the electron.  

The rate equation which was solved is given by
\begin{equation}
 \frac{dy_i}{dt}= \left(\frac{dy_i}{dt}\right)_{\rm ion} +
  \left(\frac{dy_i}{dt}\right)_{\rm rec} +
  \left(\frac{dy_i}{dt}\right)_{\rm col} +
  \left(\frac{dy_i}{dt}\right)_{\rm cha} ,
\end{equation}
where $y_i\equiv n_i/n_{\rm H}$ is the relative number density of the
$i$th species to that of hydrogen $n_{\rm H}=n_{\rm H_I}+n_{\rm H_{II}}$, and
the right hand side includes terms for the nonthermal photoionizations
(ion), thermal radiative recombinations (rec), and thermal collisional ionizations (col) of all atoms and ions, and charge transfer reactions (cha)~\citep[e.g.][]{kin1999}.  Each term is
described by
\begin{eqnarray}
 \left(\frac{dy_i}{dt}\right)_{\rm ion}&=& y_{j=(i+e^-)}
  \int_{E_{{\rm th}, j}}^\infty dE_\gamma n_\gamma (E_\gamma)
  \sigma^{\rm ion}_j (E_\gamma) c \nonumber \\
&&- y_i \int_{E_{{\rm th},i}}^\infty dE_\gamma n_\gamma (E_\gamma)
 \sigma^{\rm ion}_i (E_\gamma) c ,
\label{ionization}
\end{eqnarray}
\begin{eqnarray}
 \left(\frac{dy_{j=(i+e^-)}}{dt}\right)_{\rm rec}&=&n_{\rm H} y_i
  y_e \langle \sigma_i v \rangle^{\rm rec} \nonumber \\
&&- n_{\rm H} y_j y_e \langle \sigma_j v \rangle^{\rm rec} ,
\end{eqnarray}
\begin{eqnarray}
 \left(\frac{dy_i}{dt}\right)_{\rm col}&=&n_{\rm H} y_{j=(i+e^-)}
  y_e \langle \sigma_j v \rangle^{\rm col} \nonumber \\
&&- n_{\rm H} y_i y_e \langle \sigma_i v \rangle^{\rm col} ,
\end{eqnarray}
\begin{eqnarray}
\left(\frac{dy_i}{dt}\right)_{\rm cha}&=& \sum_{j,k} \left( n_{\rm H} y_{j} y_k \langle \sigma_{jki} v \rangle^{\rm cha}\right. \nonumber \\
&&\left.- n_{\rm H} y_i y_k \langle \sigma_{ikj} v \rangle^{\rm cha}\right) ,
\end{eqnarray}
where
$j=(i+e^-)$ is the species which is photodisintegrated into the species
$i$ and an electron,
$\langle \sigma_i v \rangle^{\rm rec}$ is the thermal average value of
the cross section times velocity for recombination.  The first and second
terms in the right hand sides of the above equations correspond to
the production and destruction terms, respectively.  In the fourth equation $\sigma_{ijk}$ is the cross section of reaction $i+j\rightarrow k+l$ with any $l$, and indexes $j$ and $k$ should be summed.

The Verner's
subroutine rrfit is used to evaluate the radiative recombination rates.
The collisional ionization by secondary electrons produced in the
photoionization of H and He is taken into account.
We adopted the fitted functions  from~\citet{shu1985} for the ratios of
energies used for secondary ionizations to the total energy produced by
the photoionization of \HI\ and \HeI.  Charge transfer rate coefficients are taken from \citet[][and updates and addenda]{kin1996}\footnote{Updates and addenda are shown at http://www-cfadc.phy.ornl.gov/astro/ps/data/.}:  FORTRAN subroutines by Jim Kingdon are used to calculate reaction rates of H and all ions of He and C.  Fitted coefficients by Rakovic et al. (2001, unpublished) for reactions of H and all ions of O are adopted.  Charge transfer reactions with He of C$^{q+}$ with $q=3$ and $4$, and O$^{q+}$ with $q=2-4$ are also included with their rate coefficients taken from the web page.

\subsection{Input Physics}

The initial abundances of H, He, C and O are given as follows.  The
helium to hydrogen number ratio is $y_{\rm He}=0.082$ as predicted in
standard big-bang nucleosynthesis model with baryon-to-photon ratio
determined from WMAP, i.e., $\eta=6.3\times 10^{-10}$~\citep{dun2009}.
$y_{\rm C}=1.4\times 10^{-4}$ and $y_{\rm O}=3.2\times 10^{-4}$ are
used, which correspond to the local ISM values~\citep{mai2005}.  All
particles are assumed to be in the form of neutral atoms.  The initial
number density is uniformly given by $n_{\rm H}=1.9\times 10^{-7}
\delta_{\rm b} (1+z)^3$~cm$^{-3}$ corresponding to the adopted $\Omega_{\rm b}$
value.  Two cases of $\delta_{\rm b}=10^3$ and $10^4$ are calculated and
presented in this paper.  We assume that there is no astrophysical
heat source before the epoch corresponding to the beginning of the
calculation.  The initial gas temperature then should be given by that
having decoupled from the CBR temperature at $z\sim 200$ and experienced
the redshift as $T\propto (1+z)^2$.  We adopt such temperature
calculated by~\citet{loe2004}, i.e., $T=2.3$~K~$[(1+z)/10]^2$.

The local ionization flux as a function of frequency $\nu$ is
\begin{equation}
 F_\nu (r)=L_{\nu_i} \exp(-\tau_\nu) \frac{1}{4\pi r^2} H(z_{\rm start}-z_i) ,
\label{eq8}
\end{equation}
where
$H(x)$ is the Heaviside's step function,
$\nu_i$ is the the frequency which redshifts to $\nu$ after running a
distance $r$.
$\nu_i$ is, therefore, the initial frequency of photon emitted at the
origin at redshift $z_i$, i.e.,
\begin{equation}
 \frac{\nu_i}{\nu}=\frac{1+z_i}{1+z} ,
\end{equation}
$z_{\rm start}=8.7$ is the initial redshift in the calculation,
$\tau_\nu(r)$ is the optical depth at time $t$ between the origin and
a given position $r$, and is properly given according to the position
$r$.  For example, when the radius $r$ is larger than
all the ionization fronts, it is
\begin{eqnarray}
 \tau_\nu(r) &\hspace{-0.8em}\simeq&\hspace{-0.8em} n_{\rm H}
  \left[\left( r-r_{\rm I}^{\rm H_I} \right) \sigma^{\rm
 ion}_{\rm H_I} (h \nu) 
 + y_{\rm He_I} \left( r-r_{\rm I}^{\rm He_I} \right) \sigma^{\rm
 ion}_{\rm He_I} (h \nu)\right.
		\nonumber \\
 &\hspace{-0.8em}&\hspace{-0.8em} \left.+ y_{\rm He_{II}} \left( r_{\rm
						    I}^{\rm He_I}
						    -r_{\rm I}^{\rm He_{II}}  \right)
  \sigma^{\rm ion}_{\rm He_{II}} (h \nu) \right],
\end{eqnarray}
where
$r_{\rm I}^i$ is the radius for the ionizing front of species $i$, i.e.,
the boundary between the ionized and un-ionized regions.  The first, second and third terms in the square bracket in the right hand side are for ionizations of \HI\ in the region of $r\gid r_{\rm I}^{\rm H_I}$, \HeI\ in the region of $r\gid r_{\rm I}^{\rm He_I}$, and \HeII\ in the \HeII\ region enclosed in $r_{\rm I}^{\rm He_{II}} \lid r \lid r_{\rm I}^{\rm He_I}$, respectively.

Dusts might have been produced by Type II supernovae even in the early universe~\citep{dwe2007}.  Recent discoveries of hot-dust free quasars at $z\sim 6$ may indicate possible differences in dust properties in early epoch of the universe~\citep{jia2006,jia2010}.  The dust abundances in the early universe is, however, still very uncertain.  The effects of dust are then not considered in this study.  In the energy region of $E_\gamma \gid 7.5$~eV (corresponding to the frequency of $\lambda \lid 0.17~\mu$m) related to the UV pumping, an estimate of mean extinction $\tau_{\rm ext}$ of the diffuse ISM with solar metallicity indicates $\tau_{\rm ext}/N_{\rm H} \lid O(10^{-23})$ cm$^2$~\citep{zub2004} where $N_{\rm H}$ is the hydrogen column density.  If abundances of dust are similar to that in the solar neighborhood, the extinction length scale would be
\begin{equation}
 l \equiv \frac{1}{\left[\tau_{\rm ext}/N_{\rm H}\right] n_{\rm H}} \lid O(0.1~{\rm Mpc}) \left(\frac{1+z}{10}\right)^{-3} \left(\frac{\delta_b}{10^3}\right)^{-1}.
\end{equation}

\subsection{Grid Size}

In order to calculate evolutions of the \HII\ region precisely, a
grid spacing should be small enough to correspond to an optically thin
distance~\citep{mad1997}.  \citet{mad1997} have set the ionizing
photon flux to be zero for energies less than 1.8 ryd, the \HeI\
ionization potential to prevent too rapid a growth of the \HI\
ionizing front.  In the present setting, the distance corresponding to optical depth of $\Delta \tau_\nu
=1$ is given by
\begin{equation}
 \Delta r\sim \frac{1}{n_{\rm H}(z) \sigma_\nu^{\rm H_I}}\gid
  10^{-3}~{\rm Mpc}~\left(\frac{1+z}{9.7}\right)^{-3}
  \delta_{\rm b}^{-1} ,
\end{equation}
where the ionization cross section of $\sigma_\nu^{\rm H_I}\ \lid 6\times
10^{-18}$~cm$^2$ [e.g., fig. 2.2 in \citet{ost1989}] was used.  The
grid size in this calculation is $\sim 0.5$~kpc.
This is optically thick for $\delta_{\rm b}=10^3$ or $10^4$.  Although
optically thin grid spacings thus could not be used in this treatment of coupled hydrodynamics and chemistry, we do not focus
on details of the evolution of ionizing front in this study.

The approximated treatment taken here is to account for an optical depth in the
photo-induced heating and reaction rates.  We input the average number
density of photon $n_\gamma(E_\gamma)$ over the grid interval
region into equations (\ref{heating}) and (\ref{ionization}).  The average
number density over the grid interval between $r$ and $r+\Delta r_{\rm
1grid}$ is given by
\begin{eqnarray}
n_\gamma(E_\gamma) c &=&\frac{\int_0^{\Delta r_{\rm 1grid}} F_\nu(r)/(h\nu)
 \exp\left[-\tau_\nu(dr^\prime)\right]dr^\prime}{\int_0^{\Delta r_{\rm 1grid}}
 dr^\prime} \cdot \frac{d\nu}{dE_\gamma}\nonumber \\
&=&\frac{F_\nu(r)}{ h^2\nu} \frac{1-\exp(-\Delta \tau_\nu)}{\Delta \tau_\nu},
\end{eqnarray}
where $\Delta \tau_\nu$ is the optical depth between $r$ and $r+\Delta
r_{\rm 1grid}$.

\subsection{Time Scales}

The time scale of light front move is
\begin{equation}
 \Delta t_{\rm l} \equiv \frac{r}{c} =3~{\rm Myr} \left(\frac{r}{\rm
						   Mpc}\right) .
\end{equation}
The time scale of recombination in the fully-ionized state is
\begin{eqnarray}
 \Delta t_{\rm rec}&\hspace{-0.8em}\equiv&\hspace{-0.8em}\frac{1}{n_e \alpha_{\rm B}} \nonumber \\
 &\hspace{-0.8em}\sim&\hspace{-0.8em} 7\times 10^2~{\rm
  Myr}~T_4^{0.845}~(1+2y_{\rm He})^{-1} \left(\frac {1+z}{9.7}\right)^{-3}
  \delta_{\rm b}^{-1}, \nonumber \\
\end{eqnarray}
where
$\alpha_{\rm B}\sim 2.6\times 10^{-13}~T_4^{-0.845}$~cm$^3$~s$^{-1}$
~\citep{mad1997} is the
recombination coefficient to the excited states of
hydrogen, and 
$T_4$ is the gas temperature in units of $10^4$~K.

The ionization time scale at around the ionization front of $\tau_\nu=0$
is
\begin{eqnarray}
 \Delta t_{\rm ion} &\equiv & \frac{n_{{\rm H}_{\rm I}}}{\dot{n_{{\rm
  H}_{\rm I}}}} = \left(\int_{13.6~{\rm eV}}^\infty n_\gamma (E_\gamma)
		   \sigma^{\rm ion}_{{\rm H}_{\rm I}} (E_\gamma) c
		   ~dE_\gamma  \right)^{-1} \nonumber \\
&\sim& 10~{\rm yr} \left(\frac{r}{0.1~{\rm Mpc}}\right)^2.
\label{eq4}
\end{eqnarray}

\subsection{Evolutions of Ionization Fronts}

The ionization time scale [equation\ (\ref{eq4})] is much smaller than the scale now treated in the hydrodynamical
calculation.  The photoionization, therefore, can be regarded to occur
instantaneously.  The total number of photon causing the ionization satisfies
\begin{eqnarray}
\lefteqn{\sum_i S_\gamma \int_0^{t-\Delta t_i\left(r_{\rm I}^{i\prime}\right)} f_i (t_\ast) dt_\ast} \nonumber\\
&\hspace{-1.em}\approx& \hspace{-1.em}\sum_i \left[\frac{4\pi \left(r_{\rm I}^{i\prime}\right)^3}{3} n^{0\prime}_{j(i)} + \int_0^t dt_\ast \int_0 ^{r_{\rm I}^{i\prime}(t)} dr^\prime \left(4\pi {r^\prime}^2 n_e n_{j(i)}^{0\prime} \alpha_{{\rm B},i}\right)\right],
\label{eq23}
\end{eqnarray}
where
$S_\gamma$ is the emission rate of ionizing photon assumed to be independent of
time here, 
$f_i (t)$ is the number fraction of photon emitted at time $t$ which is used for the ionization of species $i$,
$\alpha_{{\rm B},i}$ is the recombination coefficient to excited states of species $i$, and
two quantities, i.e., $r_{\rm I}^{i \prime}=r_{\rm I}^i/a$ and
$n^{0\prime}_{j(i)}=n^0_{j(i)}~a^3$ [$j(i)=$H for $i=$\HI\ and $j(i)=$He for $i=$\HeI\ and \HeII]
, are measured in the comoving frame.  $n_{j(i)}^0$ is the initial abundance of neutral \HI\ or \HeI.  We note that for the case of He, a two-step ionization of \HeI\ $\rightarrow$ \HeII\ $\rightarrow$ \HeIII\ need two ionizing photons per one He atom inside the \HeII\ ionization front.  
The quantity
 $\Delta t_i=[r_{\rm I}^{i \prime}/(3c)(3H_0\sqrt[]{\mathstrut
\Omega_{\rm m}}/2)^{2/3}+t_0^{1/3}]^3-t_0$ is the time it takes for the light to travel the
distance $r_{\rm I}^i$ in the matter dominated universe with $t_0$ the age
of the universe at the initial time of the calculation.  The left hand side of the equation is the total number of photons used for ionization. The first term in the square bracket in the right hand side is the number of photons required to ionize the region of radius $r_{\rm I}^i$, while the second is number of times of recombination occurring inside $r_{\rm I}^i$.

A time evolution of ionization front of species $i$, i.e., $r_{\rm I}^i$ can be described by a solution to the equation of the total number of photon causing the ionization of $i$, i.e.,
\begin{equation}
  S_\gamma \int_0^{t-\Delta t_i\left(r_{\rm I}^{i\prime}\right)} f_i^{\rm eff} (t_\ast) dt_\ast \approx \frac{4\pi \left(r_{\rm I}^{i
				       \prime}\right)^3}{3} n^{0\prime}_{j(i)},
\label{eq17}
\end{equation}
where
\begin{equation}
f_i^{\rm eff} (t) = \sum_k (f_k-\delta f_k) P_k(i,t) ~~~~~\equiv  \sum_k f_k f_k^{\rm sur} (t) P_k(i,t)
\end{equation}
is the effective fraction of photons emitted at $t$ to ionize species $i$ eventually.  
$f_k$ is the fraction of photons emitted in energy range $k$ which is specified by $E_1\lid E_\gamma < E_2$ ($k$=1), $E_2\lid E_\gamma < E_3$ ($k$=2) and  $E_3\lid E_\gamma$ ($k$=3), where $E_1=13.60$~eV, $E_2=24.59$~eV and $E_3=54.42$~eV are the threshold energies of \HI, \HeI\ and \HeII, respectively.  
$\delta f_k$ is the absorption factor by photoionization evaluated at ionization front of $i$.  
$f_k^{\rm sur}=1-\delta f_k/f_k$ is therefore, the survival fraction of photons in energy range $k$.  
$P_k(i,t)$ is the fraction of photons emitted at time $t$ to ionize species $i$ which is calculated by a prescription given in Appendix\ \ref{appendix}.

The ionization cross sections have peaks just above threshold energies and rapidly decrease as the energy increases [equations\ (\ref{eqa1}), (\ref{eqa2}) and (\ref{eqa3})].  We, therefore, approximately identify recombinations of \HII, \HeII\ and \HeIII\ as losses of photons in energy ranges of $k=1$, $2$ and $3$, respectively, as a result of subsequent ionizations by photons in the energy ranges.  The amplitude of photon flux in energy range $k$ at the ionization front is reduced by a factor $f_k^{\rm sur} (t)$ in the present treatment.  The recombination of each species then involves photons in only one energy range, and a relation holds:
\begin{equation}
f_k^{\rm sur} (t) = 1- \frac{1}{f_k S_\gamma} \int_0^{r_{\rm I}^{i\prime}(t)} dr^\prime \left(4\pi {r^\prime}^2 n_e n_{j(i)}^{0\prime} \alpha_{{\rm B},i}\right) \frac{dt_\ast}{d(t_\ast-\Delta t_{\ast i})},
\end{equation}
where 
$i=$\HI\ (for $k=1$), \HeI\ (for $k=2$) and \HeII\ (for $k=3$).  
When all photons of $E_1 \lid E_\gamma < E_2$ are used for recombination inside the \HI\ ionization front ($f_1^{\rm sur} <0$ is then derived from naive estimate), photons of $E_2 \lid E_\gamma < E_3$ are supplementarily used with $f_1^{\rm sur}$ fixed to be zero.  The ratio $dt_\ast/d(t_\ast-\Delta t_{\ast i})$ derives from the time integral range of the left hand side and the second term of the right hand side of equation (\ref{eq23}).  We evaluate ionization fronts at any $t$.  They are determined only by ionizing photons arriving at the ionization fronts by the time $t$ which should be emitted by $t-\Delta t_i$.

As for the input for $f_k^{\rm sur} (t)$ values, we assume that the temperature of ionization regions is $T_{\rm gas}=3\times 10^4$~K which corresponds to the pressure of $P\sim 1$~eV cm$^{_3}$ in \HII\ and \HeIII\ regions of $\delta_{\rm b}=10^3$ (see Fig.\ \ref{fig4}).  The number density of electron is approximately given by
\begin{eqnarray}
n_e&=&(1+2y_{\rm He})n_{\rm H}~~~~~(0\lid r < r_{\rm I}^{\rm He_{II}}) \nonumber\\
   & &~(1+y_{\rm He})n_{\rm H}~~~~~(r_{\rm I}^{\rm He_{II}}\lid r < r_{\rm I}^{\rm H_I}) \nonumber\\
   & &~~~~y_{\rm He}n_{\rm H}~~~~~~~~~~(r_{\rm I}^{\rm H_I} \lid r < r_{\rm I}^{\rm He_{I}}) \nonumber\\
   & &~~~~~0~~~~~~~~~~~~~(r_{\rm I}^{\rm He_I} \lid r) \nonumber
\end{eqnarray}
Using the fixed temperature and the electron number density given above, $f_k^{\rm sur} (t)$ is calculated.

Equation\ (\ref{eq17}) is rewritten in the form of
\begin{equation}
{r^{i\prime}_{\rm I}}^3- \frac{3 S_\gamma}{4\pi n_{j(i)}^{0\prime}} \int_0^{t-\Delta t_i(r^{i\prime}_{\rm I})} f_i^{\rm eff} (t_\ast)~dt_\ast=0.
\label{eq18}
\end{equation}
The integral is numerically estimated discretely, i.e.,
\begin{equation}
\int_0^{t-\Delta t_i(r^{i\prime}_{\rm I})} f_i^{\rm eff} (t_\ast)~dt_\ast \rightarrow \sum_{t_\ast \lid t} [f_i^{\rm eff} (t_\ast-\Delta t_{\ast i}) \delta(t_\ast-\Delta t_{\ast i})],
\end{equation}
where $f_i^{\rm eff} (t-\Delta t_i)$ is the effective fraction for photons emitted at time $t-\Delta t_i$, which arrive at the ionization front of $i$ at time $t$.  $r_{\rm I}^i (t)$ values are obtained by solving equation\ (\ref{eq18}).  $f_i^{\rm sur} (t)$ values are simultaneously obtained.  The solution $r_{\rm I}^i (t)$ is used as the evolution of the ionization front in this calculation.

\begin{figure}
\includegraphics[width=84mm]{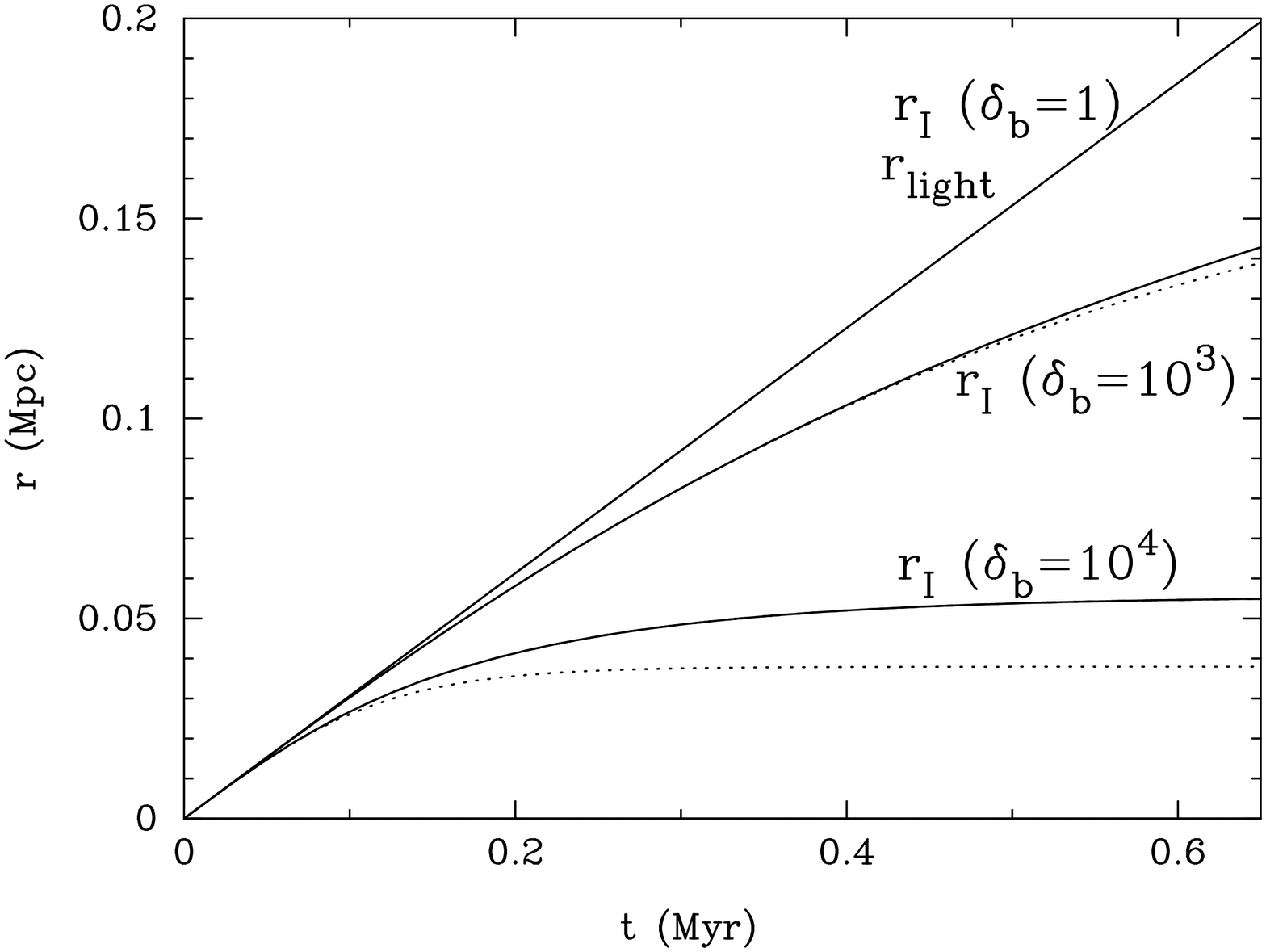}
\caption{Radius of the light front and those of the ionization fronts
 for \HI\ (solid lines), \HeI\ (dashed) and \HeII\
 (dotted).  Lines correspond to the cases of density excess values of $\delta_{\rm b}=1$,
 $10^3$ and $10^4$.}
\label{fig3}
\end{figure}

Figure\ \ref{fig3} shows the
radii of the ionization fronts for \HI\ (solid lines), \HeI\ (dashed) and \HeII\ (dotted) as well as the radius of the
light front $r_{\rm light}$.  Curves for three cases, i.e., $\delta_{\rm b}=1$, $10^3$ and
$10^4$ are drawn.  The curves for $\delta_{\rm b}=1$ almost overlap that for
the light front.  This means that there is only a narrow region where
the hydrogen exists as a neutral atom inside the light radius of $\sim
O(0.1)$~Mpc.  In this case signals of UV radiation through \CI,
\CII\ and \OI\ would be very weak since their abundances are very small after a very short time interval between the UV photon arrival and the ionization.  We then calculate the evolution of the chemical
abundances in the cases of $\delta_{\rm b}=10^3$ and $10^4$, in which relatively large regions of neutral hydrogen exist inside the light radius since it takes long time to ionize dense regions.  The density
excess values of $\delta_{\rm b}=10^3$ and $10^4$ seem somewhat high.  The
cases, however, would quite roughly simulate situations where high density clouds
exist in the line of
sight towards the UV emitting object.

\begin{figure}
\includegraphics[width=84mm]{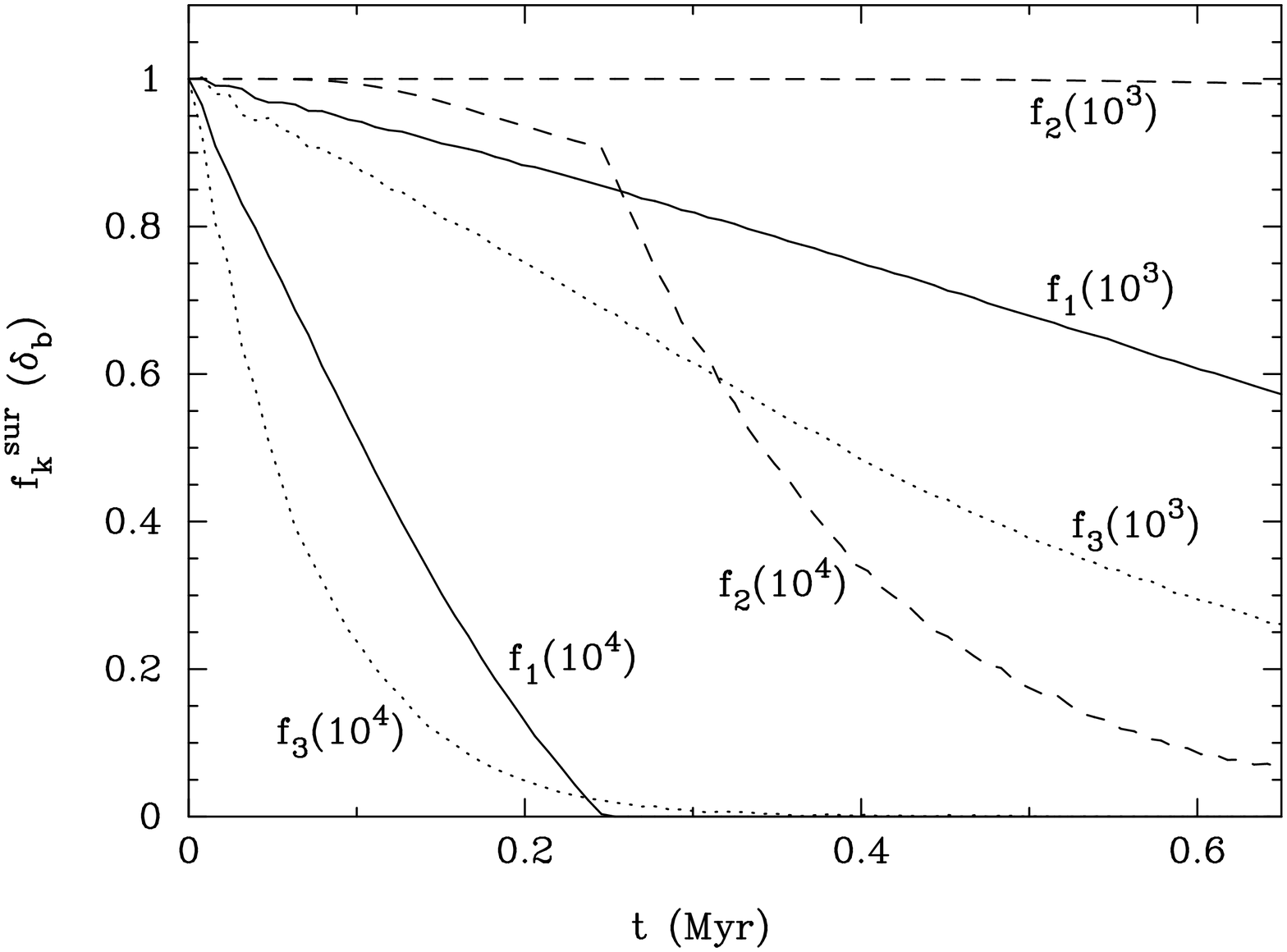}
\caption{Survival fractions of photons in energy ranges $k$ after traveling from a QSO to ionization fronts, i.e., $f_k^{\rm sur} (t)$ in the cases of density excess values of $\delta_{\rm b}=10^3$ and $10^4$.}
\label{fig_ex}
\end{figure}

Figure\ \ref{fig_ex} shows the calculated survival fractions of photons in three energy ranges after traveling from a QSO to ionization fronts, i.e., $f_k^{\rm sur} (t)$ for cases of $\delta_{\rm b}=10^3$ and $10^4$.  The three energy ranges are as follows: (1) $E_1 \lid E_\gamma < E_2$, (2) $E_2 \lid E_\gamma < E_3$ and (3) $E_3 \lid E_\gamma$.  The survival fractions are used in this calculation to take account of reduced UV fluxes at ionization fronts.

\section{Results}\label{sec4}

\subsection{Evolution of an Ionized Region}

\subsubsection{Case of $\delta_{\rm b}=10^3$}

\begin{figure}
\includegraphics[width=84mm]{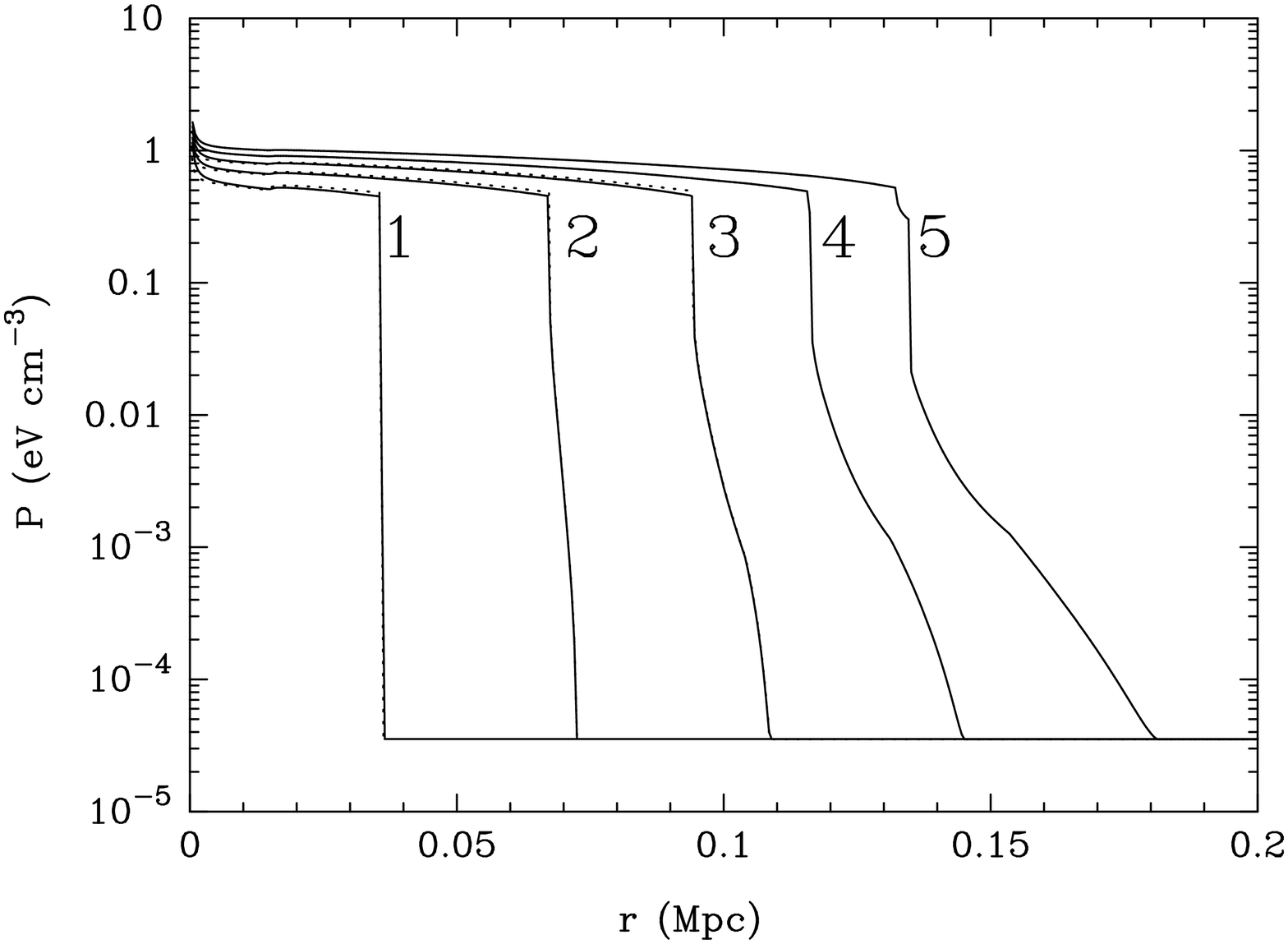}
\includegraphics[width=84mm]{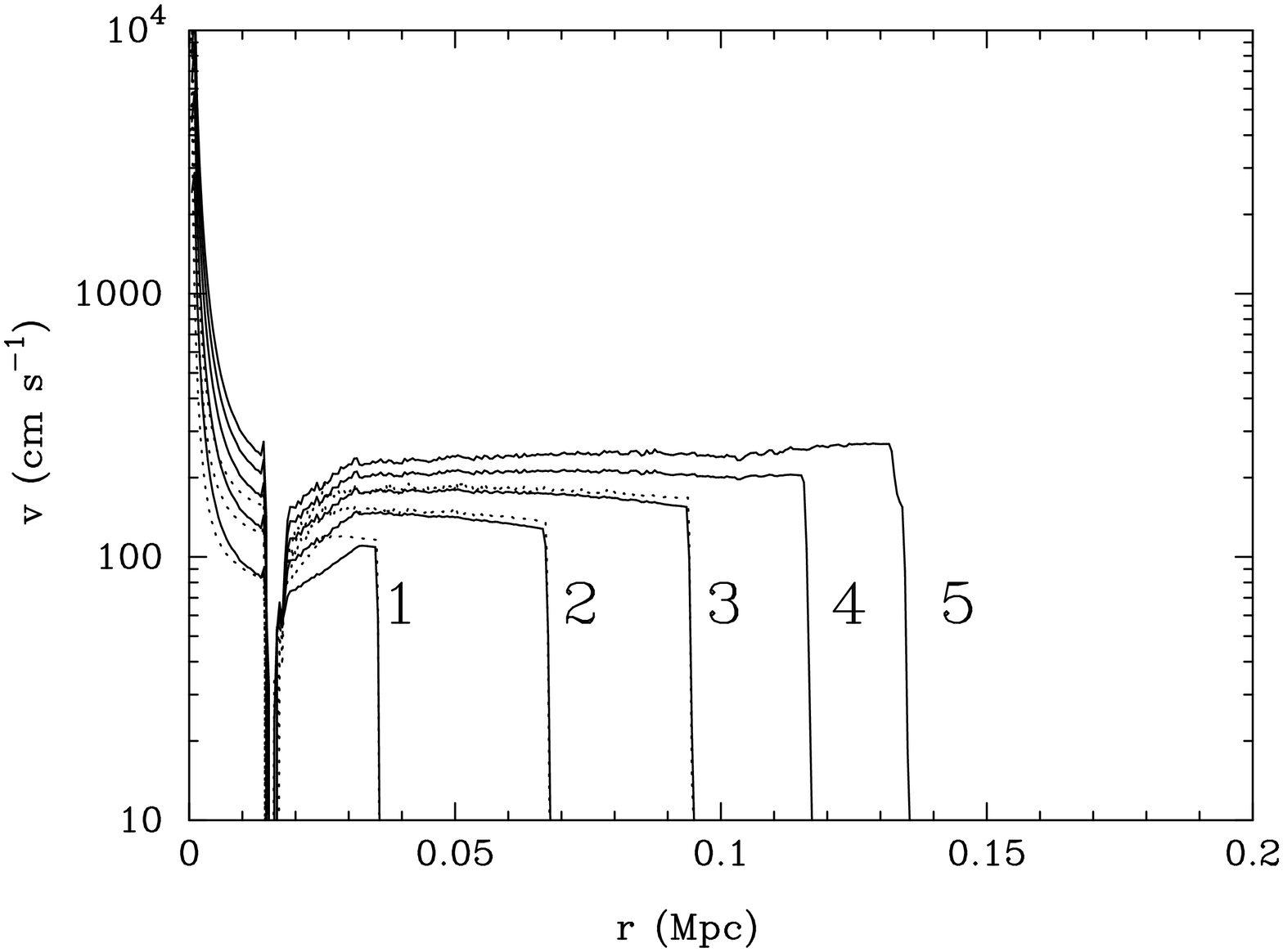}
\caption{Pressure (upper panel) and velocity (lower) as a function of
 the radius $r$ at time $t$=0.12~Myr (1), 0.24~Myr (2), 0.35 Myr (3),
 0.47~Myr (4) and 0.59~Myr (5) for the case of $\delta_{\rm b}=10^3$.  Solid (dotted) lines correspond to calculations with grid number 400 (800).}
\label{fig4}
\end{figure}

Figure\ \ref{fig4} shows the pressure (upper panel) and the velocity
(lower) as a function of the radius at time $t=0.12$~Myr (1), 0.24~Myr
(2), 0.35 Myr (3), 0.47~Myr (4) and 0.59~Myr (5).  A bump in the
pressure curve appears at $r\sim0.015$ Mpc although it is difficult to read it clearly.  This is caused by the discrete
treatment of space.  The
point corresponds to the innermost grid where a light front abundant in
energetic UV photons arrives faster than ionization fronts.  Environments
rich in energetic photons make it possible to heat up gas particles
efficiently.  Inside this bump, gas particles receive UV photons
both of relatively high and low energies simultaneously leading to a
relatively inefficient heating.  The bump causes a discontinuity in velocity.  Although these
discontinuities are unphysical, they do not affect the result of
structures of ionized region at larger scales.  We then do not
search for a solution to the bump here.

Dotted lines in Fig.\ \ref{fig4} show the same physical quantities for another calculation with a larger grid number of 800 which was performed for a shorter time.  Differences in pressure, or temperature, of the two calculations with different spatial resolutions are seen to be small.  We then do not expect so large errors in line signals derived from the calculated temperatures, as described below.  Nevertheless, more complete estimations with high resolution calculations are desired.

\begin{figure}
\includegraphics[width=84mm]{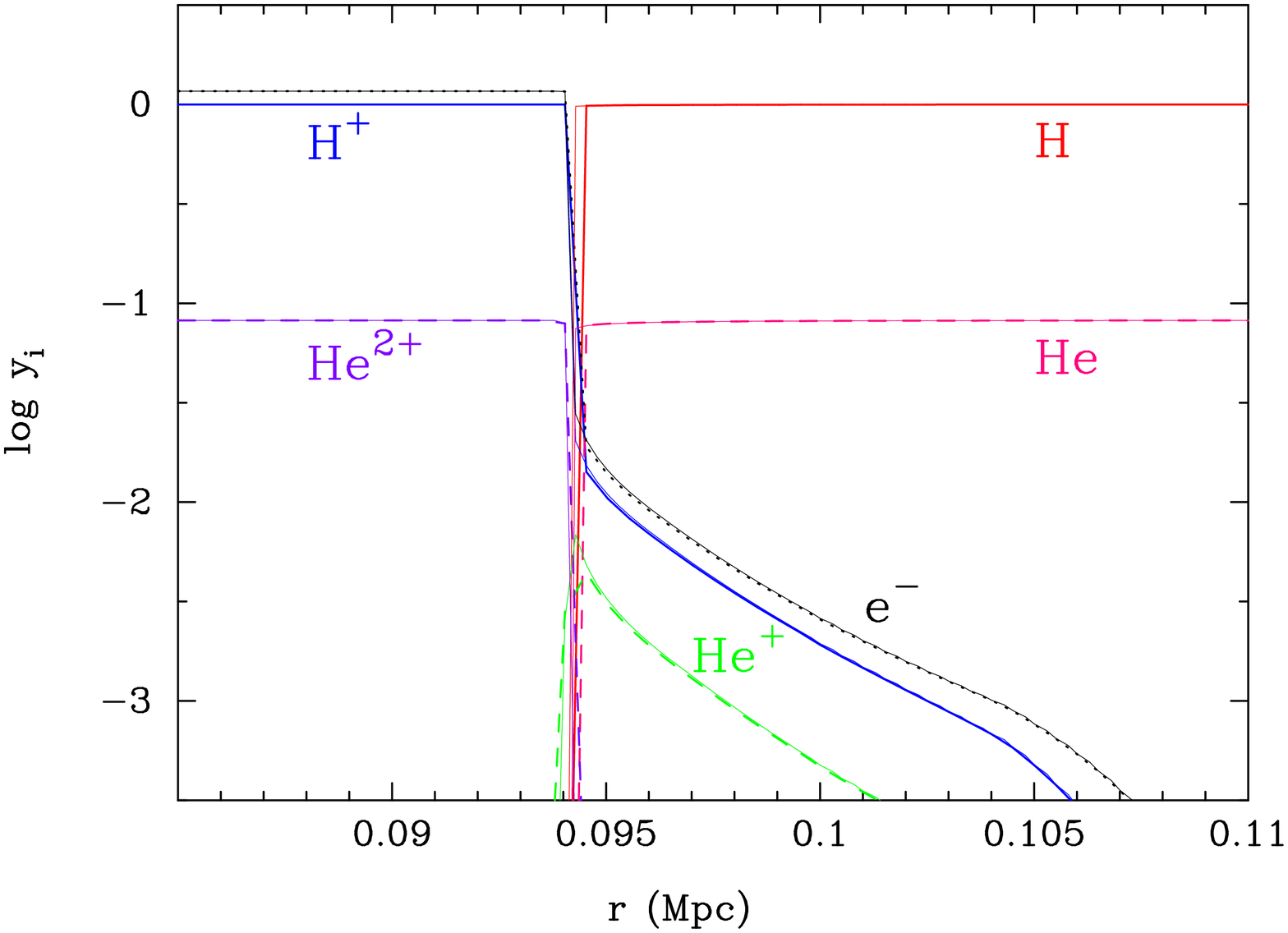}
\includegraphics[width=84mm]{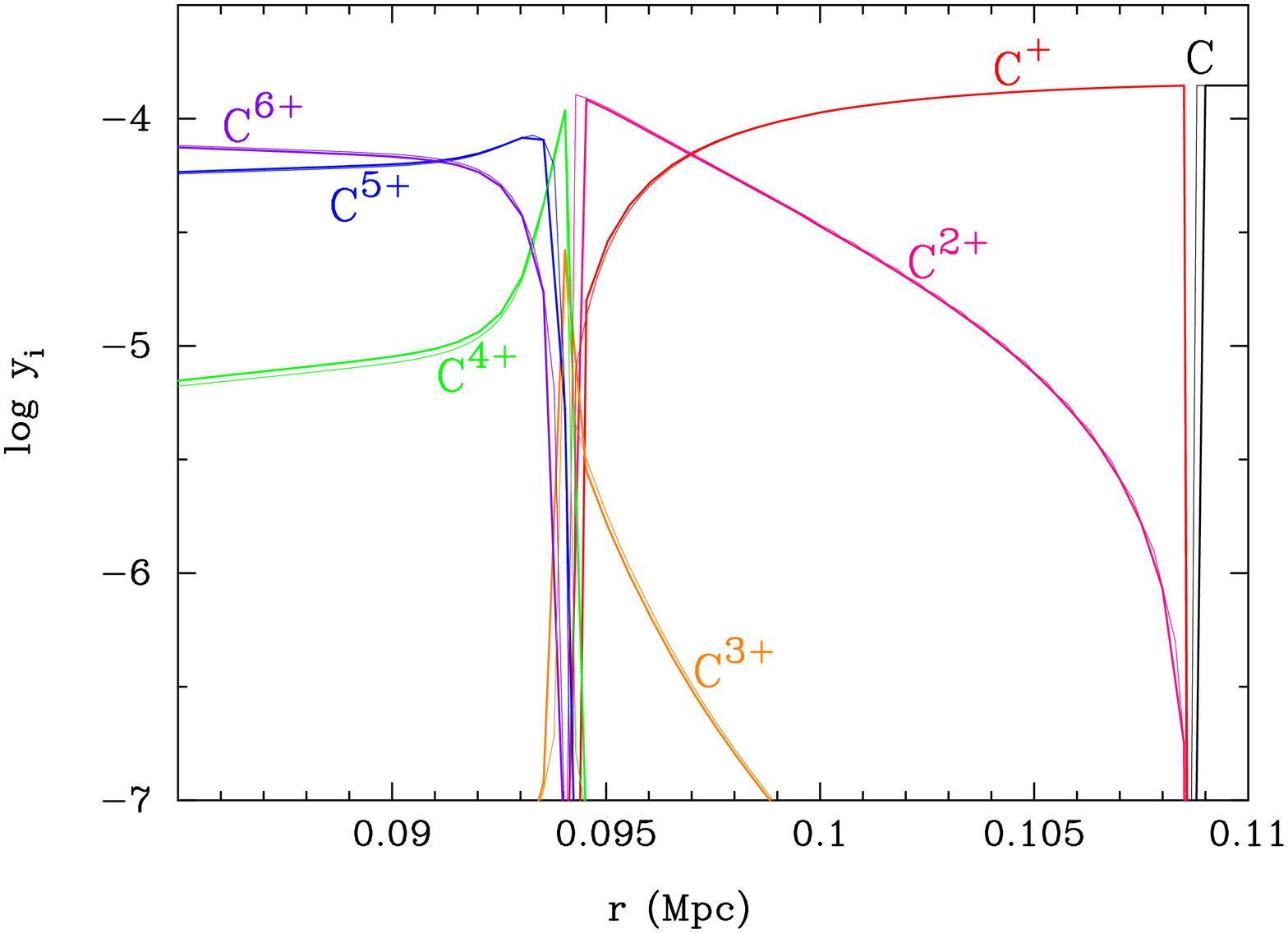}
\includegraphics[width=84mm]{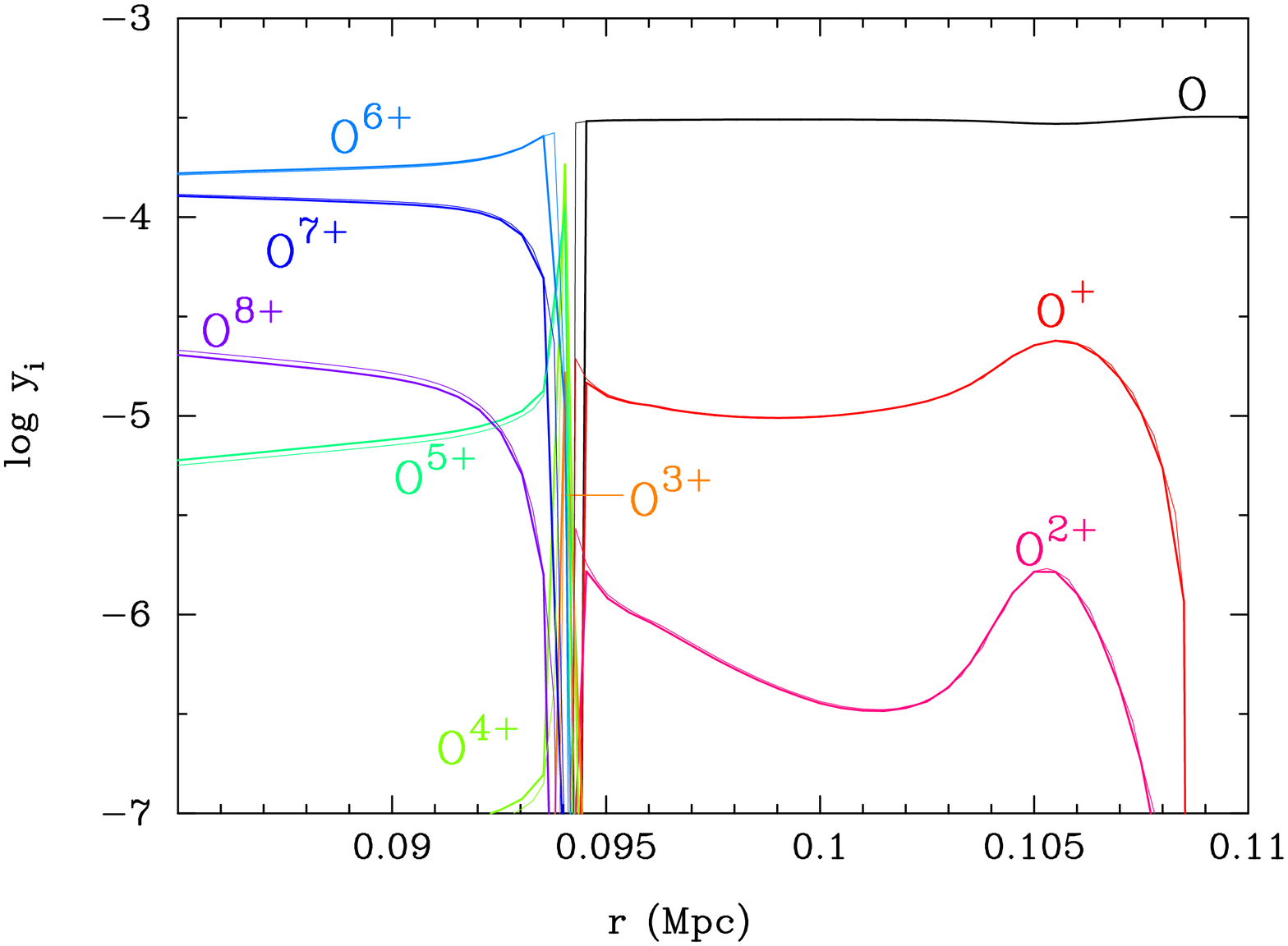}
\caption{Chemical abundances of hydrogen, helium and electron (top
 panel), carbon (middle) and oxygen (bottom) as a function of the radius
 $r$ at $t=0.35$~Myr (3) for the case of $\delta_{\rm b}=10^3$.  Thick (thin) lines correspond to calculations with grid number 400 (800).}
\label{fig5}
\end{figure}

Figure\ \ref{fig5} shows chemical abundances of hydrogen, helium and
electron (top panel), carbon (middle) and oxygen (bottom) as a function
of the radius $r$ at $t=0.35$~Myr (3).   Thin solid lines show the same physical quantities calculated with 800 grid points.

\subsubsection{Case of $\delta_{\rm b}=10^4$}

\begin{figure}
\includegraphics[width=84mm]{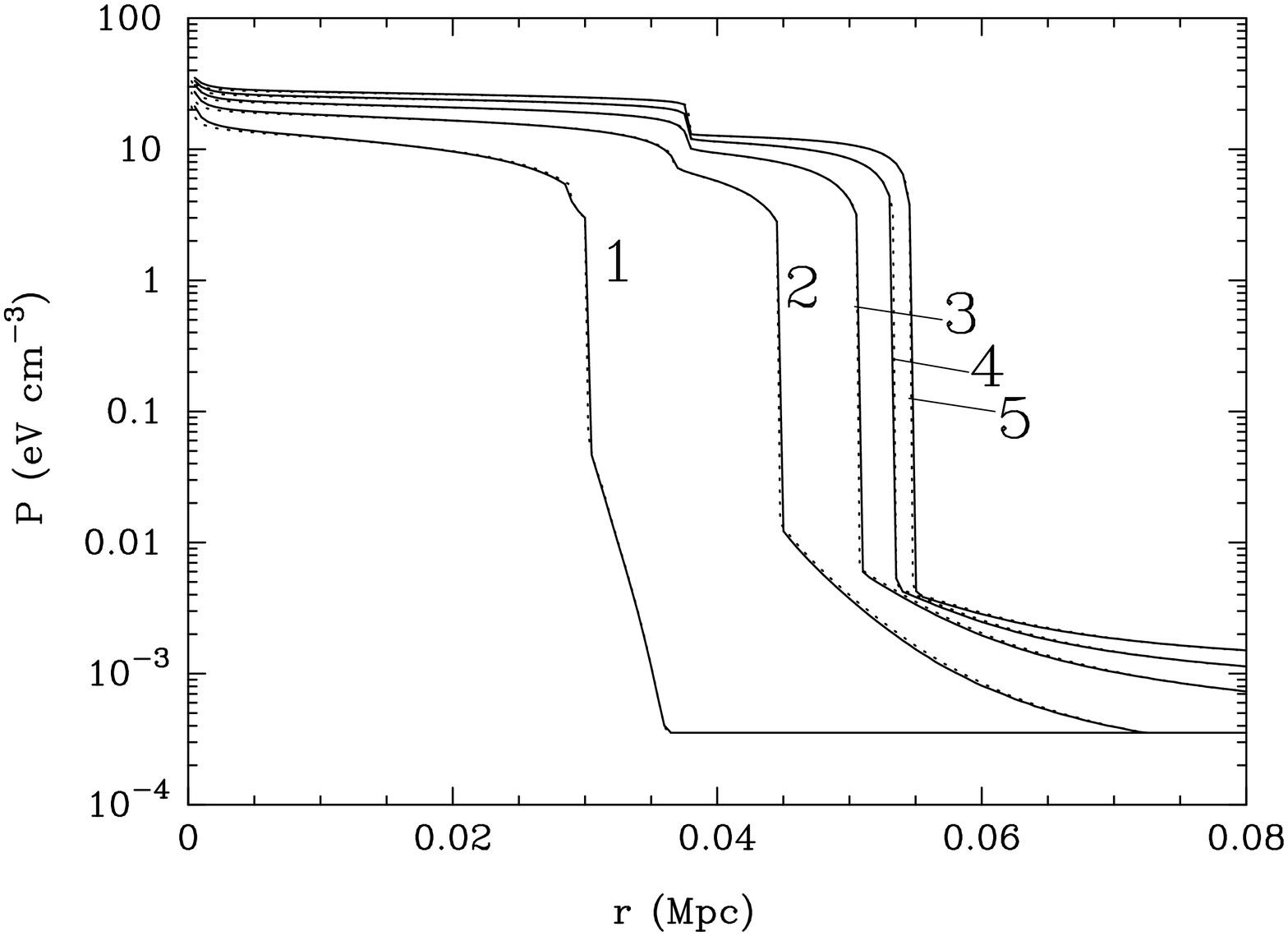}
\includegraphics[width=84mm]{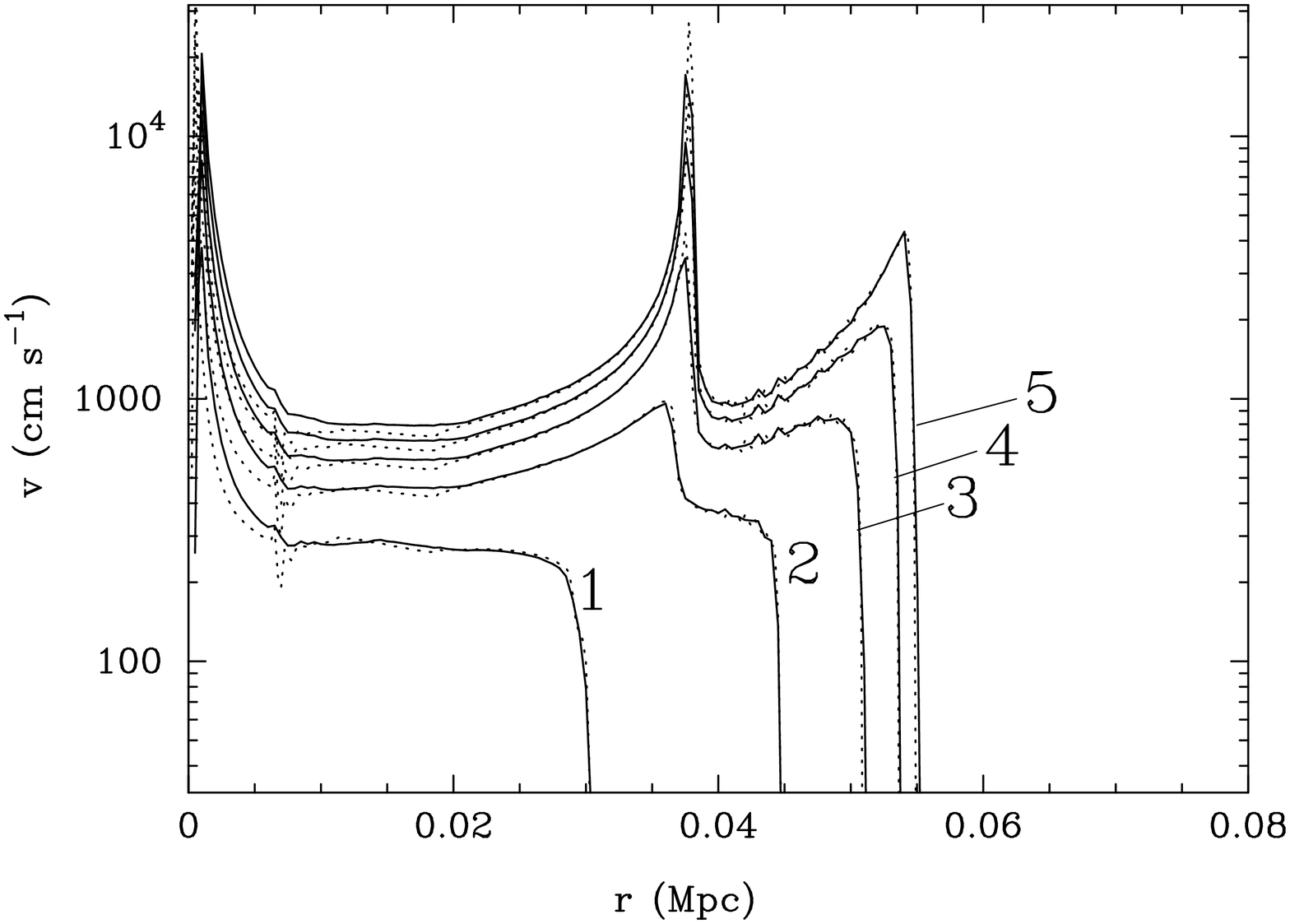}
\caption{Pressure (upper panel) and velocity (lower) as a function of
 the radius $r$ at time $t$=0.12~Myr (1), 0.24~Myr (2), 0.35 Myr (3),
 0.47~Myr (4) and 0.59~Myr (5) for the case of $\delta_{\rm b}=10^4$.  Solid (dotted) lines correspond to calculations with grid number 400 (800).}
\label{fig6}
\end{figure}

Figure\ \ref{fig6} shows the pressure (upper panel) and the velocity
(lower) as a function of the radius at time $t=0.12$~Myr (1), 0.24~Myr
(2), 0.35 Myr (3), 0.47~Myr (4) and 0.59~Myr (5).  A step
in pressure exist at $r\sim 0.04$~Mpc (the ionization front of \HeII).  Sharp decreases in pressure cause some structures in velocity which are seen in Fig.\ \ref{fig6}.  Dotted lines show the same physical quantities for another calculation with grid number of 800.

\begin{figure}
\includegraphics[width=84mm]{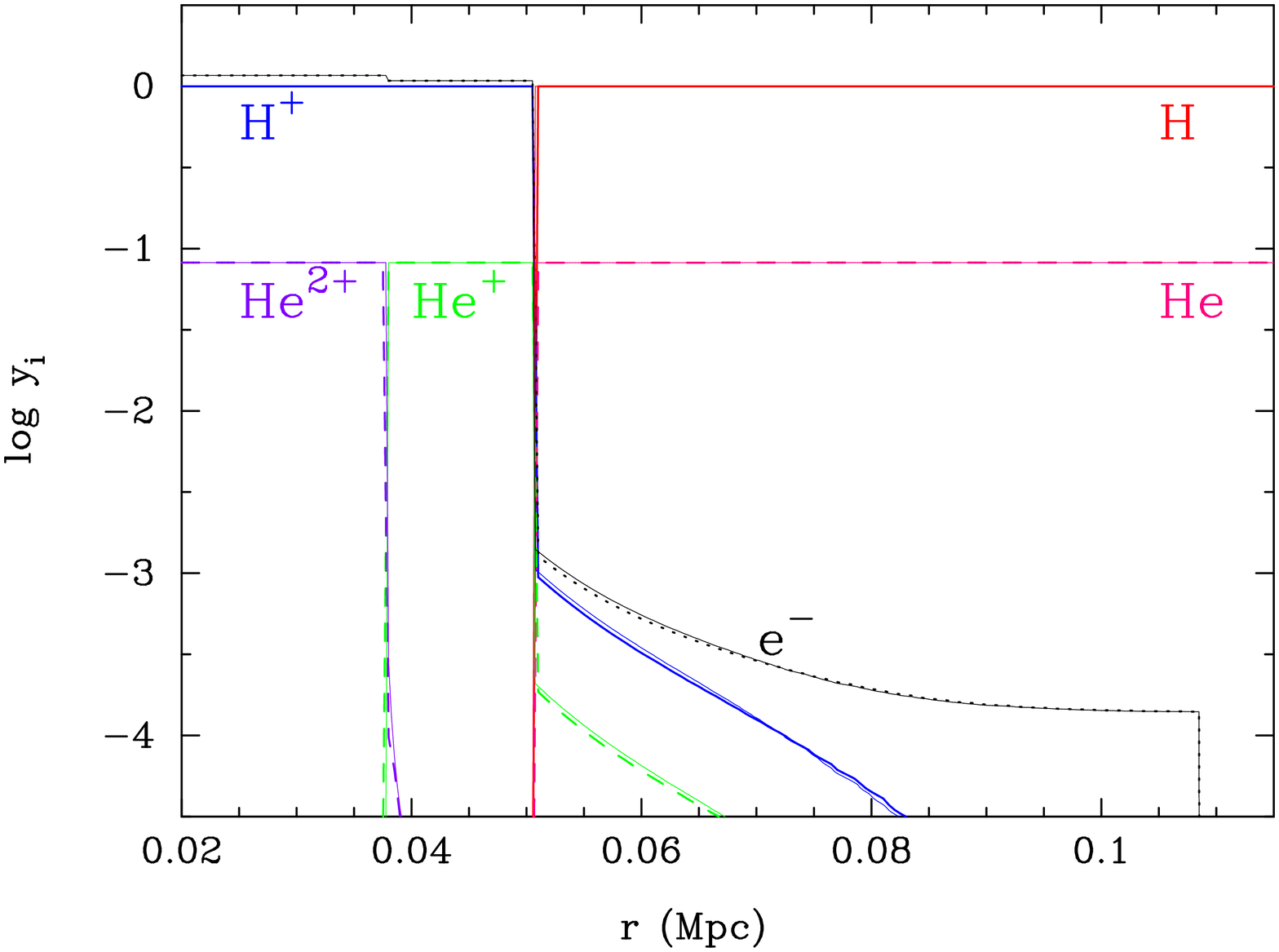}
\includegraphics[width=84mm]{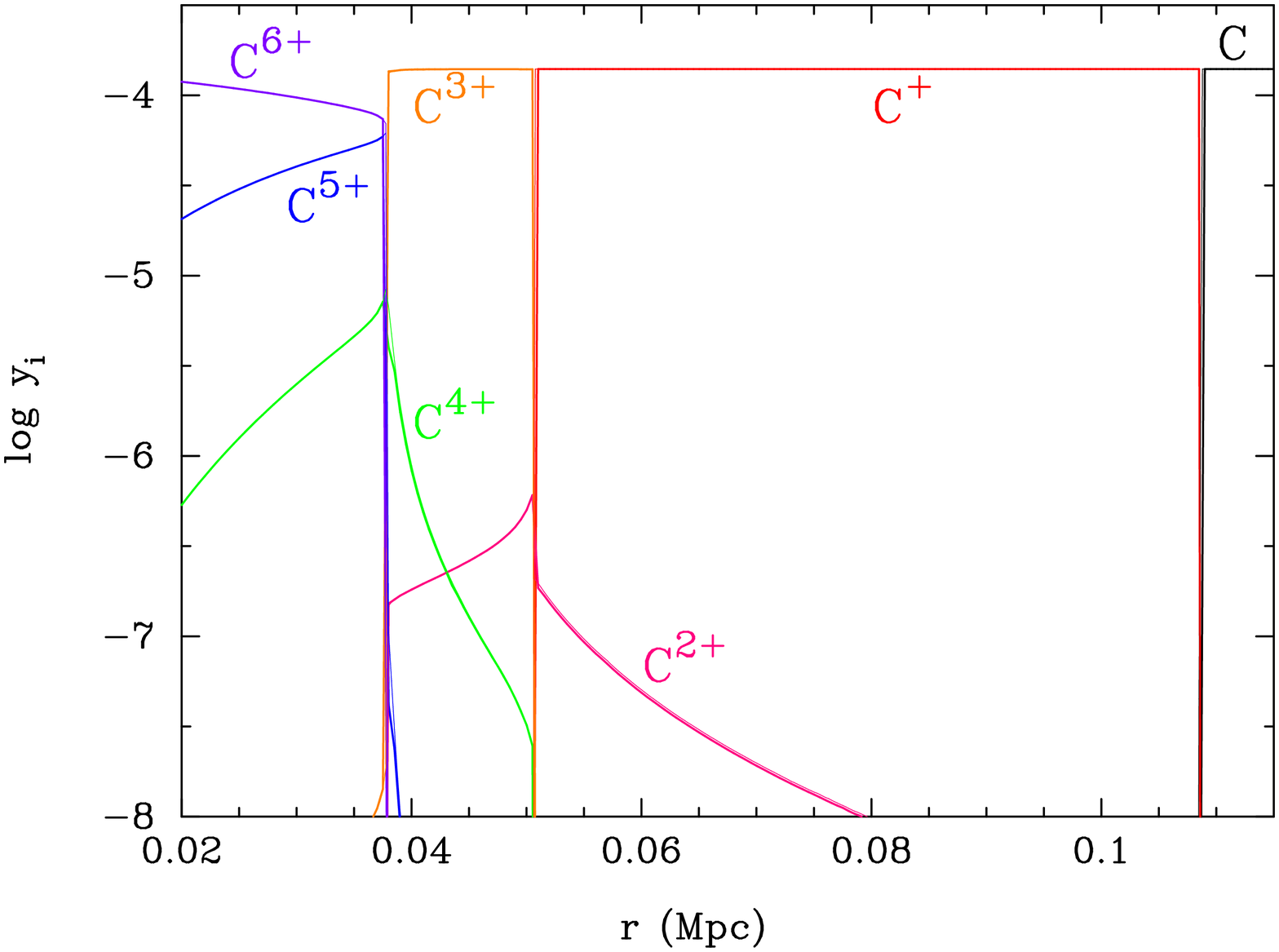}
\includegraphics[width=84mm]{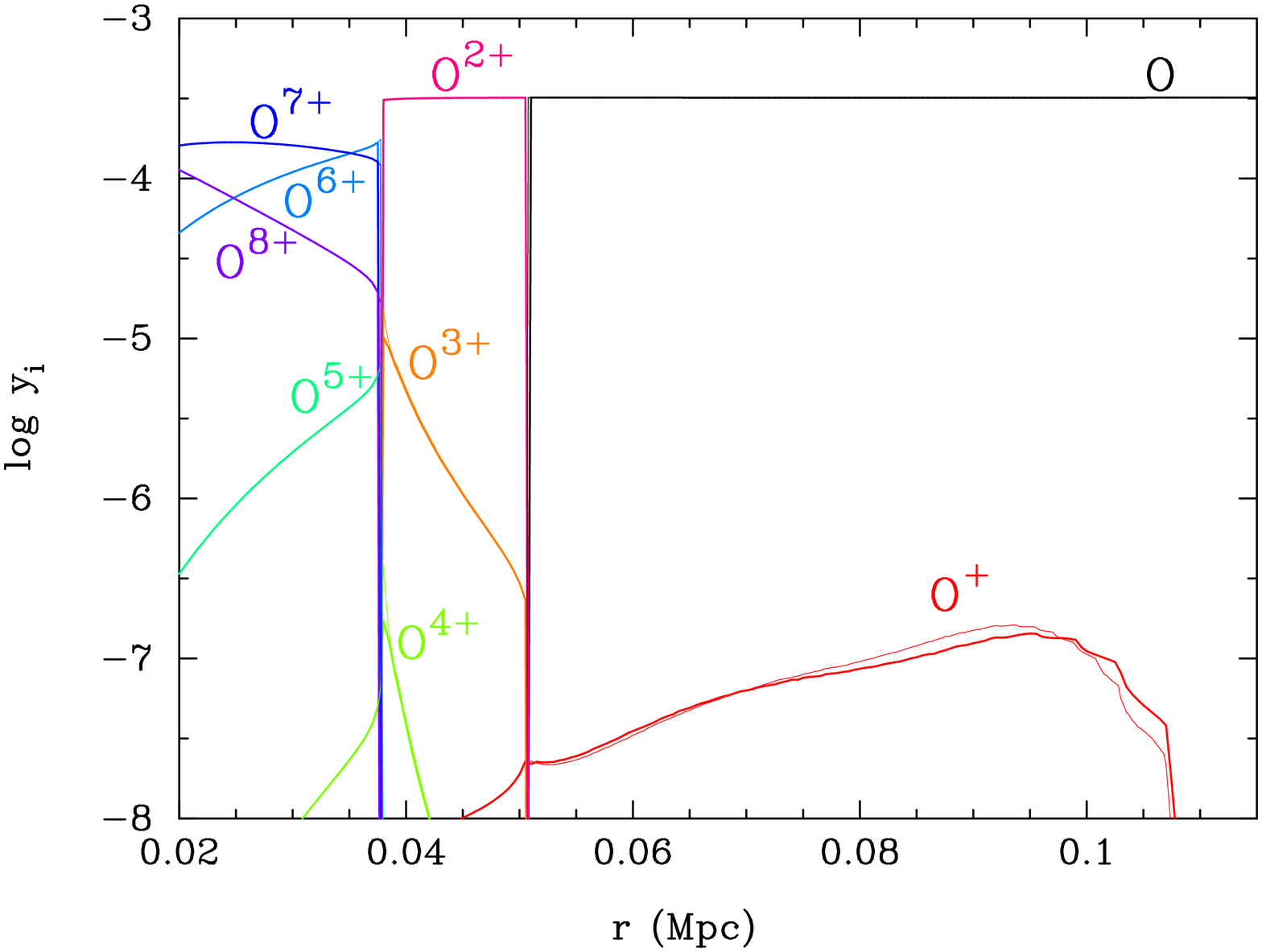}
\caption{Chemical abundances of hydrogen, helium and electron (top
 panel), carbon (middle) and oxygen (bottom) as a function of the radius
 $r$ at $t=0.35$~Myr (3) for the case of $\delta_{\rm b}=10^4$.  Thick (thin) lines correspond to calculations with grid number 400 (800).}
\label{fig7}
\end{figure}

Figure\ \ref{fig7} shows chemical abundances of hydrogen, helium and
electron (top panel), carbon (middle) and oxygen (bottom) as a function
of the radius $r$ at $t=0.35$~Myr (3).  Thin solid lines show the same physical quantities calculated with 800 grid points.

\subsubsection{Effect of UV background}\label{sec413}

A UV background can cause a heating of gas.  In addition, even a weak UV background in the epoch of the hydrogen reionization had quickly ionized neutral carbon with low ionization potential.  For example, consider a region of density excess $\delta_{\rm b}=10^4$ at $1+z=10$ exposed to a soft UV background with flux $J=10^{-21} J_{21} (E_\gamma/13.6~{\rm eV})^{-1.5}$~ergs~s$^{-1}$~cm$^{-2}$~Hz$^{-1}$~str$^{-1}$ in $E_\gamma \lid 13.6$~eV.  Assuming the gas temperature of 300~K and that the abundance of electron is equal to that of \CII, the equilibrium abundance ratio of \CII\ to \CI\ is approximately $n_{\rm C_{II}}/n_{\rm C_I}=3.8\times 10^4~J_{21}$.  Even a $J_{21}$ value as low as $10^{-4}$, which is possibly achieved very soon after the onset of Population III star formation~\citep*[see e.g.][]{hai2000} is enough to ionize most of the carbon.  In order to check effects
of UV background the following extra calculations are performed.

We assume that there are UV sources which appear $0.1H(z=8.7)^{-1}$ before the onset of the point source.  $H(z)^{-1}$ is the Hubble time at redshift $z$, and the flat $\Lambda$CDM universe has been assumed, i.e., $H(z)=H_0\sqrt[]{\mathstrut \Omega_{\rm m}(1+z)^3+\Omega_\Lambda}$.  The UV background is, therefore, switched on at $z=9.8$ corresponding to the cosmic age of $0.50$~Gyr.

The UV spectrum is roughly given by
\begin{eqnarray}
 J_\nu=\left\{ \begin{array}{l}
10^{-21} J_{21}~(E_\gamma/13.6~{\rm eV})^{-1.5}~{\rm ergs/s/Hz/str}\\
~~~~~~~~~~{\rm for}~E_\gamma \lid 13.6~{\rm eV}\\
10^{-21} J_{21} \epsilon_X (E_\gamma/13.6~{\rm eV})^{-1.5}~{\rm ergs/s/Hz/str}\\
\times \exp[-10^{22} (\sigma^{\rm ion}_{{\rm H_I}, {\rm cm}^2}+0.08\sigma^{\rm ion}_{{\rm He_I}, {\rm cm}^2})]\\
~~~~~~~~~~{\rm for}~E_\gamma > 13.6~{\rm eV},
\end{array}\right.
\end{eqnarray}
where
$\sigma^{\rm ion}_{i, {\rm cm}^2}$ is the photoionization cross section of $i$ in units of cm$^2$.
This form is adopted from~\citet{hai2000} although the modulation factor is neglected, and the spectral index is different.  The exponential factor corresponds to the absorption above ionization energy threshold of \HI\ by the neutral IGM under the assumption of a hydrogen column density of $N_{\rm H}=10^{22}$~cm$^{-2}$.  A parameter $\epsilon_X$ is the ratio of X-ray to UV flux: $\epsilon_X=0$ for pure stellar sources and $\epsilon_X\sim 1$ for typical quasar spectra.  The parameter $\epsilon_X$ would thus parametrize fractions of contributions of stars and quasars.  Two cases of $\epsilon_X=0$ (soft) and 1 (hard spectrum) with a fixed value of $J_{21}=1$ are calculated assuming the homogeneous UV backgrounds.

In both cases of $\epsilon=0$ and 1, \CI\ is completely ionized after the UV exposures over 0.1 Hubble time at $z=8.7$ (88 Myr).  In the case of $\epsilon=1$, the ionization by hard UV photons heat up the homogeneous gas.  The temperature at $z=8.7$ is then $10^3$~K.  We first calculated one-zone chemical abundances in an exposure to the UV background from $z=9.8$ to $8.7$.  We then calculated chemical abundances around the point source as a function of radius taking account of ionizations by the point source and the homogeneous UV backgrounds using results of one zone chemical evolution as input.

\begin{figure}
\includegraphics[width=84mm]{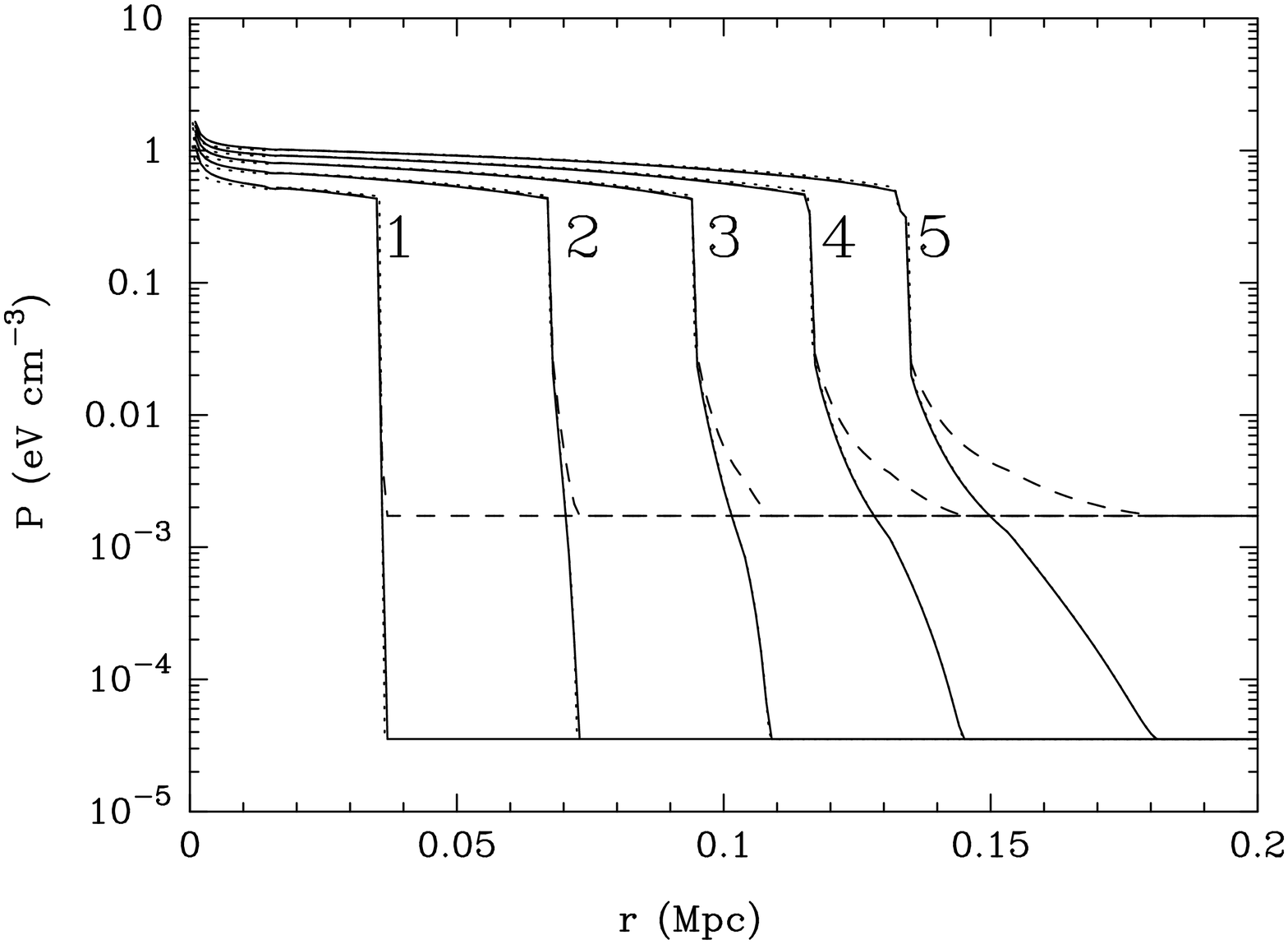}
\includegraphics[width=84mm]{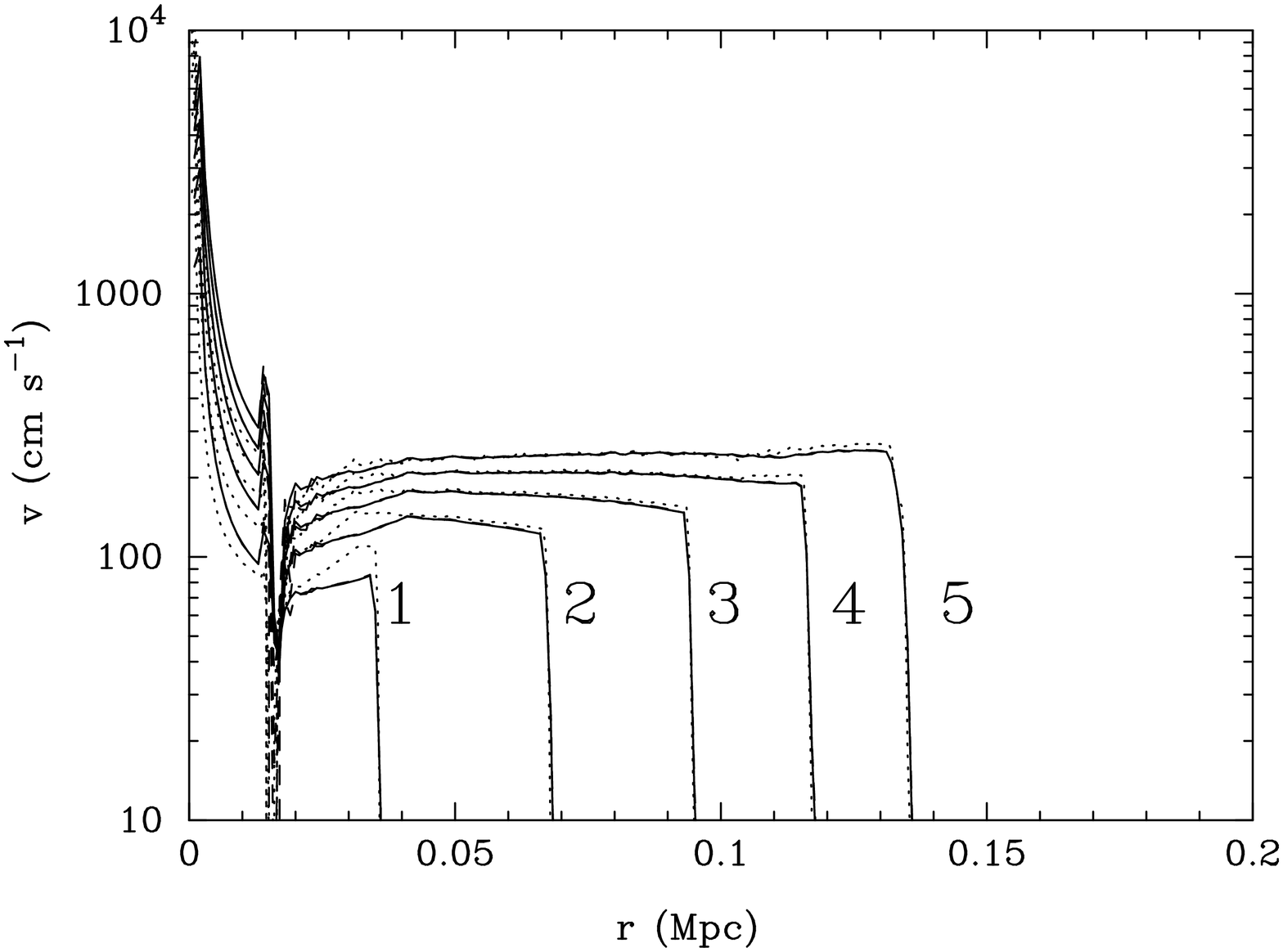}
\caption{Pressure (upper panel) and velocity (lower) as a function of
 the radius $r$ at time $t$=0.12~Myr (1), 0.24~Myr (2), 0.35 Myr (3),
 0.47~Myr (4) and 0.59~Myr (5) for the case of $\delta_{\rm b}=10^3$.  Solid, dashed and dotted lines correspond to cases of a soft, hard and no UV backgrounds, respectively.}
\label{fige1}
\end{figure}

Figure\ \ref{fige1} shows the pressure (upper panel) and the velocity
(lower) as a function of the radius at time $t=0.12$~Myr (1), 0.24~Myr
(2), 0.35 Myr (3), 0.47~Myr (4) and 0.59~Myr (5).  Solid, dashed and dotted lines correspond to cases of a soft, hard and no UV backgrounds which is the same as in Fig.\ \ref{fig4}, respectively.  Calculations for two cases with UV backgrounds are performed with 200 grid points.  There is no large difference between values in three cases except for pressure values in the case of hard UV background.

\begin{figure}
\includegraphics[width=84mm]{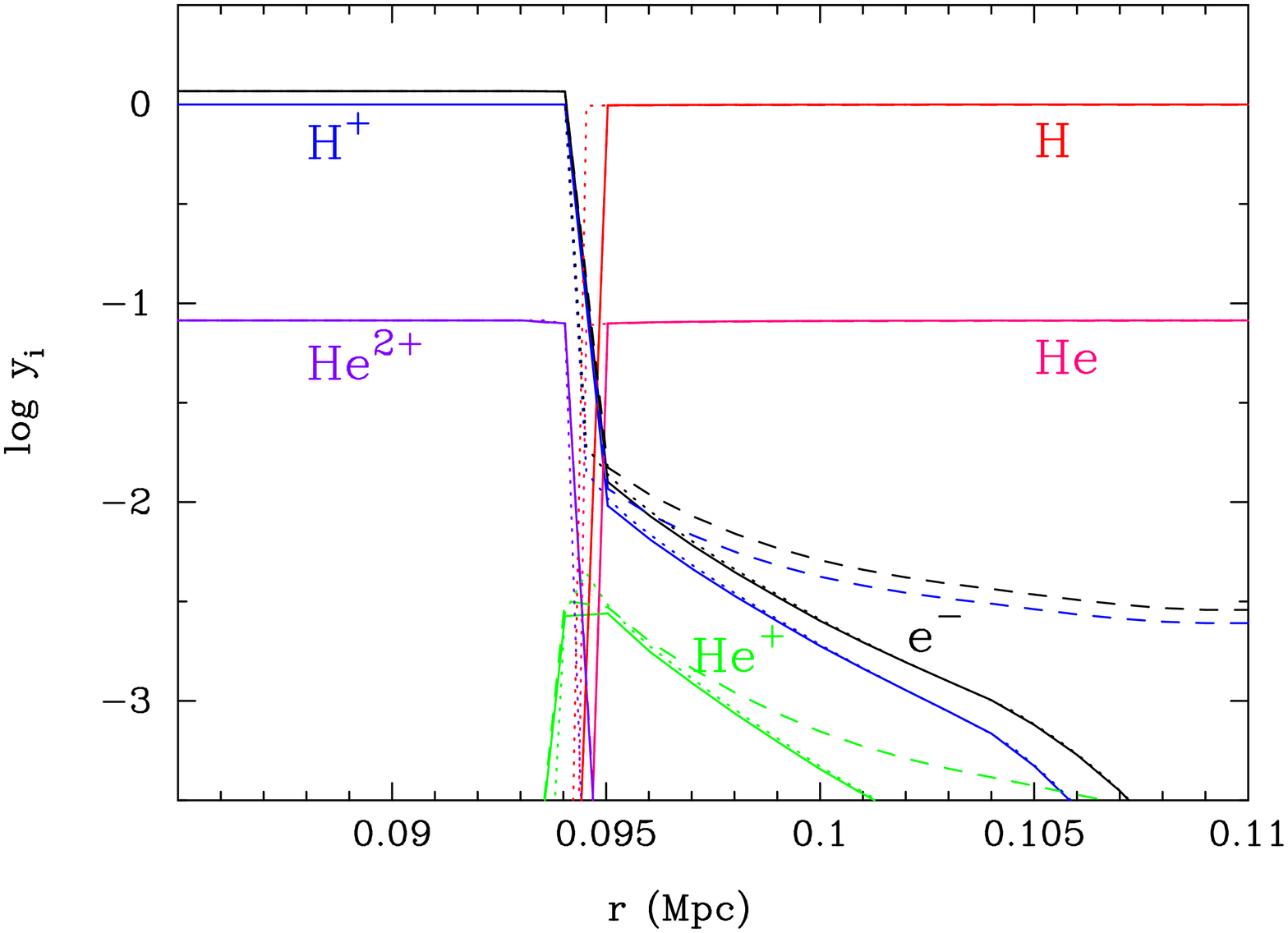}
\includegraphics[width=84mm]{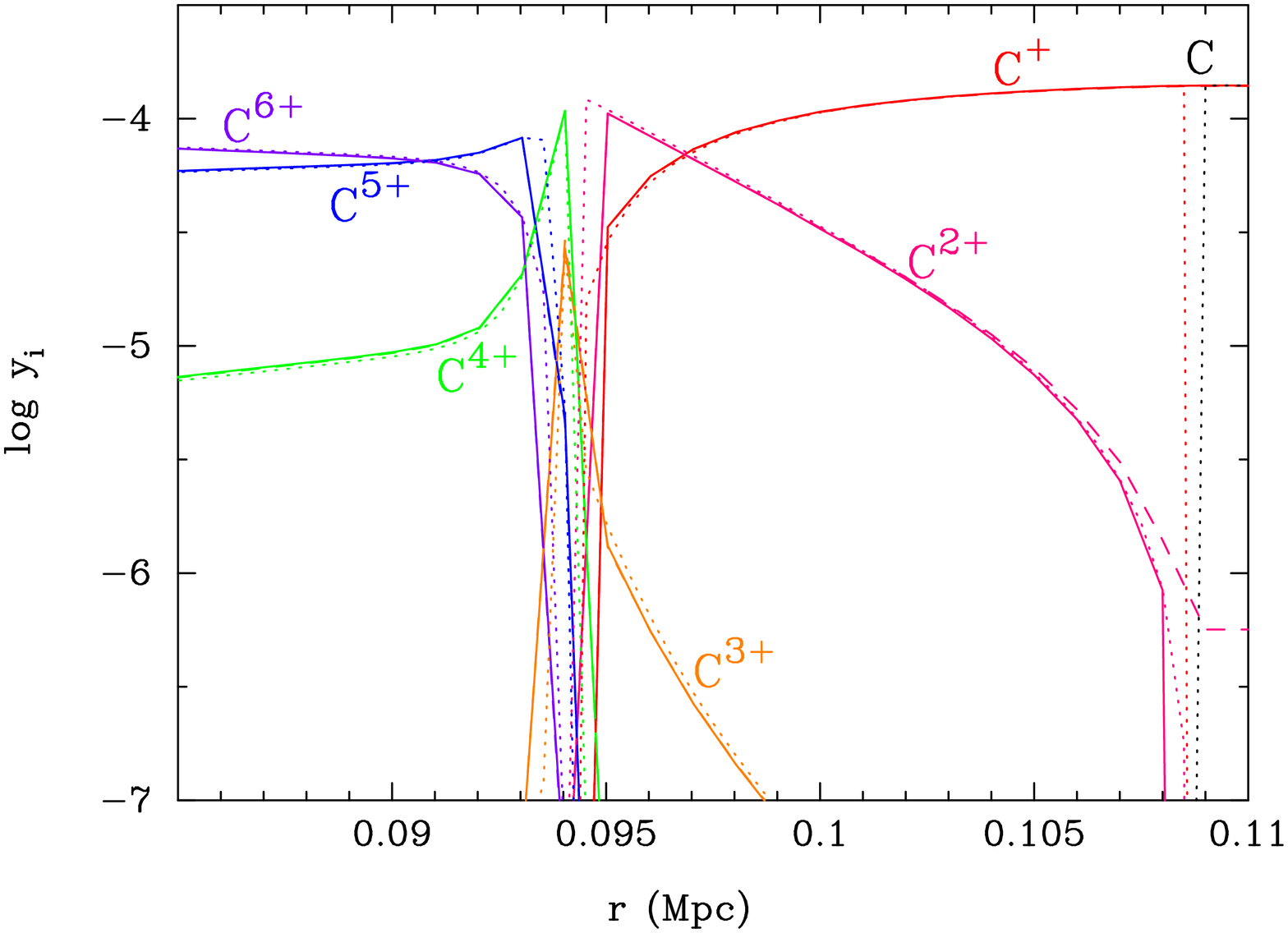}
\includegraphics[width=84mm]{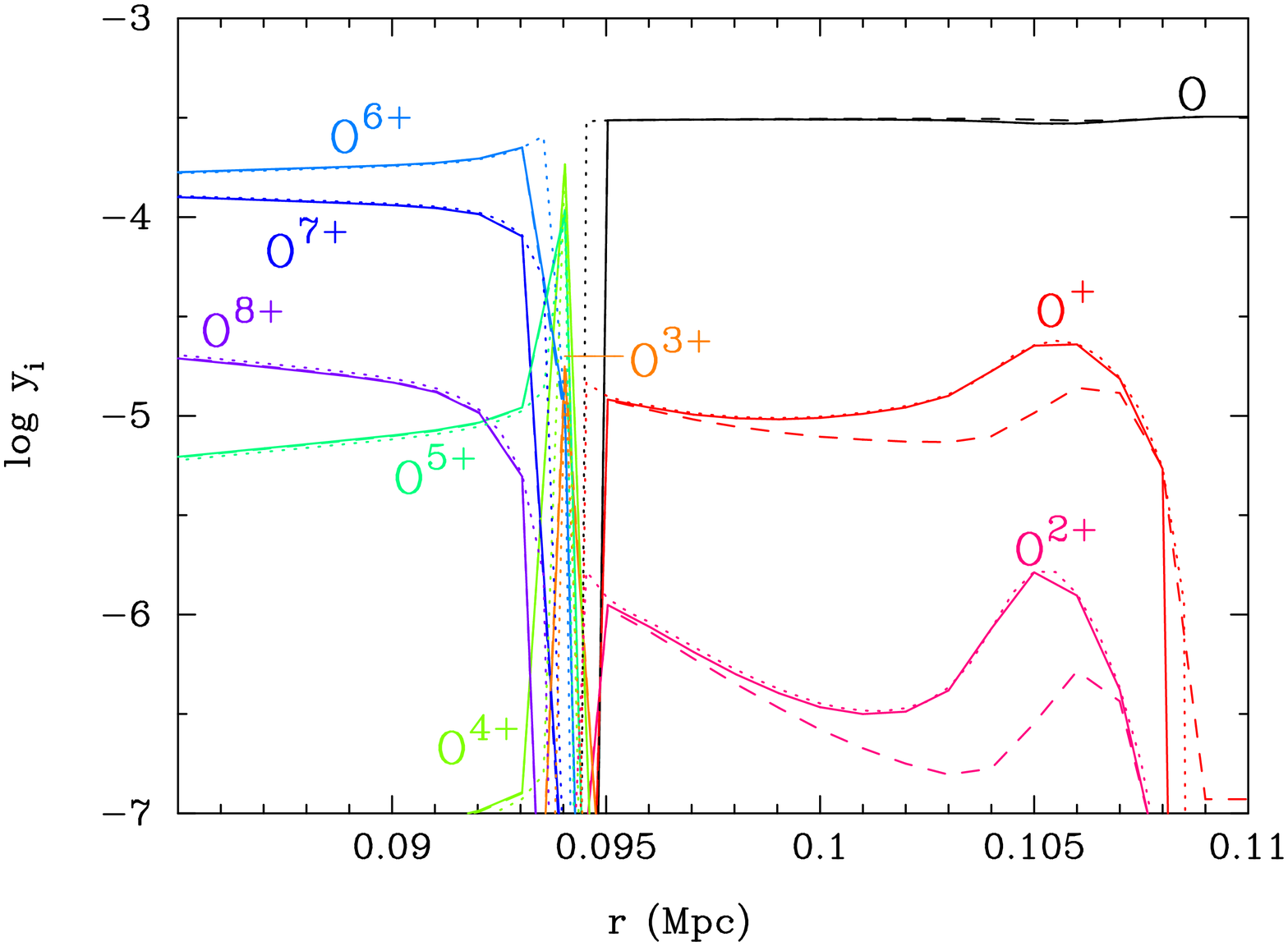}
\caption{Chemical abundances of hydrogen, helium and electron (top
 panel), carbon (middle) and oxygen (bottom) as a function of the radius
 $r$ at $t=0.35$~Myr (3) for the case of $\delta_{\rm b}=10^3$.  Solid, dashed and dotted lines correspond to cases of a soft, hard and no UV backgrounds, respectively.}
\label{fige2}
\end{figure}

Figure\ \ref{fige2} shows chemical abundances of hydrogen, helium and
electron (top panel), carbon (middle) and oxygen (bottom) as a function
of the radius $r$ at $t=0.35$~Myr (3).   Solid, dashed and dotted lines correspond to cases of a soft, hard and no UV backgrounds which is the same as in Fig.\ \ref{fig5}, respectively.  In the cases with UV backgrounds, \CI\ has already been ionized while in the case of no UV background \CI\ exists outside the light front.

\subsection{Signals through Fine Structure Transitions}

We divide signals by the UV photon pumping into two classes.  One is a
signal emitted in regions just outside of ionized regions. The other is
that of UV photons generated as more energetic than that of the line
center frequency and redshifted to the line center when they arrive at
radii larger than ionization fronts.

\subsubsection{Signals from Ionization Fronts}

The UV photons can pump up the energy levels of \CI, \CII\ and \OI\ when
they arrive in regions of abundant given
chemical species.  Since the amount of UV photon is finite, flux
densities of the pumping photons are reduced as they pump up.  Firstly,
we check if there are enough UV photons for mixing the fine structure
levels so that the UV pumping can effectively occur inside the light
front before arrivals of ionization fronts in the setting of this study.

The ionization time scale at around the ionization front (i.e.,
an optically thin region) is given by equation (\ref{eq4}), and the
precise number regarding \HI\ ionization is $\Delta t_{\rm
ion}({\rm H_I})=2.2\times 10^8~{\rm s}~[r/(0.1~{\rm Mpc})]^2$.  Minimum cross sections for
scattering off of UV photons by \Cboth\ and \OI\ required
for a scattering in the time scale of ionization, i.e., $\Delta t_{\rm
ion}({\rm H_I})$ is
\begin{eqnarray}
 \sigma&=& \frac{1}{n_{A_N} \Delta t_{\rm ion}}\nonumber \\
 &=&8.7\times 10^{-15}~{\rm cm}^2
  \left(\frac{y_{A_N}}{10^{-4}}\right)^{-1} \left(
   \frac{1+z}{9.7}\right)^{-3} \nonumber \\
&&\times \left( \frac{\delta_{\rm
   b}}{10^3}\right)^{-1} \left( \frac{r}{0.1~{\rm Mpc}}\right)^{-2}.
\end{eqnarray}
The frequency ranges around the line center of frequency $\nu_{\rm UV}$
which satisfy $\sigma \gid 10^{-14}$~cm$^2$ in the case of $T_{\rm
gas}=2.3~{\rm K}[(1+z)/10]^2$~\citep{loe2004} is $\Delta
\nu/\nu_{\rm UV}=6.2\times 10^{-7}$ for the ground state \CI\ atom,
$6.8\times 10^{-7}$ for the ground state \CII\ ion, and
$5.2\times 10^{-7}$ for the ground state \OI\ atom.  The
emission rates $S_{\rm S}$ of UV line photons which interact with C and O before
ionizing photons could ionize \HI\ are derived by
\begin{equation}
 S_{\rm S}=\frac{L_{\nu_{\rm UV}}}{h \nu_{\rm UV}}\Delta \nu.
\end{equation}
Values in the present setting are obtained using
equation\ (\ref{eq5}):  $S_{\rm S}=3.8\times 10^{52}~{\rm s}^{-1}$ for
\CI, $3.0\times 10^{52}~{\rm s}^{-1}$ for \CII, and $2.2\times 10^{52}~{\rm s}^{-1}$ for \OI.

\begin{figure}
\includegraphics[width=84mm]{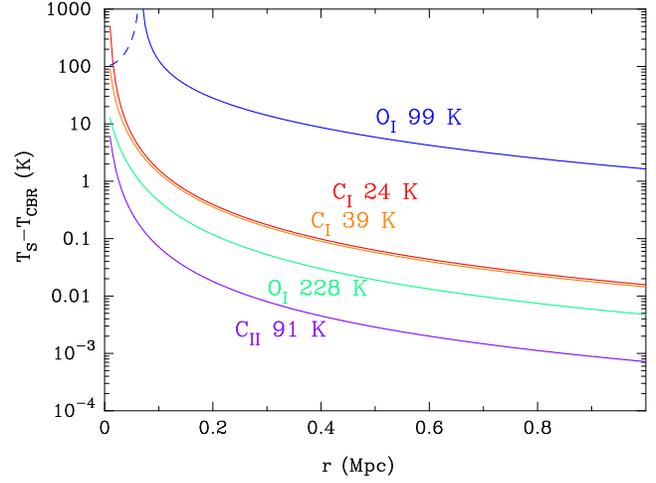}
\caption{Differences between spin temperatures and the CBR temperature
 (K) as a function of the radius $r$ (solid lines).  The value is negative
 for the \OI\ 99K line at $r\leq 0.07$~Mpc due to the large population of second
 excited state of \OI\ by an efficient pumping, and its absolute
 value is shown (dashed line).}
\label{fig8}
\end{figure}

Figure\ \ref{fig8} shows differences between spin temperatures and the
CBR temperature in units of K as a function of the radius $r$.  In drawing
this figure, losses of UV line photons by interaction with atoms and
ions are not considered.  As will be discussed below, UV line photons
are quickly used to pump up the fine structure levels.  This figure,
then, shows the maximum temperature differences adequate for only
regions right beyond the ionizing front where the loss of photons can be
neglected.

Suppose that ionizing UV photons are mainly used to ionize \HI, and
that photons of frequency around the UV pumping lines are mainly used to
excite and deexcite \Cboth\ and \OI.  We compare the emission rate of ionizing photon per hydrogen number density, i.e., $f_{\rm
H_I} S_\gamma/n_{\rm H}$ with that of line photon per differences of number
densities of excited states $i$, 
i.e., $S_S/\Delta n_{A_N}^i$.  Here
$\Delta n_{A_N}^i\equiv n_{A_N}^i(T_{\rm S})-n_{A_N}^i(T_{\rm CBR})$ was
defined, and number densities of excited state $i$, $n_{A_N}^i(T)$ are described by the steady state spin
temperature ($T_{\rm S}$) and that of CBR ($T_{\rm CBR}$), respectively [see
equations\ (\ref{eq1}) and (\ref{eq2})].
At
$r=0.1$~Mpc, $S_{\rm S}/\Delta n_{\rm C_I}^1=9.2f_{\rm H_I} S_\gamma/n_{\rm
H}$ and $S_{\rm S}/\Delta n_{\rm C_I}^2=1.4f_{\rm H_I} S_\gamma/n_{\rm H}$ for
\CI, $S_{\rm S}/\Delta n_{\rm C_{II}}^1=77f_{\rm H_I} S_\gamma/n_{\rm
H}$ for \CII, and $S_{\rm S}/\Delta n_{\rm O_I}^1=311f_{\rm H_I} S_\gamma/n_{\rm H}$
and $S_{\rm S}/\Delta n_{\rm O_I}^2=248 f_{\rm H_I} S_\gamma/n_{\rm H}$ for
\OI.  It is seen that UV photons which can scatter C and O exist
in abundances larger than ionizing photons when normalized to target
number densities.  The scattering regions,
therefore, extend outward faster than ionized regions.

Inside ionization fronts of \Cboth\ and \OI, UV line
photons propagate outward without scattering.  When they enter in region
with abundant \Cboth\ and \OI, however, they scatter the species
and change the fine structure population.  The UV line photons
effectively pump low energy level states up to high energy levels until
steady states realize.  Excited states eventually decay into low
states by magnetic dipole photon emission.  Repeating pumpings and
decays lead to net reduction of number flux of energetic UV line photon which can
pump up C and O from low energy states.

The radius scale for this scattering loss is given by those of regions where the
steady state spin temperatures can be hold by the flux from the point
source.  Balances between pumping up and spontaneous decay of excited
states in a volume $V=4\pi r_{\rm B}^2 \Delta r$, with $r_{\rm B}$
the boundary of the C and O ionization and $\Delta r$ the width, are
roughly described by
\begin{equation}
 4\pi r_{\rm B}^2 \left( \frac{F_{\nu_{\rm UV}}}{h \nu_{\rm UV}} \Delta\nu_{\rm D} -
		   \Delta r \Delta n_{A_N}^i A_{ij} \right)\sim 0,
\end{equation}
where the UV flux in the frequency range of Doppler width
$\Delta \nu_{\rm D}$ is considered in the source term which enter into
this region from the boundary $r_{\rm B}$.  The radius scale is thus
\begin{equation}
 \Delta r \sim \frac{F_{\nu_{\rm UV}}}{h \nu_{\rm UV}} \frac{\Delta\nu_{\rm D}} {\Delta n_{A_N}^i A_{ij}}.
\end{equation}
For the radius of $r=0.1$~Mpc, this equation yields
\begin{eqnarray}
 \Delta r=&\hspace{-0.8em}0.8~(\delta_{\rm b}/10^3)^{-1} T_{\rm gas, K}^{1/2}~{\rm pc}
  & {\rm (for~the~C_I~24~K~line)}, \nonumber \\
&\hspace{-0.8em} 0.02~(\delta_{\rm b}/10^3)^{-1} T_{\rm gas, K}^{1/2}~{\rm pc} &
  {\rm (C_I~39~K)}, \nonumber \\
&\hspace{-0.8em} 0.02~(\delta_{\rm b}/10^3)^{-1} T_{\rm gas, K}^{1/2}~{\rm pc} & 
 {\rm (C_{II}~91~K)}, \nonumber \\
&\hspace{-0.8em} 4 \times 10^{-3}~(\delta_{\rm b}/10^3)^{-1} T_{\rm gas,
 K}^{1/2}~{\rm  pc} & {\rm (O_I~228~K)}, \nonumber \\
&\hspace{-0.8em} 2~(\delta_{\rm b}/10^3)^{-1} T_{\rm gas, K}^{1/2}~{\rm
 pc} & {\rm (O_I~99~K)}. \nonumber
\end{eqnarray}
The UV line photons around
the line center quickly scatter C and O and decreases during
propagations through narrow regions if the density is high as in this
calculation.

\subsubsection{Signals from Outside of Ionization Fronts}\label{sec422}
\begin{enumerate}
\item Steady state abundances of excited states

Secondly, signals from excitations by redshifted UV photons are
estimated.  For simplicity, we assume that UV photons redshifted beyond
certain critical frequencies from the blue side instantaneously scatter C and O, and are used for pumping up of the species.  The number flux of the UV line photon, i.e., $F_{\nu_{\rm UV}}/(h \nu_{\rm UV})$, should satisfy
\begin{equation}
 \frac{d}{dr}\left(\frac{F_{\nu_{\rm UV}}}{h \nu_{\rm UV}}\right)= -\Delta n_{A_N}^i A_{ij} +N_{\rm eff}\frac{F_{\nu_{\rm UV}}}{h \nu_{\rm UV}}
  \frac{ \nu_{\rm UV} H(t)}{c},
\label{eq24}
\end{equation}
where
$\Delta n_{A_N}^i$ is the abundance of excited state $i$, 
$N_{\rm eff}$ is the effective number of lines which contribute to
pumping up \Cboth\ and \OI\ of fine structure levels, and
defined below.  The first term in the right hand side is for spontaneous emission of excited state $i$, while the second is for production by redshifted UV photons.  The production term is proportional to the number flux of the UV line photons and the rate of redshift per unit distance, i.e., $|d\nu/dr|\approx |d\nu/(cdt)|=\nu_{\rm UV} H(t)/c$.  The steady state for
abundances of fine structure levels would be
realized in a region between an ionization boundary and the light
front.  They are described by a balance in equation (\ref{eq24}), i.e., $d[F_{\nu_{\rm UV}}/(h \nu_{\rm UV})]/dr=0$, as
\begin{equation}
 \Delta n_{A_N}^i= N_{\rm eff}\frac{F_{\nu_{\rm UV}}}{h \nu_{\rm UV}}
  \frac{ \nu_{\rm UV} H(t)}{c A_{ij}},
\label{eq7}
\end{equation}

\item Effective number of available UV lines

UV photons described by a color temperature $T_{\rm UV}$ repeat
scattering C and O species through the UV photon pumping, and eventually
relax to the color temperature of the fine structure spin states  which
is approximately given by the CBR temperature if the UV photon pumping occurs frequently enough.  In this
process, chemical species in excited states which have been pumped up by UV
line photons can emit line photons corresponding to the fine structure
transitions.  The effective number $N_{\rm eff}$ is given as follows
taking account of the relaxation.

\begin{enumerate}
\item \CII\

For the \CII, the net
increase in the number of ions in the excited state is given by the net
increase in the number of photons corresponding to the transition
$^2$P$_{3/2}\rightarrow ^2$D$_{3/2}$ (see Table\ \ref{tab1}) after the
relaxation process through the UV scattering.  $N_{\rm eff}$
is, therefore, given by
\begin{equation}
 N_{\rm eff}=\frac{2}{1+\exp{(-T_\ast/T_{\rm CBR})}}-1,
\end{equation}
where 
the first term in right hand side is the relative number of line
in the final relaxed state, and
the second term is that in the initial state, that is equal to unity.
In the final state, initial UV photon flux for two lines, i.e.,
$^2$P$_{3/2}\rightarrow ^2$D$_{3/2}$ and $^2$P$_{1/2}\rightarrow ^2$D$_{3/2}$ is
distributed to the two lines by the equilibrium fraction of
\begin{equation}
 \frac{F_{^2{\rm P}_{3/2}}}{F_{^2{\rm P}_{1/2}}}=\exp{(T_\ast/T_{\rm CBR})},
\end{equation}
which realizes the balance between pumping rates of both directions,
i.e.,
\begin{equation}
 n_0 F_0 \sigma_{0\rightarrow u\rightarrow 1} = n_1 F_1
  \sigma_{1\rightarrow u\rightarrow 0},
\end{equation}
where
$F_0=F_{^2{\rm P}_{1/2}}$,
$F_1=F_{^2{\rm P}_{3/2}}$, and
$\sigma_{i\rightarrow u\rightarrow j}=\sigma(i\rightarrow~u) \times
A_{uj}/(\sum_k A_{uk})$ is the cross section of process $i\rightarrow
u\rightarrow j$ [cf. equation (\ref{eq6})].

\item \OI\

Similarly, effective numbers for the first and second excited states of
\OI\ are given by
\begin{equation}
 N_{\rm eff, 1}=\frac{3}{1+\exp{(-T_{\ast 2}/T_{\rm
  CBR})}\left[1+\exp{(-T_\ast /T_{\rm CBR})}\right]}-1,
\end{equation}
\begin{equation}
 N_{\rm eff, 2}=\frac{3\exp{(-T_{\ast 2}/T_{\rm
  CBR})}}{1+\exp{(-T_{\ast 2}/T_{\rm
  CBR})}\left[1+\exp{(-T_\ast /T_{\rm CBR})}\right]}-1,
\end{equation}
respectively.  Here it was assumed that the UV photon pumping of the first excited state of \OI\ is efficient as well as that of the ground state as a sufficient condition for the equilibrium UV line photon.  Since the abundance of the excited \OI\ ($^3$P$_1$) state is rather small [$g_0 n_1/(g_1 n_0)\sim \exp(-228~{\rm K}/T_{\rm CBR})=2.3\times 10^{-4}$ for $1+z=10$], this assumption is satisfied in environments of high \OI\ densities.  If the density excess is $\delta_{\rm b}\sim 10^4$, the equilibrium UV flux can realize [equation (\ref{eq21})], and a contamination from a collisional signal is still small if the gas temperature is not so high (Section\ \ref{sec212}).  If the density excess is smaller, the number of UV lines through which UV photons effectively scatter becomes smaller.  If the density of \OI\ is as small as $\delta_{\rm b}\sim1$, only the UV pumping of the ground state is efficient.  Signals of the \OI\ fine structure lines then originate from pumping of the ground state by a part of UV line photons.  In such a case of small \OI\ number density~\citep{her2007}, signals of both \OI\ 228~K and 91~K would become emissions and their intensities smaller than the estimations given here.

\item \CI\

As for \CI, two contributions through the upper
levels $^3$P$_1$ and $^3$P$_2$ exist.  The effective numbers for the first
and second excited states are then given by
\begin{eqnarray}
 N_{\rm eff, 1}&\hspace{-0.8em}=&\hspace{-0.8em}\frac{3}{1+\exp{(-T_{\ast 2}/T_{\rm
  CBR})}\left[1+\exp{(-T_\ast /T_{\rm CBR})}\right]}-1 \nonumber\\
&\hspace{-0.8em}&\hspace{-0.8em}+\frac{2}{1+\exp{(-T_{\ast 2}/T_{\rm CBR})}}-1,
\end{eqnarray}
\begin{eqnarray}
 N_{\rm eff, 2}&\hspace{-0.8em}=&\hspace{-0.8em}\frac{3\exp{(-T_{\ast 2}/T_{\rm
  CBR})}}{1+\exp{(-T_{\ast 2}/T_{\rm
  CBR})}\left[1+\exp{(-T_\ast /T_{\rm CBR})}\right]}-1 \nonumber\\
&\hspace{-0.8em}&\hspace{-0.8em}+\frac{2\exp{(-T_{\ast 2}/T_{\rm CBR})}}{1+\exp{(-T_{\ast 2}/T_{\rm CBR})}}-1,
\end{eqnarray}
respectively.  The first and second lines in right hand sides of the
above two equations correspond to the contributions of UV pumping
through the $^3$P$_1$ and $^3$P$_2$ levels, respectively.  
Since the number density of excited state \CI\ ($^3$P$_1$) is large [$g_0 n_1/(g_1 n_0)\sim \exp(-24~{\rm K}/T_{\rm CBR})=0.41$ for $1+z=10$], the equilibrium abundances realize relatively easily.

\end{enumerate}

\item Scattering rate

A steady state scattering rate [cf. equation (\ref{eq13}) for the rate derived neglecting the scattering loss] in the region outside of ionization front is estimated as follows:  The balance between a rate of scattering and that of depletion of UV line photons whose frequencies are redshifted over critical frequencies is described as
\begin{equation}
 n_{A_N}^i P_\nu= N_{\rm eff}\frac{F_{\nu_{\rm UV}}}{h \nu_{\rm UV}}\frac{\nu_{\rm UV} H(t)}{c}.
\end{equation}
We obtain the steady state $P_\nu$ value from this equation, i.e.,
\begin{eqnarray}
 P_\nu&\hspace{-0.8em}=& \hspace{-0.8em}\frac{1}{n_{A_N}^i} N_{\rm eff}\frac{L_\nu/(h\nu)}{4\pi r^2} \frac{\nu_{\rm UV} H(t)}{c}\nonumber\\
&\hspace{-0.8em}=&\hspace{-0.8em}3.3\times 10^{-10}~{\rm s}^{-1} \left(\frac{\lambda_{lu}}{\lambda_\alpha}\right)^{\alpha_{\rm S}} N_{\rm eff}\left(\frac{1+z}{10}\right)^{-3} \nonumber\\
&\hspace{-0.8em}&\hspace{-0.8em}\times \left(\frac{\delta_{\rm b} Y_{A_N}^i}{10^{-4}}\right)^{-1} \frac{\left(\nu L_\nu\right)_{\alpha, 47}}{(r_{\rm Mpc}/0.1)^2} \left(\frac{H(z)}{3.7\times 10^{-17}~{\rm s}^{-1}}\right).
\label{eq22}
\end{eqnarray}
See Section\ \ref{sec212} for effects of UV photons on spin temperatures.

\item Signals through the lowest-energy UV lines

Figure\ \ref{fig9} shows the differential antenna temperatures, i.e.,
the brightness temperature $T_{\rm b}$ minus CBR temperature in units of mK as a
function of the radius.  For this figure we assume that the abundances
of \Cboth\ and \OI\ are their maximum values, i.e.,
$y_{\rm C_I}=1.4\times 10^{-4}$, $y_{\rm C_{II}}=1.4\times 10^{-4}$, and
$y_{\rm O_I}=3.2\times 10^{-4}$, respectively.  The differential antenna temperatures are derived with the following equation which is satisfied when $T_{\rm b}-T_{\rm CBR} \ll T_{\rm CBR}$:
\begin{equation}
 \Delta\equiv \frac{T_{\rm b}-T_{\rm CBR}}{T_{\rm CBR}}=\frac{\Delta I_\nu}{B_\nu (T_{\rm CBR})} \frac{\exp(T_\ast/T_{\rm CBR})-1}{(T_\ast/T_{\rm CBR})\exp(T_\ast/T_{\rm CBR})}.
\end{equation}
  Since the differential
antenna temperature is rather small, signals of these magnitudes would not
be detected with existing radio telescope or the planned Atacama Large
Millimeter/submillimeter Array (ALMA).
\begin{figure}
\includegraphics[width=84mm]{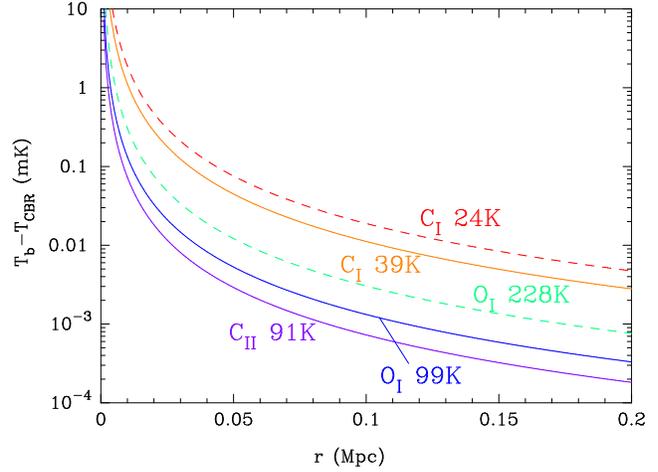}
\caption{Differential antenna temperatures (mK) as a function of the
 radius.  Solid curves are for positive values, while dashed curves are
 for absolute values of negative ones.  The respective curves are derived under the assumption that
 the elements (C and O) are completely in the chemical species (\CI,
 \CII\ and \OI), and therefore indicate maximum values for respective
 lines.}
\label{fig9}
\end{figure}

\item Signals through all available UV lines

So far only the effects of lines shown in Fig.\ \ref{fig1} are analyzed.
There are, however, many other lines of
energies less than the ionization threshold of \HI\ which can mix
fine structure
levels.  We then include effects of those lines.  Lines available for
pumping up the C and O species in the fine structure levels should
satisfy a requirement:  the time scale of UV pumping of C or O by photons of line
center frequencies $\Delta t_{\rm pump}$ is shorter than that of redshift
$\Delta t_{\rm red}$.  If this is not satisfied, only a part ($\sim \Delta t_{\rm red}/\Delta t_{\rm pump}$) of the UV line photons can scatter C or O before their frequencies are removed from the line center and they become inert.  Using $\Delta t_{\rm pump}=\sqrt[]{\mathstrut \pi}\Delta \nu_{\rm D}/(n_{A_N}\sigma)$~\citep{ryb1979} and
$\Delta t_{\rm red}=\Delta \nu_{\rm
D}/[\nu_{\rm UV} H(t)]$,
and defining $\bar{A}$ by $\sigma\sim \lambda^2/(8\pi)\bar{A}$ [cf. equation\ (\ref{eq6})], a relation is obtained, i.e,
\begin{eqnarray}
 \frac{\Delta t_{\rm pump}}{\Delta t_{\rm red}}&=&\nu_{{\rm UV}, 15}^3 \left(\frac{1+z}{10}\right)^{-3/2} \left(\frac{\delta_{\rm b} Y_{A_N}}{10^{-4}}\right)^{-1}
  \nonumber \\
 && \times \left(\frac{3.2\times 10^6~{\rm
  s}^{-1}}{\bar{A}}\right),
\label{eq21}
\end{eqnarray}
where
$\nu_{{\rm UV}, 15}$ is the line center frequency in units of
$10^{15}$~s$^{-1}$.  The lines contributing to the pumping up of fine
structure levels should have Einstein $A$-coefficient enough large in order to
satisfy $\Delta t_{\rm pump}/\Delta t_{\rm red}<1$.  This condition tends to be met in environments of high $\delta_{\rm b}Y_{A_N}$ values.  If the two timescales are nearly identical, i.e., $\Delta t_{\rm pump}\sim \Delta t_{\rm red}$, the depletion at the UV pumping of UV line photons balances with the production via cosmological redshift.  In such a situation, the steady state UV line flux and the resulting line scattering rate [equation (\ref{eq22})] are relatively large.

Assuming that all UV lines of $A>10^6$~s$^{-1}$ contribute to the pumping
up, for example, signals through the C and O species are calculated using atomic data
from~\citet{nis2008}.  Figure\ \ref{fig10} shows differential antenna
temperatures when contributions from all UV lines of $A>10^6$~s$^{-1}$
are included.  The signals for \CI\ and \OI\ are enhanced
relative to curves in Fig.\ \ref{fig9} since many additional strong lines exist.  On
the other hand, the signal for \CII\ is not enhanced much since
only two upper states, one of which is the $^2$D$_{3/2}$ shown in Fig.\
\ref{fig1}, are available.
\begin{figure}
\includegraphics[width=84mm]{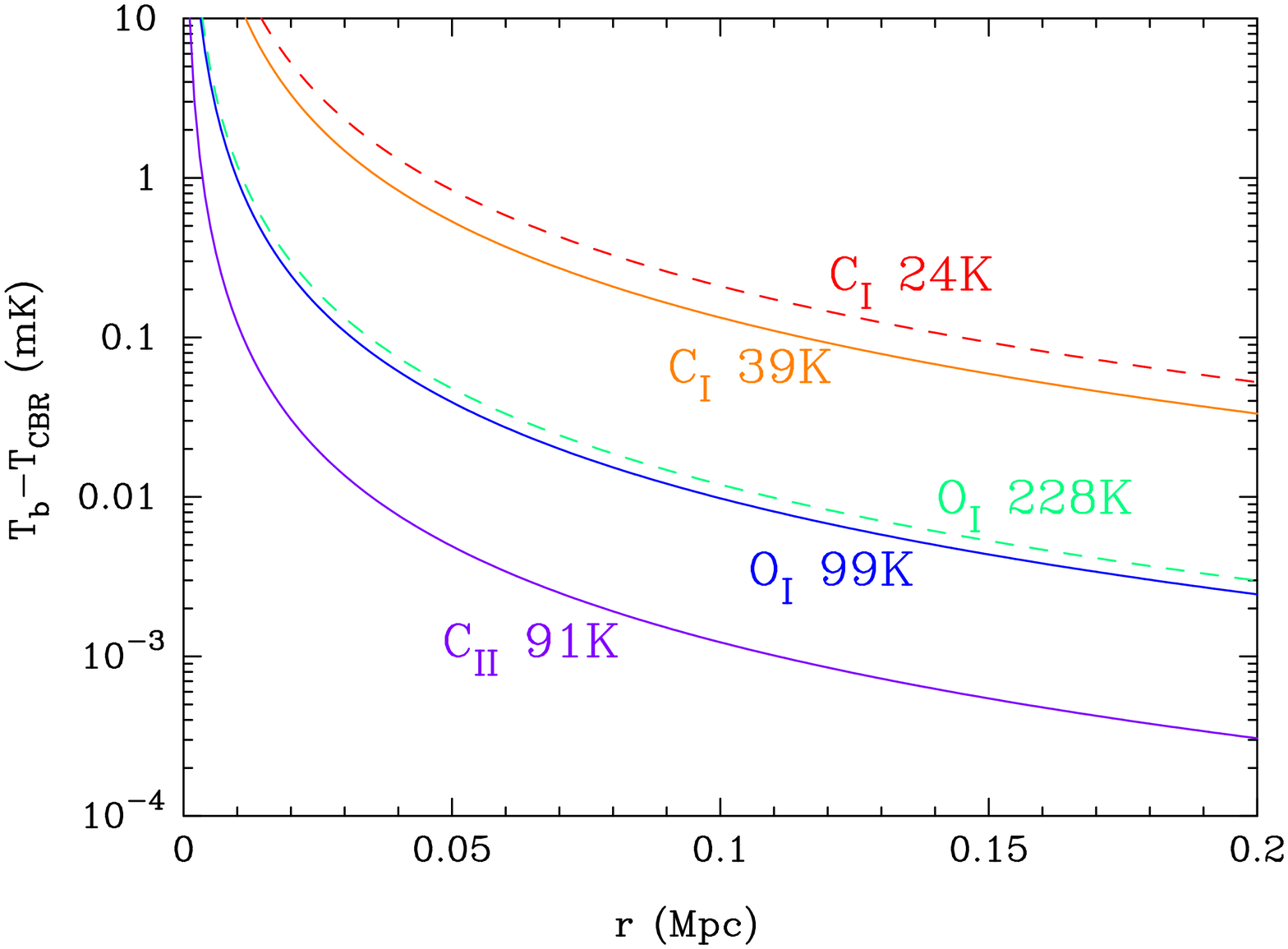}
\caption{Same as in Fig\ \ref{fig9} but including the effects of all UV lines
 of spontaneous emission rates larger than 10$^6$ s$^{-1}$.}
\label{fig10}
\end{figure}

\end{enumerate}

\subsection{Flux Densities of the Point Source}

Total flux densities, i.e., $S=\int S_\nu~d\nu$, from IGM via the fine structure line emission or
absorption are estimated with the calculated results
of the time evolution of the chemical structure.  We suppose that a
point source starts lighting at time $t=0$ at redshift $z=8.7$.  The
comoving distance to the source, $r_{\rm S}$, is defined by
\begin{equation}
 r_{\rm S}=\int_{t_{\rm S}}^{t_0} \frac{c
  dt}{a(t)}=\frac{c}{H_0}\int_{a_{\rm S}}^1 \frac{da}{\sqrt[]{\mathstrut
  \Omega_{\rm m} a +(1-\Omega_{\rm m}) a^4}},
\end{equation}
where
$a_{\rm S}$ and $t_{\rm S}$ are the scale factor and the corresponding
time of the universe when the source exists, and
$t_0$ is the age of the present universe.

We introduce two
dimensional Cartesian coordinates.  The source and the observer are
assumed to locate at the origin and $x=\infty$.  The flux density in the redshifted fine structure frequency which is
measured on the earth is given by
\begin{equation}
 S_{\rm fs}^i\approx \frac{1}{(1+z)^4} \frac{h\nu_0 A_{ij}}{r_{\rm S}^2}
  \int_0^{r_{\rm D}} \hspace{-0.8em}y dy \int_{-x_{\rm max}}^{x_{\rm max}} \hspace{-1.5em}dx~\Delta n_{A_N}^i (r,t+x/c),
\label{eq12}
\end{equation}
where 
$r_{\rm D}\sim 0.2$~Mpc is the radius of the calculation domain,
$x_{\rm max}=\sqrt[]{\mathstrut r_{\rm D}^2-y^2}$, and
$r=\sqrt[]{\mathstrut x^2+y^2}$ is the radius from the source.
We assume that the UV lines with emission rates of $A\gid 10^6$ s$^{-1}$
contribute to the UV pumping of C or O when the abundance is enough
large, i.e., $Y_{A_N}\gid 10^{-4}$.  The value of $\Delta n_{A_N}^i
(r,t+x/c)$ is thus given by equation (\ref{eq7}) if $Y_{A_N}(r,t+x/c)$
is larger than $10^{-4}$.  The value is set to be zero otherwise.  Note
that the effect of the time delay is included in $\Delta n_{A_N}^i
(r,t+x/c)$ since ionization fronts propagate at sub-light speeds.

\begin{figure}
\includegraphics[width=84mm]{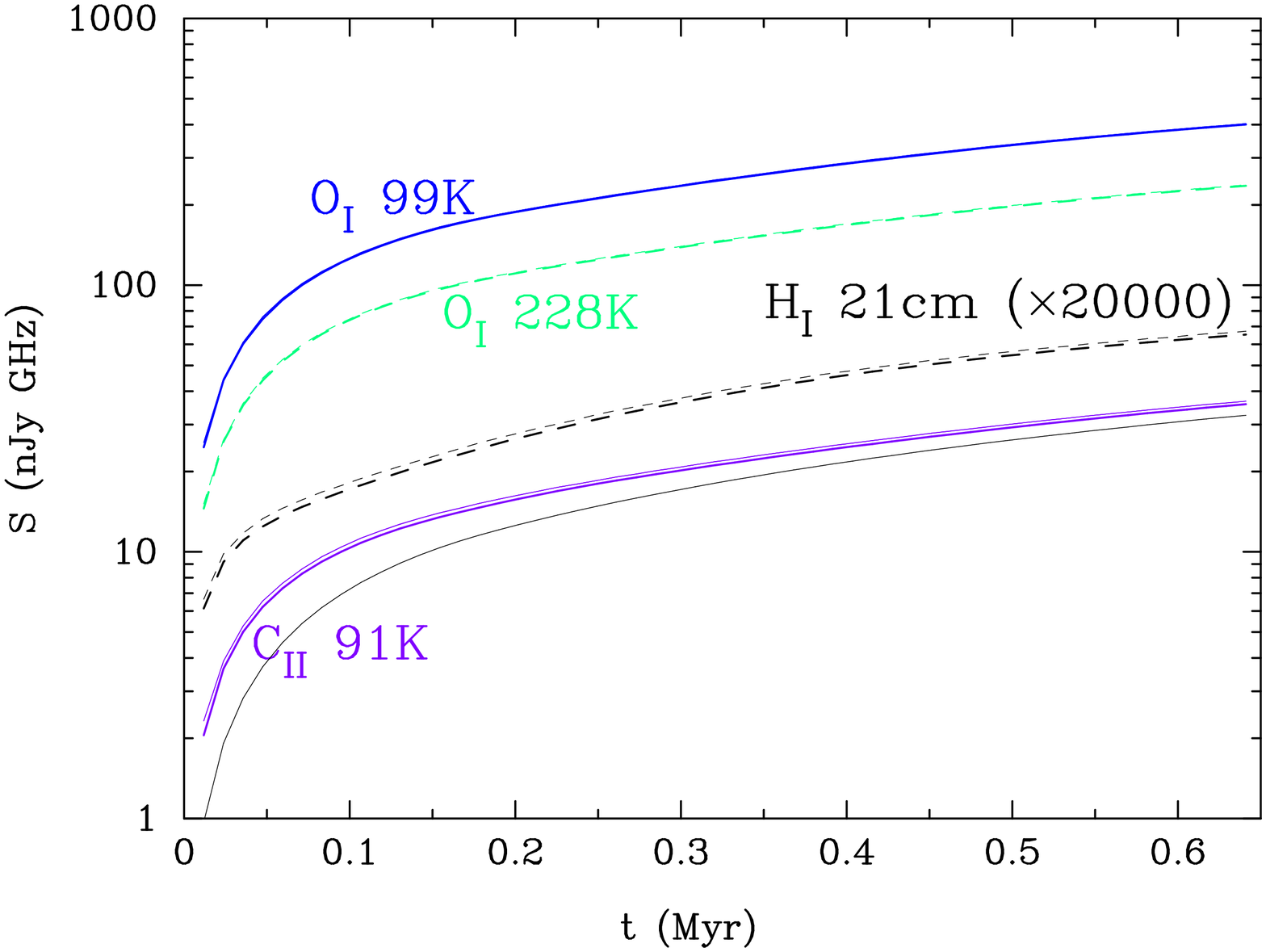}
\caption{Total flux densities (nJy GHz) emitted through the C$_{\rm II}$
 and O$_{\rm I}$
 fine structure lines and the \HI\ 21 cm line in the calculated region
 for the case of $\delta_{\rm b}=10^3$.  The values which we would measure at
 $z=0$ are shown as a function of the time from the start of the
 hydrodynamical calculation.  Solid lines correspond to emissions, while
 a dashed line is for absorption.  Thick lines correspond to the standard case of no UV background, while thin lines correspond to two cases of a soft and hard UV backgrounds.}
\label{fig11}
\end{figure}

Figure\ \ref{fig11} shows the total flux densities emitted through the \CII\
and \OI\ fine structure lines in the calculated domain of $r\lid
0.2$~Mpc in units of nJy GHz.  The calculated result for the case of
$\delta_{\rm b}=10^3$ is used for this estimation.  As the light
propagates outwards, the volume of the region experiencing the UV pumping
of C and O increases.  The emissions (solid lines) and the
absorption (dashed line) triggered by redshifted UV photons are then
enhanced.  For this figure we assume that the chemical abundances as a
function of the radius are fixed after the time of the end of
calculation, i.e., $t_{\rm end}$ at values of $t_{\rm end}$.  This is
because we do not have results of $t > t_{\rm end}$ which are necessary
to calculate the total flux densities due to the time delay effect in
equation (\ref{eq12}).  No \CI\ region exists inside the light radius in
our calculation.  There are large amounts of ionizing photons of \CI\
escaping from shielding since the threshold energy of \CI\ (11.3~eV) is
lower than that of \HI\ (13.6~eV).  Thick lines correspond to the standard case of no UV background, while thin lines correspond to two cases of a soft and hard UV backgrounds (Section\ \ref{sec413}).

In Fig.\ \ref{fig11}, the total flux density for the \HI\ 21 cm line is
also shown.  The flux density is given by
\begin{equation}
 S_{21}\approx \frac{1}{(1+z)^4} \frac{h\nu_{21} A_{21}}{r_{\rm S}^2}
  \int_0^{r_{\rm D}} \hspace{-0.8em}y dy \int_{-x_{\rm max}}^{x_{\rm max}} \hspace{-1.5em}dx~\Delta n_{\rm H}^1 (r,t+x/c),
\label{eq14}
\end{equation}
where
$\nu_{21}=1.4$~GHz is the frequency corresponding to the 21 cm line
transition,
$A_{21}=2.9\times 10^{-15}$~s$^{-1}$ is the spontaneous emission rate of the hyperfine
structure transition, and
$\Delta n_{\rm H}^1$ is the difference of the number abundance of
the excited level of \HI\ hyperfine structure (n$_{\rm H}^1$) in the presence of the UV
radiation field from that without the UV field.

\HI\ regions which are irradiated with UV photons are heated
predominantly via photoionization.  The heating rate is
given by equation (\ref{eq16}).  The gas temperature
in the \HI\ region with enough UV flux would, therefore, be higher while that in region with strongly shielded UV flux would be lower than the CBR temperature.  Note that the Ly$\alpha$ scattering provides a negligible heating \citep[e.g.,][]{che2004}.

The spin temperature of the \HI\ hyperfine structure, i.e., $T_{\rm S}$,
is usually much larger than the transition temperature $T_\ast\equiv
h\nu_{21}/k = 0.068$~K.  The abundance of the excited state is then given by
\begin{equation}
n_{\rm H}^1=\frac{3\exp(-T_\ast/T_{\rm S})}{1+3\exp(-T_\ast/T_{\rm S})}
 n_{\rm H_I}\approx \frac{3}{4}\left(1-\frac{1}{4}\frac{T_\ast}{T_{\rm
				 S}}\right) n_{\rm H_I}.\nonumber
\end{equation}

We define $T_{\rm S}$ and $T_{\rm BG}$ as the spin temperature in the
UV radiation fields and that without it.  Although the effect of the collisional
excitation of the hyperfine structure line is not negligible in regions
of high densities, we here neglect the effect in order to derive the
maximum effect of the Ly$\alpha$ pumping on the 21 cm emission.  Under
this assumption, when
there is no UV radiation field, the spin temperature is purely described
by the CBR temperature, i.e., $T_{\rm BG}=T_{\rm CBR}$.  Since the spin
temperature in the \HI\ region at radius $r\la O(0.1~{\rm Mpc})$ is
very close to the \HI\ gas temperature~\citep{mad1997}, a situation of
$T_{\rm S}>T_{\rm BG}$ is realized if the UV radiation field is not severely shielded.  From this equation, we obtain
\begin{equation}
\Delta n_{\rm H}^1 \approx \frac{3}{16} \frac{T_\ast}{T_{\rm CBR}} n_{\rm H_I}
\label{eq15}
\end{equation}
On the other hand, if the UV radiation field is shielded, an environment of $T_{\rm S}<T_{\rm BG}$ leads to
\begin{equation}
\Delta n_{\rm H}^1 \approx -\frac{3}{16} \frac{T_\ast}{T_{\rm S}} n_{\rm H_I}
\label{eq15}
\end{equation}

The line in Fig.\ \ref{fig11} was drawn assuming equations (\ref{eq14}) and using the calculated results of the \HI\ abundance and the gas temperature as a function of the time and the radius.  There is an important difference between the standard, soft UV and hard UV cases.  In the hard UV case the gas temperature has been heated up by the UV background.  The spin temperature of 21 cm line is then larger than the CBR temperature, i.e., $T_{\rm CBR}=2.725(1+z)$~K.  The 21 cm signal is, therefore, an emission, and its amplitude is a factor $\sim T_{\rm CBR}/T_{\rm S}$ smaller than that of absorption in cases of $T_{\rm CBR}>T_{\rm S}$~\citep{sco1990,mad1997}.

\begin{figure}
\includegraphics[width=84mm]{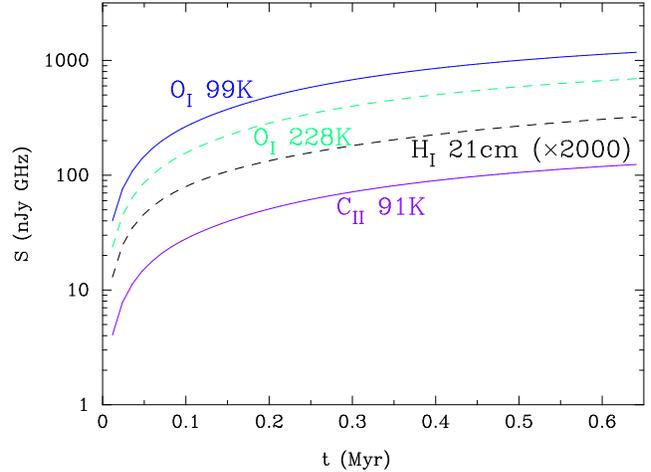}
\caption{Same as in Fig\ \ref{fig11} for the case of $\delta_{\rm b}=10^4$.}
\label{fig12}
\end{figure}

Figure\ \ref{fig12} shows the total flux densities emitted through the \CII\
and \OI\ fine structure lines and the \HI\ 21 cm line in the calculated
domain of $r\lid 0.2$~Mpc for the case of $\delta_{\rm b}=10^4$.  The high density
of the matter causes a slow expansion of ionized regions.  There are larger volume of \CII\ and \OI\ regions which are affected by
the UV pumping than in the case of $\delta_{\rm b}=10^3$.  The flux densities
are thus enhanced in this high density case.

\subsection{Effect of Collisional Excitation}

In dense regions, an effect of collisional excitation on spin
temperatures of fine structure transitions can not be neglected.  For
example, our calculations are performed under the assumption that the
number density of hydrogen is $n_{\rm H}\sim 2\times 10^{-1}
(\delta_{\rm b}/10^3) [(1+z)/10]^3$~cm$^{-3}$.  Collisional excitations and
deexcitations of \CI\ and \CII\ proceed by interactions of
those species mainly with surrounding \HI\ and electron.  \OI\ atoms can interact with surrounding
\HI\ and electron to get excited or deexcited (see Figs.~\ref{fig5}
and \ref{fig7}).  The deexcitation rates taken from~\citep{hol1989} are listed in Table\ \ref{tab1}.  Collisions contribute
to spin temperatures with their efficiency proportional to collisional
(de)excitation rates, i.e., 
\begin{eqnarray}
C_{ij}&\sim & 2\times 10^{-11}~{\rm s}^{-1} \left(\frac{\delta_{\rm b}}{10^3}\right)
\left(\frac{1+z}{10}\right)^3 y_{\rm target} \nonumber\\
&&\times \left(\frac{\gamma_{ij}}{10^{-10}~{\rm cm}^{-3}~{\rm s}^{-1}}\right).  
\end{eqnarray}
These rates
are less than spontaneous emission rates of the fine
structure transitions, and possibly of similar magnitudes to UV pumping rates [equations (\ref{eq13}) and (\ref{eq22})].  If the gas density of the observed region were high, kinetic
temperatures of \Cboth\ and \OI\ can be imprinted on the
spin temperatures through the collisional excitation (Section\ \ref{sec212}).

Signals originating from collisions need to be small in order to detect
signals from the pure UV pumping related to ionizing sources of the universe.
The best site to look for is a region which meets both of the following
two conditions.

1. a large region of moderate density.  It does not emit signals of collisional
(de)excitations.  The signals from this region separate in frequency from
those from surrounding dense regions where spin temperatures include
large contribution of collisions.

2. a region where ionizing UV photons are effectively shielded and
non-ionizing UV photons are not shielded and exist abundantly.  \Cboth\
and \OI\ can exist without being ionized, and a UV pumping by redshifted
UV photons is operative there.

\subsection{Observational Constraint on Physical Properties}

When one detect a line signal from a region whose physical condition is not known in advance from any different observation, it is not clear whether the signal originated from a UV dominated region or a collision dominated region.  Signals of different lines are, therefore, necessary to estimate the physical condition.  A line signal is determined by a physical condition of observed region specified by the density, kinetic temperature and environmental UV flux (absolute value and spectrum).  Conversely, a signal of a certain metal line could provide a constraint on physical parameters to satisfy a condition which possibly realizes the detected signal.  If signals of more than one different lines are detected from the same region, respective constraints on the parameters are obtained since a ratio between excitation rates by UV photons and collision are different for respective transitions corresponding to different lines.  Ideally parameters can then be determined and one can estimate whether the signal is from UV photons or from collisions (for individual lines).

We describe an example of how to constrain physical parameters of observed regions for demonstration assuming that line signals of \CII\ 91 K, \OI\ 228K and 99K were detected.  We note that a more precise estimation of line signals is highly desirable.  Since the precise propagation of UV photons which excite \CI, \CII\ and \OI\ is not addressed in this study, it should be studied.  For the moment, we assume that all UV line photons experience effective scattering of \CII\ and \OI\ although this assumption would not hold for low density region (see Section\ \ref{sec422}).

We suppose that three line signals at $z=8.7$ are detected:  $T_{\rm b}-T_{\rm CBR}=(1.2\pm0.4)\times 10^{-3}$~mK for \CII\ 91 K, $-(1.2\pm0.4)\times 10^{-2}$~mK for \OI\ 228K and $(9.8\pm3.3)\times 10^{-3}$~mK for \OI\ 99K.  Parameters to be constrained are, for example, $n_{\rm H_I}$, $y_{\rm C_{II}}$, $y_{\rm O_I}$, the UV flux at the Ly$\alpha$ frequency, i.e., $F_\alpha$, and the spectral index, i.e., $\alpha_{\rm S}$.  The dependence of signals on spectral index (or UV color temperature $T_{\rm UV}$) is relatively small (see Fig.\ \ref{fige3}--\ref{fig2}).  We then neglect it and fix the value to be $\alpha_{\rm S}=3/2$.  The carbon and oxygen abundances, i.e., $y_{\rm C_{II}}$ and $y_{\rm O_I}$, should be estimated anyhow since the values and the ratio between them in high redshifts are not know precisely.  Detailed theoretical studies on the cosmic chemical evolution may help this estimation~\citep[e.g.,][]{kob2007}.  We assume that the abundances, $y_{\rm C_{II}}$ and $y_{\rm O_I}$, are the same as those of present ISM.  The physical parameters in the present setting is then $n_{\rm H_I}$, $T_{\rm gas}$ and $F_\alpha$.  Effects of the UV pumping (Section\ \ref{sec422}) and the collision (Section\ \ref{sec212}) are calculated.

\begin{figure}
\includegraphics[width=84mm]{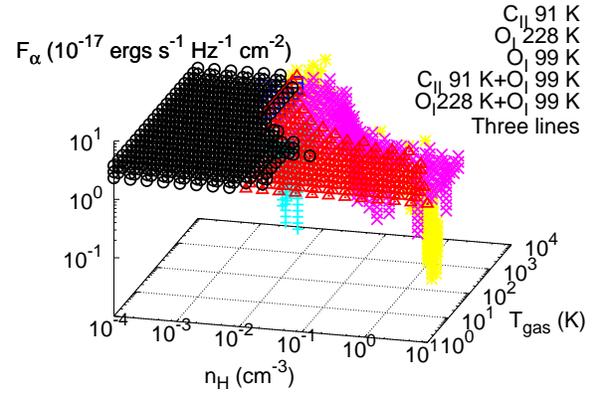}
\caption{An example of estimation of physical parameters, i.e., the \HI\ number density ($n_{\rm H}$), the gas temperature ($T_{\rm gas}$) and the UV flux at Ly$\alpha$ frequency ($F_\alpha$).  The marks of $+$, $\ast$ and $\times$ correspond to parameter sets which predict line signals consistent with only one observed line signal of \CII\ 91 K, \OI\ 228 K and \OI\ 99 K, respectively.  The squares and triangles correspond to parameter sets which predicts two right signals (\CII\ 91 K and \OI\ 99 K) and (\OI\ 228 K and 99K), respectively.  The circles correspond to parameter sets which rightly predicts all three line signals.  See text for the assumed observational signals and an explanation of adopted model.}
\label{fige6}
\end{figure}

Figure\ \ref{fige6} shows a parameter region in the 3D space of $n_{\rm H}$, $T_{\rm gas}$ and $F_\alpha$ in which magnitudes of observed signals are reproduced.  The marks of $+$, $\ast$ and $\times$ correspond to parameter sets which fail to predict observed line signals except for only one line signal of \CII\ 91 K, \OI\ 228 K and \OI\ 99 K, respectively.  The squares and triangles correspond to parameter sets which predicts two right signals (\CII\ 91 K and \OI\ 99 K) and (\OI\ 228 K and 99K), respectively.  The circles correspond to parameter sets which rightly predicts all three line signals.

\subsection{Redshift Dependence of Signals}

Signals are different in magnitude between sources of different
redshifts.  As seen in equation\ (\ref{eq7}), the line emission is
proportional to the UV flux $F_{\nu_{\rm UV}}\propto r^{-2}$ [cf. equation
(\ref{eq8})] and the Hubble expansion rate $H(z)$.  When one fixes the
angle $\theta$ between a source and a position $r$ away from it in the
direction perpendicular to the line of sight, there is a relation between
the redshift and the position, i.e.,
\begin{equation}
 \theta=\frac{r(1+z)}{r_{\rm S}}.
\end{equation}

\begin{figure}
\includegraphics[width=84mm]{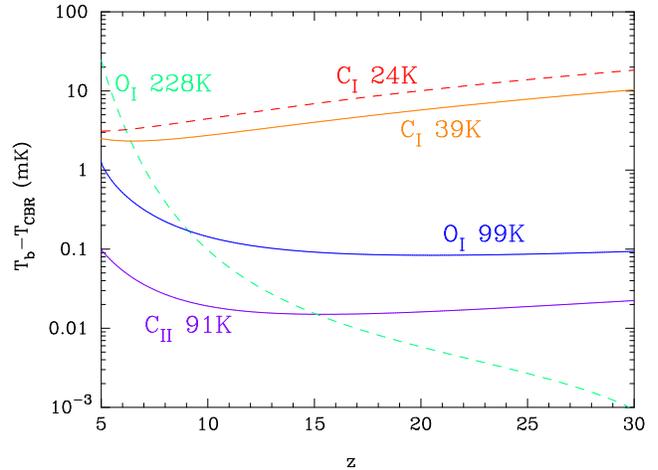}
\caption{Differential antenna temperatures (mK) at the fixed angle
 $\theta=5\arcsec$ in the sky from the direction to the point source as a function
 of the redshift.}
\label{fig13}
\end{figure}

Figure\ \ref{fig13} shows the differential antenna temperatures in units
of mK at the fixed angle of $\theta=5\arcsec$ from the point source of
the present setting as a function of the redshift under the assumption that the
observed region is abundant in \Cboth\ and \OI\ in
drawing respective lines.

The shape of curves in this figure is understood as follows.  If a
difference between the spin temperature and the CBR temperature is very small,
i.e., $\Delta T_{\rm S} \equiv T_{\rm S}-T_{\rm CBR} \ll T_{\rm CBR}$,
then it is approximately given by
\begin{equation}
T_{\rm b}- T_{\rm CBR} =\Delta T_{\rm S} \tau.
\end{equation}
In the case of two level states, for example, if the population fraction
of the ground state is dominant and the contribution of the UV pumping
to the excitation is small, then the spin temperature is given by
\begin{equation}
\Delta T_{\rm S} \sim \frac{\Delta n_1}{n_0} \frac{g_0}{g_1}
 \frac{T_{\rm CBR}^2}{T_\ast} \exp\left(\frac{T_\ast}{T_{\rm
				   CBR}}\right).
\end{equation}
The differential abundance of the excited state is proportional to the
Hubble expansion rate, i.e., $\Delta n_1\propto H(t)\propto (1+z)^{3/2}$
[equation (\ref{eq7})].  The abundance of the ground state scales as
$n_0\propto (1+z)^3$, while the optical depth scales as $\tau\propto
(1+z)^{3/2}$ [equation (\ref{eq10})].  Ultimately there is a rough
scaling of
\begin{equation}
T_{\rm b}- T_{\rm CBR} \propto (1+z)^2 \exp\{T_\ast/[2.725(1+z)~{\rm
 K}]\}.
\label{eq11}
\end{equation}
At low redshift the CBR temperature is low, and the differential
temperature increases with the decreasing redshift.  At high redshift
the differential temperature increases with the increasing redshift
mainly through the $(1+z)^2$ factor in equation (\ref{eq11}) and the fact
that the physical scale corresponding to a fixed angular scale of object
at higher redshift is smaller in the $\Lambda$CDM model adopted in the
present study.

\subsection{Detectability}

\begin{figure}
\includegraphics[width=84mm]{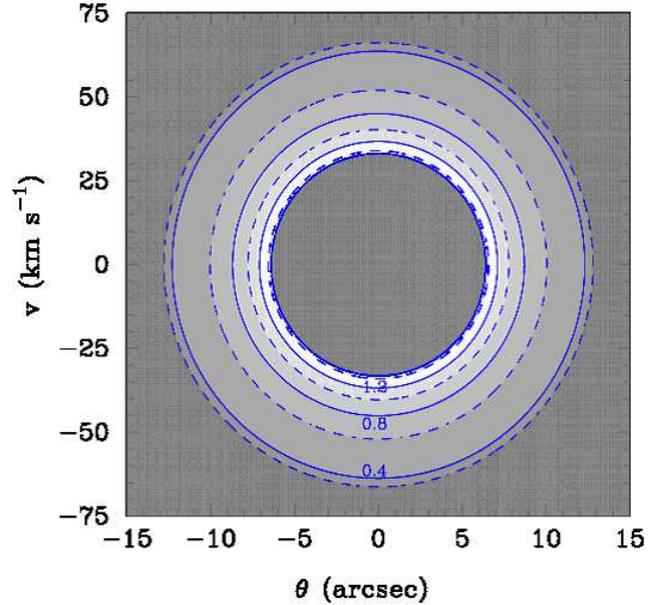}
\caption{Contour map of the differential antenna temperature (mK) of
 the \CI\ 38~K line as a function of the angle (arcsec) and the
 velocity shift (km s$^{-1}$) when a neutral \CI\ region affected
 by the UV pumping exists in the region of 0.03~Mpc $\leq r \leq$ 0.06~Mpc.}
\label{fig14}
\end{figure}

Signals through the \CI\ lines are largest as seen in Fig.\
\ref{fig10}.  If the signals of UV photons are not contaminated by other
signals at around the same redshift, we would detect them\footnote{We
do not study an effect of IR radiation of QSO itself on a total signal.
This contribution should be subtracted by some means in an
analysis of observational data.}.  The signal
predicted in this calculation is $\sim 1$~mK at 0.05 Mpc (11$\arcsec$ in
the case of $z=8.7$) if a neutral \CI\ region exists at the distance.  This is much larger than the background noise measured with
WMAP in a frequency
range of $\nu \ga 100$~GHz of $T_{\rm bg} \la 0.02$~mK mainly
from vibrational dust emission~\citep{gol2010}.  Figure\ \ref{fig14}
shows a simulation of a signal of UV photons emitted in the early
universe at $z=8.7$ through the \CI\ 38K line.  The contour map
of the differential antenna temperatures in units of mK is drawn on the
plane of the angle (arcsec) and the velocity shifts (km s$^{-1}$).  For
this figure it is assumed that a neutral \CI\ region affected by
the UV pumping exists in the region of 0.03~Mpc $\lid r \lid$ 0.06~Mpc
from the point source.

Concerning the detectability, a signal of $\sim 1$~mK
will be seen with the ALMA.  For example, a 1~mK signal emitted through the \CI\
38~K line at redshift $z=6.3$ will be seen by the 12m Array using the Receiver Band of No. 3 for observations
of extended sources at one sigma level if the following parameters are selected:  effective
bandwidth of 5.3~MHz corresponding to the scale of $\sim 0.02$~Mpc,
beamsize of 3$\arcsec$75, and
exposure time of 7.6~hr.  The ALMA Sensitivity
Calculator\footnote{http://www.eso.org/sci/facilities/alma/observing/tools/etc/.}
was used for this estimation.  Signals from \CII\ and \OI\
regions will be detected by future observations to come after the ALMA.

\section{Conclusions}\label{sec5}

The reionization history of the universe at early epoch of redshift $z
\sim 10$ is not know precisely yet.  We study signals of the
cosmological reionization which would have been left on the cosmic
background radiation (CBR) through a series of excitations of fine
structure levels of \Cboth\ and \OI\ and following
emissions (absorption) of line photons.  Since the reionization of the universe would have
proceeded inhomogeneously in space, regions of neutral or low ionization
states irradiated by non-ionizing ultraviolet (UV) photons naturally
exist during the reionization epoch.  Non-ionizing UV photons
interact with C and O to excite them to unstable excited states.  This
excitation followed by decays of the excited states leads to an
excitation of fine structure levels.  Since the UV photons as a source
of the reionization produce signals through fine structure lines in
\CI, \CII\ or \OI\ regions, some information on the
reionization may be obtained by observations for these lines.

Essentially, one ionizing photon can produce one line photon for a fine
structure transition by scattering with C and O species.  Strong signals
are, therefore, emitted at locations near strong sources of UV emission
with large flux of UV photon.  We then assume a strong point source such as quasi-stellar object (QSO) which
starts lighting in the early universe, and calculate the evolution of
ionized region utilizing a non-equilibrium chemical reaction network.  A
rough picture of ionization is shown by this calculation that \Cboth\ and
\OI\ regions irradiated by nonionizing UV line photons can exist at
locations where ionizing UV photons are effectively shielded by dense
\HI\ regions.

There are two classes of UV photons available for the UV pumping of fine
structure levels:

1. photons of frequencies just around the fine structure
lines at the boundary between ionized region and non-ionized region for
\Cboth\ and \OI,

2. photons which are emitted as more energetic photons at point source
and redshifted toward the transition energies of fine structures in
\Cboth\ and \OI\ regions.

At ionization boundaries, UV photons which can excite the fine
structure levels are immediately lost since they are used by the UV
pumping.  It is then predicted that very
small regions emitting line photons energized by UV
sources would exist at ionization boundaries.  However, the signals of
the latter class of UV photons is expected to be stronger than those of
the first.

Outside the ionization boundaries, redshifted UV photons can leave their
signatures possibly over wide region.  The ratios between
UV intensities of respective fine structure lines are changed in the
relaxation process through scattering with the C and O species, and are affected
by the spin states of C and O.  Neutral \HI\ regions near strong
UV sources emit signals of UV photons available until the relaxation is
completed.

Those signals of UV photons could be contaminated by signals of
collisional excitations if densities in observed regions were large.
Regions of intermediate densities might be good sites to investigate for
signals of the reionization since an effective UV pumping is possible in
denser region where an effect of collisional excitation is also larger.

The dependence of magnitudes of signals on the source redshift is shown,
and the detectability of such signals is discussed taking the Atacama Large
Millimeter/submillimeter Array (ALMA) as an example.  Although the
magnitudes of signals depend on physical environments of observed points,
which are roughly described by many fixed parameters in this study,
detections of fine structure line of \Cboth\
and \OI\ might be possible with the ALMA or future projects.

\section*{Acknowledgments}

We are grateful to H. Hanayama, K. Saigo, S. Kondo and
B. Hatsukade for instructive suggestions, and P. Stancil for information on charge transfer reaction coefficients.  We appreciate the
Coordinated Astronomical Numerical Software (CANS) project by R. Matsumoto and T. Yokoyama et al. supported by ACT-JST project
distributing the numerical code of hydrodynamics which we used.  This
research has made use of the VizieR catalogue access tool, CDS,
Strasbourg, France.  This work is supported by Grant-in-Aid for JSPS
Fellows No.21.6817 (Kusakabe) and Grant-in-Aid for Scientific Research from the
Ministry of Education, Science, Sports, and Culture (MEXT), Japan,
No.22540267 and No.21111006 (Kawasaki) and also by World Premier International
Research Center Initiative (WPI Initiative), MEXT, Japan.

\appendix

\section[]{Number fractions of photon ionizing H\,{\sevensize\bf I}, H{\lowercase{e}}\,{\sevensize\bf I} and H{\lowercase{e}\,{\sevensize\bf II}}}\label{appendix}

When a \HII\ region expands, ionization regions of \HII,
\HeIII\ and \HeII\ form in the order of size, smallest
first~\citep{mad1994,mad1997}.  In dense regions of $\delta_{\rm b}=10^3$ and $10^4$, however, the order of ionization fronts is found to be \HeIII,
\HII\ and \HeII\ from the treatment including effects of finite optical depths and recombination processes in ionized regions.  The cross sections for photoionization
reactions are given by
\begin{equation}
\sigma_{\rm H_I}(\nu)=6.30\times 10^{-18}
 \left(\frac{h\nu}{E_1}\right)^{-3}~{\rm cm}^2,
\label{eqa1}
\end{equation}
\begin{eqnarray}
\sigma_{\rm He_I}(\nu)&\hspace{-0.8em}=&\hspace{-0.8em}7.42\times 10^{-18}\nonumber \\
&&\hspace{-0.8em}\left[1.66\left(\frac{h\nu}{E_2}\right)^{-2.05}-
  0.66\left(\frac{h\nu}{E_2}\right)^{-3.05} \right]~{\rm cm}^2,
\label{eqa2}
\end{eqnarray}
\begin{equation}
\sigma_{\rm He_{II}}(\nu)=1.575\times 10^{-18}
 \left(\frac{h\nu}{E_3}\right)^{-3}~{\rm cm}^2,
\label{eqa3}
\end{equation}
where
$E_1=13.60$~eV, $E_2=24.59$~eV and $E_3=54.42$~eV are the threshold
energies of \HI, \HeI\ and \HeII,
respectively~\citep{nak2001}\footnote{Note that their equation (B5)
contains minor typos of wrong signs in indexes.}.

Ionizing photons of energies $E_3 \lid E_\gamma$ can react with \HI, \HeI\ and \HeII.  The probabilities of photons of energy $E_\gamma$ to react with \HI\ and \HeI\ are
\begin{eqnarray}
g_1(E_\gamma)&=&{\rm e}^{-\tau_1} \left[(1-{\rm e}^{-\tau_2}) \frac{n_{\rm H}\sigma_{\rm H_I}}{n_{\rm H}\sigma_{\rm H_I}+ n_{\rm He}\sigma_{\rm He_{II}}} \right.\nonumber\\
&&\hspace{2.9em}\left. + {\rm e}^{-\tau_2} \frac{n_{\rm H}\sigma_{\rm H_I}}{n_{\rm H}\sigma_{\rm H_I}+ n_{\rm He}\sigma_{\rm He_{I}}}\right],
\label{eqa4}
\end{eqnarray}
and 
\begin{equation}
g_2(E_\gamma)={\rm e}^{-\tau_1} {\rm e}^{-\tau_2} \frac{n_{\rm He}\sigma_{\rm He_I}}{n_{\rm H}\sigma_{\rm H_I}+ n_{\rm He}\sigma_{\rm He_{I}}},
\label{eqa5}
\end{equation}
where
$\tau_1 (\nu) \equiv n_{\rm He}\sigma_{\rm He_{II}} (\nu) [r_{\rm I}^{\rm H_I} (t) -r_{\rm I}^{\rm He_{II}} (t-\Delta t_1)]$ is the optical depth between the ionization fronts of \HI\ and \HeII, and $\Delta t_1$ is the interval from the time when the light leaves at $r_{\rm I}^{\rm He_{II}}$ to that when it arrives at $r_{\rm I}^{\rm H_I}$, and satisfies the relation (neglecting the effect of cosmic expansion), i.e, 
\begin{equation}
r_{\rm I}^{\rm He_{II}} (t-\Delta t_1)\approx r_{\rm I}^{\rm H_I} (t) -c \Delta t_1.
\end{equation}
Similarly, $\tau_2\equiv [n_{\rm H}\sigma_{\rm H_I} (\nu) + n_{\rm He}\sigma_{\rm He_{II}} (\nu)] [r_{\rm I}^{\rm He_I} (t) -r_{\rm I}^{\rm H_I} (t-\Delta t_2)]$ is the optical depth between the ionization fronts of \HeI\ and \HI.  $\Delta t_2$ satisfies
\begin{equation}
r_{\rm I}^{\rm H_I} (t-\Delta t_2)\approx r_{\rm I}^{\rm He_I} (t) -c \Delta t_2.
\end{equation}
The reaction rate is proportional to the product of the photon flux, the number density of target and the cross section, i.e., $\propto F(E_\gamma) n_{\rm target} \sigma(E_\gamma)$.  Photons emitted at $r=0$ are absorbed inside the \HI\ ionization front by a factor of ${\rm e}^{-\tau_1}$ [right hand side (RHS) of equation (\ref{eqa4})].  Similarly, the factor, i.e., ${\rm e}^{-\tau_1}{\rm e}^{-\tau_2}$ in equation (\ref{eqa5}), is for absorption inside the \HeI\ ionization front.  Fraction parts in RHSs of the equations show the fractions of photons which are used for ionizations of \HI\ and \HeI, respectively.  The first and second terms in the square bracket of equation (\ref{eqa4}) correspond to the fraction for \HI\ ionization in $r_{\rm I}^{\rm H_I}\lid r < r_{\rm I}^{\rm He_I}$, and $r_{\rm I}^{\rm He_I} \lid r$, respectively, while the term in equation (\ref{eqa5}) is for \HeI\ ionization which occurs only in $r_{\rm I}^{\rm He_I} \lid r$.

The total fraction of photons in this energy range to react with \HI\ is
\begin{equation}
P_3({\rm H_I})=\frac{\int_ {E_3}^\infty \left[L_\nu(E_\gamma)/E_\gamma\right]
 g_1(E_\gamma) dE_\gamma}{\int_ {E_3}^\infty\left[L_\nu(E_\gamma)/E_\gamma\right]
 dE_\gamma},
\end{equation}
where $L_\nu \propto \nu^{-\alpha_S}$ with $\alpha_S=3/2$ is assumed in
this paper.  The fraction to react with \HeI\ is
\begin{equation}
P_3({\rm He_I})=\frac{\int_ {E_3}^\infty \left[L_\nu(E_\gamma)/E_\gamma\right]
 g_2(E_\gamma) dE_\gamma}{\int_ {E_3}^\infty \left[L_\nu(E_\gamma)/E_\gamma\right]
 dE_\gamma},
\end{equation}
The fraction to react with \HeII\ is
$P_3({\rm He_{II}})=1-P_3({\rm H_I})-P_3({\rm He_I})$ accordingly.

Photons of energies $E_2 \lid E_\gamma < E_3$ can react with \HI\ and \HeI.  The probability of photons to react with \HI\ is
\begin{equation}
g_3(E_\gamma)=\left[ 1- \exp(-\tau_3) \right] +\exp(-\tau_3) \frac{n_{\rm H}\sigma_{\rm H_I}}{n_{\rm H}\sigma_{\rm H_I}+ n_{\rm He}\sigma_{\rm He_I}},
\end{equation}
where 
the first term in the right hand side is the probability to react inside the \HeI\ ionization front, and the second is that to react outside.  
$\tau_3\equiv n_{\rm H}\sigma_{\rm H_I} (\nu) [r_{\rm I}^{\rm He_I} (t) -r_{\rm I}^{\rm H_I} (t-\Delta t_3)]$ is the optical depth between the ionization fronts of \HeI\ and \HI.  $\Delta t_3$ is determined by
\begin{equation}
r_{\rm I}^{\rm H_I} (t-\Delta t_3)\approx r_{\rm I}^{\rm He_I} (t) -c \Delta t_3.
\end{equation}
The total fraction of photons in this energy range to react with \HI\ is
\begin{equation}
P_2({\rm H_I})=\frac{\int_ {E_2}^{E_3}\left[L_\nu(E_\gamma)/E_\gamma\right]
 g_3(E_\gamma) dE_\gamma}{\int_ {E_2}^{E_3}\left[L_\nu(E_\gamma)/E_\gamma\right]
 dE_\gamma}.
\end{equation}
The fraction to react with \HeI\ is $P_2({\rm He_I})=1-P_2({\rm H_I})$.

Finally photons of energies $E_1 \lid
E_\gamma < E_2$ can ionize only \HI, and all are used for the
\HI\ ionization.

The fractions of ionizing photons in three energy ranges are $f_1=0.589$
$(E_1 \lid E_\gamma < E_2)$, $f_2=0.286$ $(E_2 \lid E_\gamma < E_3)$
and $f_3=0.125$ $(E_3 \lid E_\gamma)$ under the assumption of $\alpha_S=3/2$.  The fractions to ionize \HI,
\HeI\ and \HeII\ are then estimated as follows: $f_{\rm H_I}=f_1+f_2 P_2({\rm H_I})+f_3 P_3({\rm H_I})$, $f_{\rm He_I}=f_2 P_2({\rm
He_I})+f_3 P_3({\rm He_I})$ and  $f_{\rm He_{II}}=f_3 P_3({\rm He_{II}})$.  Generally the fraction is given by $f_i=\sum_{k=1}^3 f_k P_k(i)$ with $k$ the index for energy ranges.

\bsp

\label{lastpage}

\end{document}